\documentclass[10pt]{article}
\pdfoutput=1
\usepackage{amsmath}
\usepackage{amssymb}

\usepackage[pdftex,breaklinks=true,colorlinks=true,linkcolor=blue,citecolor=blue,urlcolor=blue,
filecolor=blue,pdffitwindow,backref=true,pagebackref=false,bookmarks=true,bookmarksopen=true,
bookmarksnumbered=true]{hyperref}

\usepackage[scaled]{helvet}
\usepackage[T1]{fontenc}
\usepackage{lineno}

\usepackage{microtype}
\DisableLigatures[f]{encoding = *, family = * }

\usepackage{rotating}

\usepackage[labelfont=bf,labelsep=period,justification=raggedright, font=small]{caption}

\usepackage[style=chicago-authordate, 
	natbib=true, 
	sorting=nyt, 
	backend=bibtex, 
	maxbibnames=9,
	minbibnames=3, 
	maxcitenames=2,
	backref=False, 
	url=false, 
	useprefix=true,
	firstinits=true,
	terseinits=true]{biblatex}
\DeclareLanguageMapping{english}{cms-english}

\DeclareNameAlias{default}{last-first}
\DeclareNameAlias{sortname}{last-first}

\renewbibmacro*{name:last-first}[4]{%
  \ifuseprefix
    {\usebibmacro{name:delim}{#3#1}%
     \usebibmacro{name:hook}{#3#1}%
     \ifblank{#3}{}{%
       \ifcapital
         {\mkbibnameprefix{\MakeCapital{#3}}\isdot}
     {\mkbibnameprefix{#3}\isdot}%
       \ifpunctmark{'}{}{\bibnamedelimc}}%
     \mkbibnamelast{#1}\isdot
     \ifblank{#4}{}{\bibnamedelimd\mkbibnameaffix{#4}\isdot}%
     \ifblank{#2}{}{\bibnamedelimd\mkbibnamefirst{#2}\isdot}}
    {\usebibmacro{name:delim}{#1}%
     \usebibmacro{name:hook}{#1}%
     \mkbibnamelast{#1}\isdot
     \ifblank{#4}{}{\bibnamedelimd\mkbibnameaffix{#4}\isdot}%
     \ifblank{#2}{}{\bibnamedelimd\mkbibnamefirst{#2}\isdot}%
     \ifblank{#3}{}{\bibnamedelimd\mkbibnameprefix{#3}\isdot}}}

\defbibheading{bibliography}[\refname]{%
  \section*{#1}%
  \markboth{}{}
}

\makeatletter
\renewcommand{\fnum@figure}{Fig \thefigure}
\renewcommand{\@biblabel}[1]{\quad#1.}
\AtBeginDocument{\toggletrue{blx@useprefix}}
\renewbibmacro*{begentry}{\midsentence} 
\makeatother

\date{}

\usepackage{polski}
\usepackage[english]{babel}
\usepackage[usenames, dvipsnames]{color}
\usepackage{soul}
\usepackage[plain]{fancyref}
\usepackage{colortbl}
\usepackage{tabularx}
\usepackage{lipsum}
\usepackage{float}
\floatstyle{plaintop}
\restylefloat{table}

\Frefformat{plain}{\fancyreffiglabelprefix}{Fig~#1}
\newcommand*{\Frefsecshortname}{Section}%
\Frefformat{plain}{\fancyrefseclabelprefix}{\Frefsecshortname~#1}
\Frefformat{plain}{\fancyrefeqlabelprefix}{Eq~(#1)}

\newcommand{\modelhdr}[3]{%
   \multicolumn{#1}{|l|}{%
     \color{white}\cellcolor[gray]{0.0}%
     \textbf{\makebox[0pt]{#2}\hspace{0.5\linewidth}\makebox[0pt][c]{#3}}%
   }%
}
\newcommand{\parameterhdr}[3]{%
   \multicolumn{#1}{|l|}{%
     \color{black}\cellcolor[gray]{0.8}%
     \textbf{\makebox[0pt]{#2}\hspace{0.5\linewidth}\makebox[0pt][c]{#3}}%
   }%
}
\def\tabspace{0.5ex}

\begin{document}


\begin{flushleft}
{\Large
\textbf{Hybrid scheme for modeling local field potentials from point-neuron networks}
}
\\
\vspace{5mm}
\textbf{Espen Hagen}$^{1,2,\dagger,\ast}$, 
\textbf{David Dahmen}$^{1,\dagger}$, 
\textbf{Maria L. Stavrinou}$^{2}$,
\textbf{Henrik Lind{\'e}n}$^{3,4}$,
\textbf{Tom Tetzlaff}$^{1}$,
\textbf{Sacha J. van Albada}$^{1}$,
\textbf{Sonja Gr{\"u}n}$^{1,5}$,
\textbf{Markus Diesmann}$^{1,6,7}$,
\textbf{Gaute T. Einevoll}$^{2,8,\ast}$
\\
\vspace{2mm}
{\bf{1}} Inst. of Neuroscience and Medicine (INM-6) and Inst. for Advanced Simulation (IAS-6) and JARA BRAIN Inst.~I, J\"{u}lich Research Centre, J\"{u}lich, Germany
\\
{\bf{2}} Dept.~of Mathematical Sciences and Technology, Norwegian University of Life Sciences, 
  {\AA}s, Norway
\\
{\bf{3}} Dept.~of Neuroscience and Pharmacology, University of Copenhagen, 
  Copenhagen, Denmark
\\
{\bf{4}} Dept.~of Computational Biology, School of Computer Science and Communication, 
  Royal Institute of Technology, Stockholm, Sweden
\\
{\bf{5}} Theoretical Systems Neurobiology, RWTH Aachen University, Aachen, Germany
\\
{\bf{6}} Dept.~of Psychiatry, Psychotherapy and Psychosomatics, Medical Faculty, RWTH Aachen University, Aachen, Germany
\\
{\bf{7}} Dept. of Physics, Faculty 1, RWTH Aachen University, Aachen, Germany
\\
{\bf{8}} Dept.~of Physics, University of Oslo, Oslo, Norway
\\
{\bf $\dagger$} Equal contribution
\\
\vspace{2mm}
$\ast$ E-mail: Corresponding authors: \href{mailto:e.hagen@fz-juelich.de}{e.hagen@fz-juelich.de}, \href{mailto:gaute.einevoll@nmbu.no}{gaute.einevoll@nmbu.no}
\end{flushleft}

\section*{Abstract}
\label{section:abstract}

Due to rapid advances in multielectrode recording technology, 
the local field potential (LFP) has again
become a popular measure of neuronal activity in both basic research and clinical applications.
Proper understanding of the LFP requires detailed mathematical modeling incorporating 
the anatomical and electrophysiological features of neurons near the recording electrode,
as well as synaptic inputs from the entire network. 
Here we propose a hybrid modeling scheme combining the efficiency of 
commonly used simplified
point-neuron network models with the biophysical principles underlying LFP generation by real neurons.
The scheme can be used with an arbitrary number of point-neuron network populations.
The LFP predictions rely on populations of network-equivalent, anatomically reconstructed  multicompartment neuron models with layer-specific synaptic connectivity.
The present scheme allows for a full separation of the network dynamics simulation and LFP generation.
For illustration, we apply the scheme to a full-scale cortical network model for a $\sim$1~mm$^2$ patch of primary visual cortex and predict laminar LFPs for different network states, assess the relative LFP contribution from different laminar populations, and investigate the role of synaptic input correlations and neuron density on the LFP. 
The generic nature of the hybrid scheme and its publicly available implementation in \texttt{hybridLFPy} 
form the basis for LFP predictions from other point-neuron network models, as well as 
extensions of the current application to larger circuitry and additional biological detail.

\section*{Author summary}
\label{section:author}

The recording of extracellular potentials inside the brain is among the most commonly used measures of neural activity.
 While the high-frequency part of the signal measures neural action potentials, the low-frequency part (local field potential, LFP)  carries information from thousands of neurons and is difficult to interpret. 
The interpretation of the LFP has been hampered by the lack of a good `forward modeling scheme', 
that is, a scheme providing a link between activity in candidate network models and the resulting LFP signal. 
While many models of neural network dynamics are based on simplified point neurons such as the leaky integrate-and-fire (LIF) neuron model, 
point neurons do not generate LFPs per se.  

Here we describe a new hybrid modeling scheme overcoming this limitation, where the network spiking dynamics is modeled by means of point-neuron networks, 
while the LFP is subsequently computed from the resulting spike trains according to the biophysical principles underlying LFP generation in real neurons.
For illustration, we apply the scheme to a full-scale cortical network model 
of a 1~mm$^2$ patch 
of primary visual cortex comprising 78,000 neurons and explore how the different cortical populations contribute to the LFP and how the signal depends on network state and other system properties.


\section{Introduction}
\label{sec:introduction}

The local field potential (LFP), the low-frequency component
($\lesssim$500~Hz) of the extracellular potential recorded in
the brain, is commonly used as a measure of neuronal activity
\citep{Buzsaki2012, Einevoll2013}. 
The LFP originates from transmembrane currents \citep{Nicholson1975},  and
at the single-cell level the biophysical origin of such extracellular potentials is well understood (see, e.g.,
\citet{Rall1968,Holt1999,Buzsaki2012,Einevoll2013}). 
However, the interpretation of the LFP remains  difficult due to the large number of neurons
contributing to the recorded signal. In neocortex, for example, the measured
LFP is typically generated by thousands or even millions of neurons  near the recording electrode
\citep{kajikawa11_847, Linden2011, Leski2013}.
Moreover, the LFP reflects synaptic input also generated by remote populations, e.g., 
inputs from other cortical or subcortical areas in addition to local network interactions~\citep{Herreras2015}.
A thorough theoretical description of the LFP therefore needs to account not only for the
anatomical and electrophysiological features of neurons in the
vicinity of the recording electrode, but also for the entire large-scale 
neuronal circuitry generating synaptic
input to these cells.

Modeling large-scale neural-network dynamics with individual spiking neurons 
is challenging due to the memory required to represent the large number of synapses.  
With current technology and using the largest supercomputers available today,
simulations of neural networks comprising up to $10^9$ neurons and
$10^{13}$ synapses (roughly corresponding to the size of a cat brain)
are feasible for simplified model neurons~\citep{Diesmann13_8,Kunkel2014}.
Typically, these simplified
models neglect the spatial aspects of neuronal morphologies and
describe neurons as points in space (point-neuron models).  Despite
their simplicity, point-neuron-network models explain a variety of
salient features of neural activity observed \emph{in vivo}, such as
spike-train irregularity \citep{Softky1993,Vreeswijk1996, Amit1997, Shadlen1998},
membrane-potential fluctuations \citep{Destexhe1999}, asynchronous
firing \citep{Ecker2010,Renart2010,Ostojic2014}, correlations in neural activity \citep{Gentet2010,Okun2008,Helias2013}, 
self-sustained activity \citep{Ohbayashi2003a,Kriener2014} 
and realistic firing rates across laminar cortical populations \citep{Potjans2014}.
Point-neuron networks are amenable to mathematical analysis (see, e.g.,
\citet{Brunel2000,Deco2008,Tetzlaff2012,Helias2013,Kamps2013,Schuecker2015})
and can be efficiently evaluated numerically
\citep{Plesser07_672,Brette07_349, Helias2012, Kunkel2014}. 
The mechanisms governing networks of biophysically detailed multicompartment model neurons, in contrast, 
are less accessible to analysis and these models are more prone to overfitting.
Existing multicompartment neuron network models accounting for realistic cell
morphologies are restricted to sizes of $\sim10^4$--$10^5$ neurons
\citep{Hines2008j,Reimann2013,Migliore2014,Markram2015}.
Large-scale models are, however, necessary to include 
contributions to the LFP from distant populations
in situations where the spatial reach of the LFP is known to be large 
\citep{Linden2011, Leski2013}.  

Although point-neuron networks capture many features of \emph{in vivo} spiking
activity, they fail to predict extracellular potentials that 
result from transmembrane currents
distributed across the cell surface.  According to Kirchhoff's law of
current conservation, the sum of all transmembrane currents, including
all ionic and capacitive currents, must be zero for each neuron.
In a point-neuron model, all transmembrane currents are collapsed in
a single point in space.  The net transmembrane current, and hence the
extracellular potential, therefore vanishes.  Only the spatial
separation between current sinks and sources leads to a non-zero
extracellular potential~\citep{Pettersen2012,Einevoll2013}. 
\emph{A priori}, the prediction of extracellular
potentials therefore requires spatially extended neuron models
accounting for the spatial distribution of transmembrane 
currents, commonly handled using multicompartment neuron models \citep{DeSchutter2009}. 
In several previous studies
\citep{Bazhenov2001,Hill2005,Ursino2006,Mazzoni2008,Mazzoni2010,Mazzoni2011},
the activity of point-neuron networks (e.g.~population firing rates,
synaptic currents, membrane potentials) has nevertheless been used as
a proxy for the LFP when comparing with experiments. 
In a recent study comparing different candidate proxies, it was found that a suitably chosen
sum of synaptic currents could provide a good LFP proxy, but only for the case when the LFP is generated
from transmembrane currents of 
a single population of pyramidal neurons~\citep{Mazzoni2015}. 
In cortex, however, several populations 
in general contribute to the LFP, and there are spatial cancellation effects when positive LFP contributions from one population overlap in space
with negative LFP contributions from other populations. 
This effect cannot be accounted for by a simple LFP proxy.

\begin{figure}[!h]
\includegraphics[width=\textwidth]{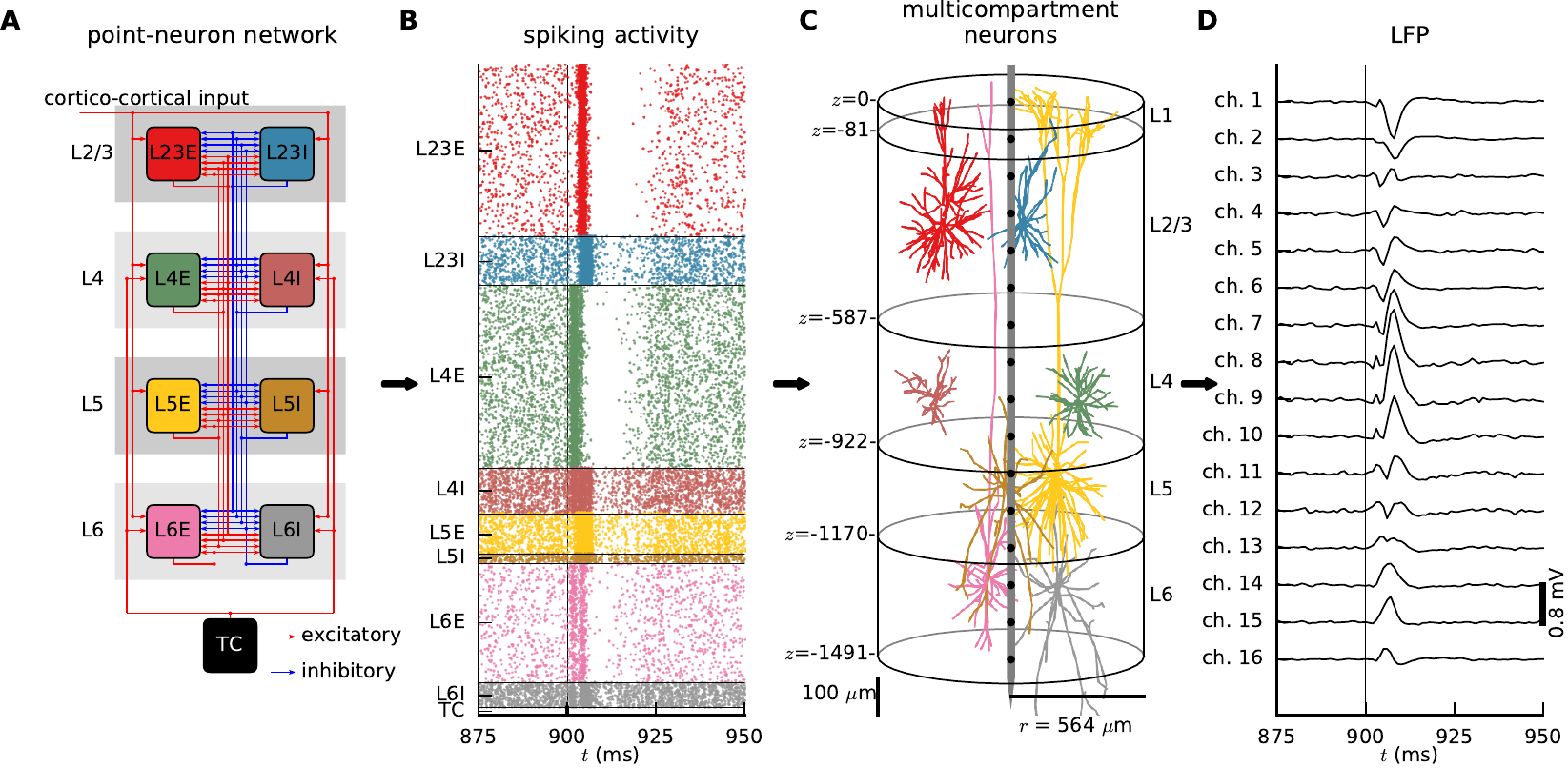}
\caption{\textbf{Overview of the hybrid LFP modeling scheme for  a cortical microcircuit model.} 
  \textbf{A)}~Sketch of the point-neuron network representing a
  1~mm$^2$ patch of early sensory cortex (adapted from \citet{Potjans2014}). The
  network consists of 8 populations of leaky integrate-and-fire (LIF) neurons, representing
  excitatory (E) and inhibitory neurons (I) in cortical layers 2/3,
  4, 5 and 6.  External input is provided by a population of
  thalamo-cortical (TC) neurons and cortico-cortical afferents.  The
  color coding of neuron populations is used consistently throughout
  this paper.  Red arrows: excitatory connections. Blue arrows:
  inhibitory connections.
  See \Fref{tab:1}-\ref{tab:2}, \ref{tab:5}-\ref{tab:6} for details on the network model.
  \textbf{B)}~Spontaneous ($t<900$~ms) and stimulus-evoked spiking
  activity (synchronous firing of TC neurons at time $t=900$~ms, denoted by thin vertical line)
  generated by the point-neuron network model shown in panel A, sampled
  from all neurons in each population.  Each dot represents the
  spike time of a particular neuron.
  \textbf{C)}~Populations of LFP-generating multicompartment model neurons  
with reconstructed, layer- and cell-type specific morphologies. Cells are distributed within a
  cylinder spanning the cortex. Layer boundaries are marked by
  horizontal black lines (at depths $z$ relative to cortex surface
  $z=0$).  Only one representative neuron for each population is
  shown (see Fig.~4 for a detailed overview of cell types and morphologies). 
  Sketch of a laminar recording electrode (gray) with 16 
  contacts separated by $100~\mu$m (black dots).
  \textbf{D)}~Depth-resolved LFP traces predicted by the model (cf. \Fref{tab:3}-\ref{tab:4}).
  Note that channel 1 is at the pial surface, so that channel 2 corresponds to 
  a cortical depth of 100~$\mu$m and so forth.
}
\label{fig:1}
\end{figure}

In this article, we present a hybrid modeling scheme which combines the simplicity and efficiency of point-neuron
network models and the biophysical principles underlying LFP generation captured by multicompartment neuron models with 
anatomically reconstructed morphologies. 
The scheme allows for arbitrary numbers of LFP-contributing populations, 
and directly incorporates spatial cancellation effects. 
Further, the spatially extended LFP-generating neurons assure that effects from intrinsic dendritic filtering of synaptic inputs are included in the predicted 
LFP~\citep{Linden2010}. 
The scheme assumes that the spiking activity of the
neural network (\Fref{fig:1}B) generating the synaptic input reflected
in the LFP is well described by a point-neuron network model
(\Fref{fig:1}A).
The network spiking activity serves as synaptic input to a population of
mutually unconnected multicompartment model neurons with realistic
morphologies positioned in 3D space 
(\Fref{fig:1}C) and is thereby
translated into a distribution of transmembrane currents and, hence,
an LFP (\Fref{fig:1}D). Thus each multicompartment model neuron
has its equivalent in the point-neuron network and receives input spikes from 
the same presynaptic neurons as this point-neuron equivalent.

In the proposed hybrid modeling scheme, the LFP stems from the presynaptic spiking activity,
but does not affect the spike-generation dynamics. 
Thus, the modeling of the spike trains and the LFP generation are
separated so that 
the effects of the spatial and electrophysiological properties of the postsynaptic (multicompartment) neurons on the LFP
can be investigated independently of the spike-generation dynamics.
Due to the linearity of Maxwell's equations 
and  volume conduction theory linking transmembrane currents to extracellular 
potentials~\citep{Pettersen2012,Einevoll2013},
the compound LFP results from the linear superposition of
all single-cell LFPs generated by the collection of neurons in the
multicompartment model neuron population~\citep{Einevoll2013a}. 
Note that this linear
superposition principle applies even for nonlinear cell
dynamics (e.g.,~nonlinear synaptic integration, action-potential
generation, active conductances) as in \citep{Reimann2013}. 
As ephaptic interactions \citep{Anastassiou2011} are neglected, the 
LFP contribution from each multicompartment model neuron can be treated
independently from the others.
The computational hybrid LFP scheme proposed here exploits
the methodological and conceptual advantages due to
the independence of the contributions to the LFP from each multicompartment model neuron:  
The evaluation of the LFPs becomes
`embarrassingly parallel' (see \citealp{Foster1995}) and
simulations of the multicompartment model neuron dynamics can be easily
distributed in parallel across many compute units (i.e., CPUs). 
Although tailored towards use on high-performance computing facilities, 
the hybrid simulation can in principle be run on a single laptop.

The hybrid scheme 
predicts spatially and temporally resolved neural
activity at various scales: spikes, synaptic currents,
membrane potentials, current-source densities (CSD, see e.g.,
\citet{Nicholson1975,Pettersen2006, Pettersen2008}), and LFPs. It therefore allows for investigation
of relationships between different measures of neural activity.
Thus, although point-neuron networks until now only have connected to \emph{in vivo} experiments via measurement of spikes, 
single-neuron membrane potentials and currents,
the present hybrid scheme allows for comparison of model predictions also with measured LFPs (and associated CSDs). 

As an illustration, we apply the hybrid scheme to a multi-layered
point-neuron network model of an early sensory cortical microcircuit
\citep{Potjans2014}. 
We thereby demonstrate how to obtain LFP
predictions from point-neuron network models using additional
spatial connectivity information from anatomical data \citep{Binzegger2004, Izhikevich2008}. 
The example illustrates how the hybrid scheme can be used to
examine the relation between single-neuron and population signals, i.e., spikes and LFPs,
the effect of network dynamics on the LFP, and the interpretation of the LFP in
terms of underlying laminar neuron populations. We further use the example to
demonstrate that synaptic-input correlations result in a non-trivial dependence of the LFP on the neuron density.
Correct LFP predictions can therefore only be obtained by accounting
for realistic neuron densities.

The network model of \citet{Potjans2014} is chosen here since it has a minimum level of detail in the sense that individual neurons have simplified leaky integrate-and-fire (LIF) dynamics,  but still represents a cortical column with full density of neurons and connections. The connectivity in such a full-scale circuit alone
suffices to explain realistic firing rates across populations as well as propagation of activity through layers \citep{Potjans2014}. 
Applicability of the scheme is, however, not restricted to this model 
as it in principle can be used for all network models generating spikes.

In  Methods (\Fref{sec:methods}), we detail the components of the hybrid
scheme and their application to the cortical microcircuit model:
The point-neuron-network model (\Fref{sec:2.1}), 
the populations of multicompartment neurons (\Fref{sec:2.2}), the synaptic
connectivity of the point-neuron network and the
multicompartment model neuron populations (\Fref{sec:2.3}), and the
biophysical forward-modeling scheme of extracellular potentials
(\Fref{sec:2.4}). 
We further describe the analysis of the  data generated by the simulations, as well
as the \texttt{hybridLFPy} software implementation
(\Fref{sec:2.5}). 
In Results (\Fref{sec:results}), we apply the hybrid
scheme to the cortical microcircuit model of Potjans \& Diesmann
\citep{Potjans2014} 
and study the effects of network dynamics on the LFP, 
the contributions of individual cortical subpopulations to the LFP, 
the role of correlations and neuron density, and how well the LFP can be predicted from
population firing rates (rather than from individual spikes).
In Discussion (\Fref{sec:discussion}), 
we outline implications of our work, and in particular future applications and extensions of the hybrid LFP modeling scheme.


\section{Methods: Hybrid LFP modeling scheme}
\label{sec:methods}

\subsection{Point-neuron network model}
\label{sec:2.1}

The point-neuron network model is a key component of the hybrid scheme. 
The hybrid scheme enables LFP predictions from network models
with an arbitrary number of populations and thus permits application to a large class of networks
with arbitrarily complex single-neuron and synapse dynamics. The example network of 'spike-generators' used here is,
except for some minor adjustments (see below), the multi-layered model of a cortical microcircuit published by \citet{Potjans2014}.
The model is implemented and included in \texttt{NEST} (\url{http://www.nest-simulator.org}, \citet{eppler_2015_32969}) and was recently made freely available (\url{http://www.opensourcebrain.org/projects/potjansdiesmann2014}).

The network model describes $1$ mm$^2$ of primary sensory cortex and 
consists of four layers with one excitatory (E) and one inhibitory (I) neuron
population each, as illustrated in \Fref{fig:1}A. 
The network receives modulated thalamic input in addition to stationary external input. 
Whereas the neuron (leaky-integrate-and-fire) and synapse (static, exponential-current-based) model 
are intentionally left simple, the focus of this network implementation is 
on the complex connectivity (see \Fref{sec:2.3}) 
which integrates multiple sources of anatomical and electrophysiological data \citep[and references therein]{Potjans2014}.
Apart from the layer identity, the model does not explicitly account for cell positions.
For the full network description, see \Fref{tab:1},\ref{tab:2} and \ref{tab:5}. 
The microcircuit model reproduced experimentally observed distributions of firing rates across populations
and propagation of activity across layers \citep{Potjans2014}. It thus forms a suitable starting point for LFP predictions in a cortical column.

The stationary thalamic Poisson input and cortico-cortical input to the microcircuit present in the original model of ~\citet{Potjans2014}, 
are here replaced by DC currents. DC input slightly increases the degree of synchrony (see, e.g., \citet[Fig. 5]{Brunel2000}), 
but retains network dynamics and firing rate distributions
across populations as in \citet{Potjans2014}. 
The Potjans \& Diesmann network shows slightly synchronous behavior due to
the E-I network of layer 4 being close to the synchronous irregular (SI) regime \citep{Brunel2000}.  
In order to reduce synchrony, we here increased the average synaptic weight from neurons in population L4I (inhibitory) to L4E 
(excitatory) neurons by 12.5\%,
resulting in attenuated oscillations in layer 4. 
Taking advantage of the fact that point-neuron networks are amenable for theoretical analysis, we derived
modified weights based on predictions from dynamical mean-field theory applied to the microcircuit model \citep{Bos2015}.
Moreover, we found that high-frequency network oscillations seen for Gaussian synaptic weight distributions are reduced 
when using lognormally distributed synaptic weights~\citep{Sarid2007, Iyer2013, Teramae2014}. 
This  made 
the dynamics more similar to experimental observations \citep{Song2005, Buzsaki2014}, and we thus also chose this for our network.
Henceforth, we refer to our modified network as the `reference network'.
Modulated activity of each thalamo-cortical (TC) neuron in the external thalamic population
was modeled as synchronous spikes or as independent 
non-stationary Poisson processes with sinusoidally oscillating rate profiles (cf. \Fref{eq:ac_modulation}).

\subsection{Populations of multicompartment model neurons}
\label{sec:2.2}
\begin{figure}[!h]
\includegraphics[width=\textwidth]{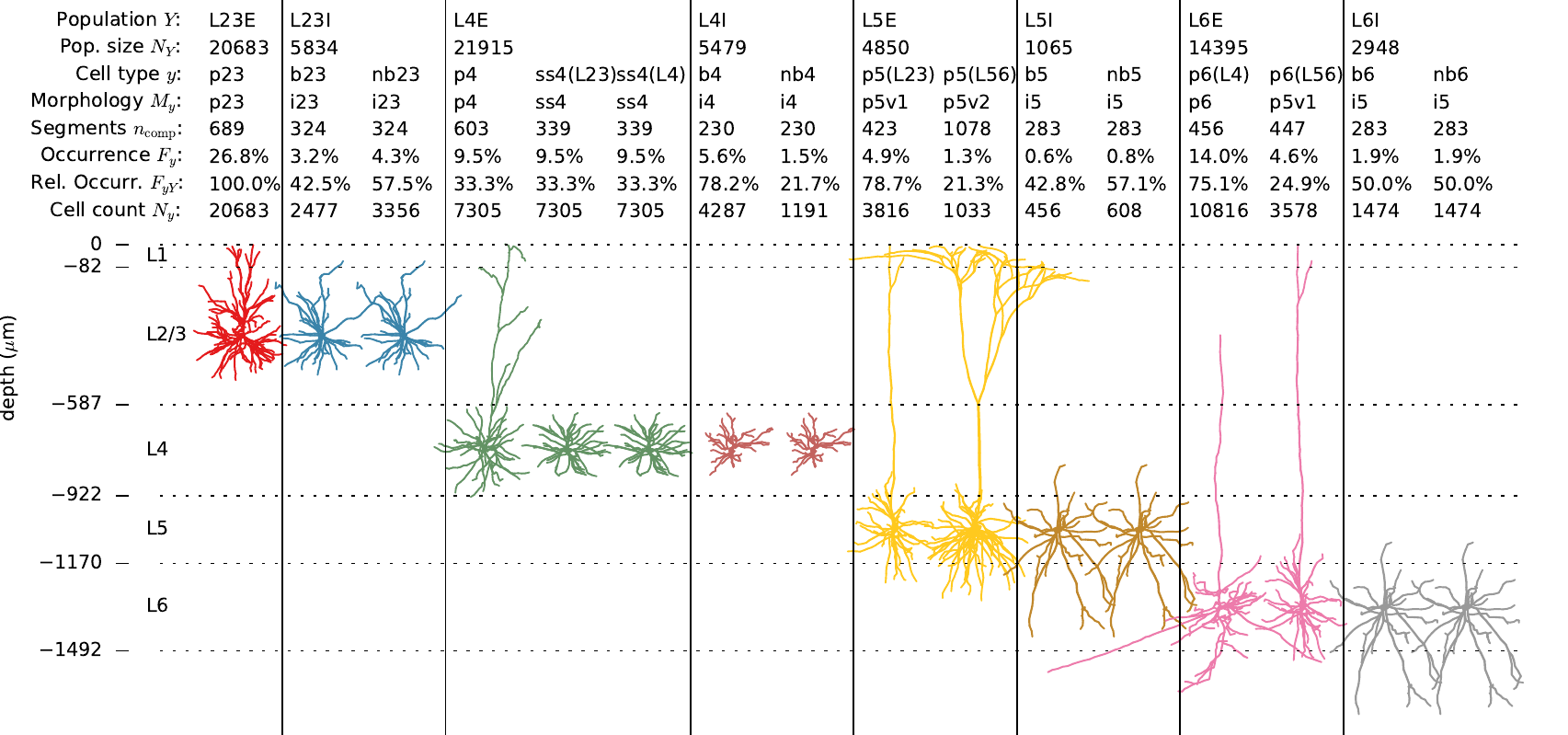}
\caption{\textbf{Cell types and morphologies of the multicompartment-neuron populations.}
The 8 cortical populations $Y$ of size $N_Y$ in the  microcircuit network model are represented by 16 subpopulations of cell type $y$
with detailed morphologies $M_y$ \citep{Binzegger2004, Izhikevich2008}.
Neuron reconstructions are obtained from cat visual cortex and cat somatosensory cortex  
(source: NeuroMorpho.org \citep{Ascoli2007}, \citet{Contreras1997, Mainen1996, Kisvarday1992, Stepanyants2008}, cf. \Fref{tab:7}). 
Each morphology $M_y$ is here shown in relation to the layer boundaries (horizontal lines). 
Colors distinguish between network populations as in \Fref{fig:1}. 
The number of compartments ($n_\text{comp}$), frequencies of occurrence ($F_y$), relative occurrence ($F_{yY}$) and cell count ($N_y$) is given for each cell type $y \in Y$.
}
\label{fig:4}
\end{figure}

Cancellation effects from positive and negative contributions to extracellular potentials and effects of intrinsic dendritic filtering can only be captured with spatially extended multicompartment neuron models \citep{Einevoll2013}. 
In the hybrid scheme, extracellular potentials are estimated 
from the spiking activity in the point-neuron network (cf. \Fref{sec:2.1}) through 
synaptic activation of populations of multicompartment model neurons (`LFP generators'). 
In principle, these mutually unconnected model neurons mirror their network counterparts and receive inputs from exactly the same point neurons. 
In addition to the description of the point-neuron network model, 
different types of spatial information are thus needed to predict LFPs. 
For one, detailed dendritic morphologies are required for each individual network population (\Fref{fig:4}).
Further, the positions of neurons and synaptic connections must also be specified, 
as well as the separation of network populations into morphologically distinct cell types
(\Fref{fig:4} and \ref{fig:3}).

Availability of detailed cell-type specific connectivity of neural circuits,
especially including information about synapse positions, is limited due to the 
substantial experimental effort involved. 
However, several ongoing large-scale neuroscience projects \citep{Kandel2013} address this issue and detailed connectomes are beginning to become publicly available \citep{Jiang2015, Reimann2015, Markram2015}.
In the present example application we used the connection probabilities as given by \citet{Izhikevich2008} derived from \citet{Binzegger2004}. 
Note that the point-neuron network connectivity was partially derived from the same data \citep{Potjans2014}.
Quantitative data was provided for the number of connections in five cortical layers 
(layer 1 (L1), layers 2 and 3 grouped into a joint layer 2/3 (L2/3), and layers 4 (L4), 5 (L5) and 6(L6)) 
between 17 cortical cell types, 
cortico-cortical connections from other areas, and two thalamo-cortical relay cell types.
We follow the nomenclature  of \citet{Izhikevich2008}, 
where $y=\text{p23}$ denotes pyramidal cell types in layer 2/3, $y=\text{b23}$ 
and $y=\text{nb23}$ basket interneurons and non-basket interneurons within the same layer, 
$y=\text{ss4(L23)}$ spiny stellate cells in layer 4 with targets mainly within layer 2/3,
$y=\text{p4}$ layer 4 pyramidal cells and so forth. 
Out of the 17 covered intracortical cell types 
only the $y=\text{nb1}$ cell type is not associated with any point-neuron network population in our scheme.
To account for the lack of layer 1 in our model, we renormalized the connection probabilities
for the remaining 16 cortical cell types including the two thalamo-cortical (TC) relay-cell types,  such that the occurrences $F_y$ of all cell types $y$ summed to 100\% 
as given in \Fref{tab:8}.
Further, we assumed that the excitatory point-neuron network populations within one layer are composed of pyramidal cells and spiny stellate cells if 
both are present in the layer, and that inhibitory network populations encompass both types of interneurons.
This results in the grouping of cell types $y$ into postsynaptic populations $Y$ illustrated in \Fref{fig:4}.
The neuron count $N_y$ of each cell type is then trivially computed from the frequency of occurrence $F_y$ as given in \Fref{tab:8} and \Fref{fig:4}. 

Inclusion of cell-type and layer-specific connections in the present hybrid scheme
has some implications for 
how we proceed with setting up equivalent populations consisting of morphologically detailed model neurons. 
Different cell types belonging to a particular population may have different spatial distributions of synapses, 
or the populations may consist of different morphological classes of neurons \citep{Kisvarday1992, Nowak2003, Stepanyants2008}. 
An example is layer 4  in which spiny stellate cells lack apical dendrites, 
while pyramidal cells have apical dendrites extending into layer 1. 
To incorporate some of this morphological diversity we considered altogether 16 cell types for the
8 cortical network populations. These cell types are described below, 
with cell- and layer-specific connectivity derived in \Fref{sec:2.3}.

For each of the 16 cell types, 
we acquired representative morphological reconstructions of predominantly cat visual cortex neurons from several sources \citep{Contreras1997, Kisvarday1992, Mainen1996, Stepanyants2008} (cf. Fig. \ref{fig:4}, Tab. \ref{tab:7})). 
Morphology files were obtained either from 
\href{http://neuromorpho.org}{NeuroMorpho.org} 
\citep{Ascoli2007} or through personal communication with the authors. 
Constrained by layer boundary depths 
\citep[Fig. 3]{Stepanyants2008} and laminar connectivities (cf. \Fref{sec:2.3}) we applied an intermediate preprocessing step to our pyramidal cell morphologies. 
Assuming that the soma compartments of each cell type were centered in their corresponding layer, 
and noting that the layer-specific connectivity (cf. \Fref{tab:8}) implies connections to layer 1,
we stretched the apical dendrites along the axis perpendicular to the cortical surface such that they reached the pial surface. 
The only exception was the p6(L4) morphology, 
which we extended to reach the center of layer 2/3 in accordance with \citet{Stepanyants2008} and the observation that
Tab. \ref{tab:8} predicts zero connections within layer 1 and very few connections in layer 2/3
to the p6(L4) cell type. 
Due to lack of available morphologies of sufficient reconstruction quality,  
certain cell types were represented by the same neuron morphology. 
Interneuron types and spiny stellate cells in a given layer
shared morphologies, 
the same interneuron morphology was reused in both layer 5 and 6, 
and finally the p5(L23) and p6(L56) cell morphology were similar except for the stretching of the apical dendrites.

Preserving the laminar cell density under 1~mm$^2$ surface area
of the point-neuron network model, 
we created for each postsynaptic cell type $y$ model populations where
somas were assigned random locations within cylindrical slabs with radius $r$=564~$\mu$m and thickness $h$=50~$\mu$m, 
each centered in their respective layer (illustrated in \Fref{fig:1}D).
Regardless of the vertical offset of the soma of pyramidal cells, 
postsynaptic target dendrites were therefore present 
within the $\sim$80~$\mu$m thick \citep{Stepanyants2008} uppermost layer 1 
except for cell type p6(L4).
For simplicity, each cell type was represented by a single reconstructed morphology in the present application.
The full specification of the populations is 
given in \Fref{tab:3}.

Each neuron is modeled using the
multicompartmental, passive cable formalism \citep{Rall1964, Rall2009, DeSchutter2009}, 
describing the changes in membrane voltage and the associated transmembrane currents throughout all parts of the neuron geometry (cf. \Fref{sec:2.4} and \Fref{tab:4}). 
We used (non-plastic) exponential current-based synapses as in the point-neuron network model (cf. \Fref{sec:2.1} and \Fref{tab:2}).
\Fref{tab:4}-\ref{tab:6} summarize parameters relevant for the synapse models and passive parameters of the multicompartment models.

\subsection{Spatial synaptic connectivity}
\label{sec:2.3}

A full description of the connectivity in networks of multicompartment model neurons requires a 3-dimensional (3D) representation, 
for example in the form of sparse $N_X \times N_Y \times n_\text{comp}$ matrices 
of synaptic weights and spike-transmission delays between presynaptic neurons $i\in[1,N_X]$ and compartments $n\in[1,n_\text{comp}]$ of postsynaptic cells $j\in[1,N_Y]$. 
Here, $N_X$ and $N_Y$ denote the number of pre- and postsynaptic neurons in population $X$ and $Y$, respectively, and $n_\text{comp}$ the number of compartments of the postsynaptic cell.
In point-neuron networks, in contrast, connectivity is by definition only 2-dimensional (2D) as the cell morphology is collapsed into a single point and, consequently, 
the specificity of synapse locations on the postsynaptic morphology is ignored.
In the proposed hybrid modeling scheme, the connectivity within the point-neuron network is consistent with the connectivity between point neurons and multicompartment model neurons. 
Ideally, each multicompartment model neuron has its equivalent in the point-neuron network and receives inputs from exactly the same presynaptic sources as its point-neuron counterpart. 
Synapses should be positioned on the dendritic tree according to anatomical data, and synaptic weights and time constants should be adapted such that the
somatic membrane potential or somatic current match the point-neuron counterparts. 
Such mapping between point neurons and passive multicompartment neurons is feasible~\citep{Koch1985,Wybo2013,Wybo2015}.

In the current application of the hybrid scheme to the cortical microcircuit model, 
we make the simplest approximation to the mapping problem and
fixed the current amplitudes $I_{ji,\text{max}}$ and synaptic time constants as in the network model, 
with compartment specificity of connections dependent on compartment surface area (see \Fref{tab:4}).
We further preserve only the statistics of connections (average number of inputs, distribution of spike-transmission delays) for each pair of pre- and postsynaptic neuron populations, 
exploiting that connections between network populations are drawn randomly with fixed probabilities. 
Finally, we simplify the positioning of synapses to a layer-specificity of connections.
The activation times of each synapse are then 
given by the spike train of a randomly drawn point neuron in the network model, 
with random delays consistent with the delay distribution in the network (\Fref{tab:2} and \ref{tab:4}).

\begin{figure}[!h]
\includegraphics[width=0.33\textwidth]{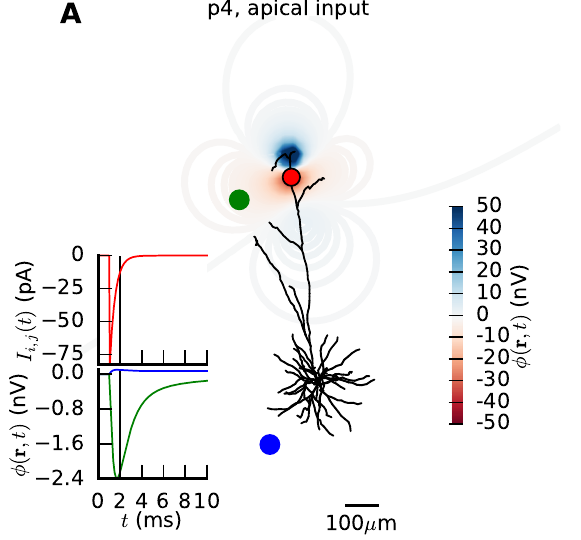}
\includegraphics[width=0.33\textwidth]{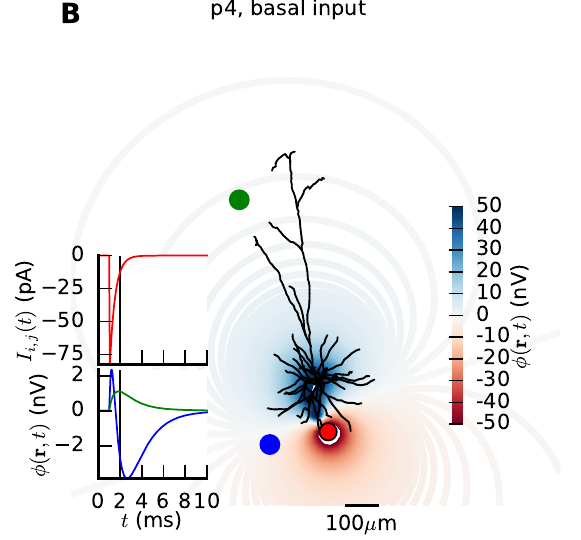}
\includegraphics[width=0.33\textwidth]{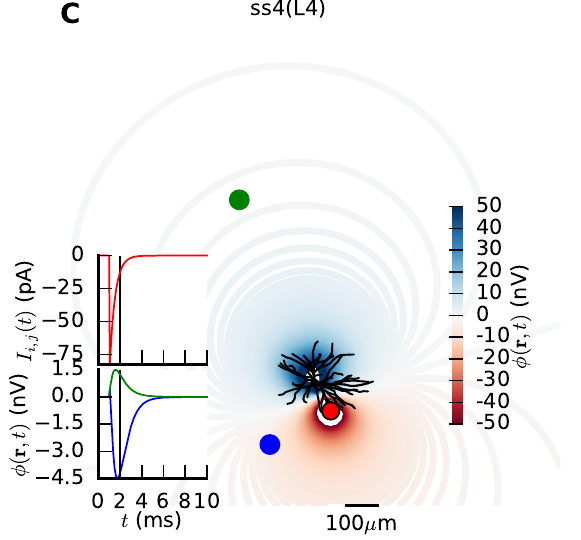}
\caption{
\textbf{Example LFP responses from single-synapse activations of layer 4 neurons.}
\textbf{A})~Illustration of the non-trivial relationship between  apical synaptic input (red circle) 
onto a reconstructed morphology (black) of a pyramidal cell in layer 4 and the corresponding
extracellular potential. The exponential synaptic input current $I_{i, j}(t)$ (upper inset) 
results in deflections in the extracellular potential $\phi(\mathbf{r},t)$ 
here shown as time courses at two locations in proximity to the input site and the basal dendrites 
(green and blue circle, respectively; lower inset).
The color-coded isolines show the magnitude of the scalar extracellular potential at $t=2$~ms (vertical black line in insets)  in the vicinity of the cell. Negative signs are indicated by dashed lines.
\textbf{B})~Same as in panel A, however with the synaptic input current relocated 
to a basal dendrite,
resulting in an extracellular potential with a different spatiotemporal signature 
less dependent on the geometry of the apical dendritic tree. 
At the location denoted by the blue circle, the extracellular potential 
changes sign with time due to interactions between signal propagation in the passive model neuron and volume conduction.
\textbf{C})~Same as panels B and C for a spiny stellate cell in layer 4 receiving an excitatory synaptic input on a basal dendrite.  
}
\label{fig:2}
\end{figure}

In \Fref{sec:2.3.1}, we first show how to derive a 2D point-neuron connectivity from a given 3D multicompartment-neuron connectivity and describe the case where the complexity of the point-neuron network is further reduced by pooling cell types. 
In \Fref{sec:2.3.2} we describe the opposite procedure, connecting an existing (published) point-neuron network with a predefined 2D connectivity to a population of multicompartment model neurons such that the resulting 3D connectivity is as consistent as possible with anatomical datasets accounting for the compartment specificity of connections (for example, the layer specificity of connections as in the anatomical data published by \citet{Binzegger2004}).
The procedures outlined below, allow a reduction of complexity within the point-neuron network while accounting for the full diversity in cell types and synapse locations for multicompartment-neuron populations which is essential for predicting extracellular potentials~(\Fref{fig:2}).

\subsubsection{Construction of point-neuron network connectivity}
\label{sec:2.3.1}

\begin{figure}[!h]
\centering
\includegraphics[width=\textwidth]{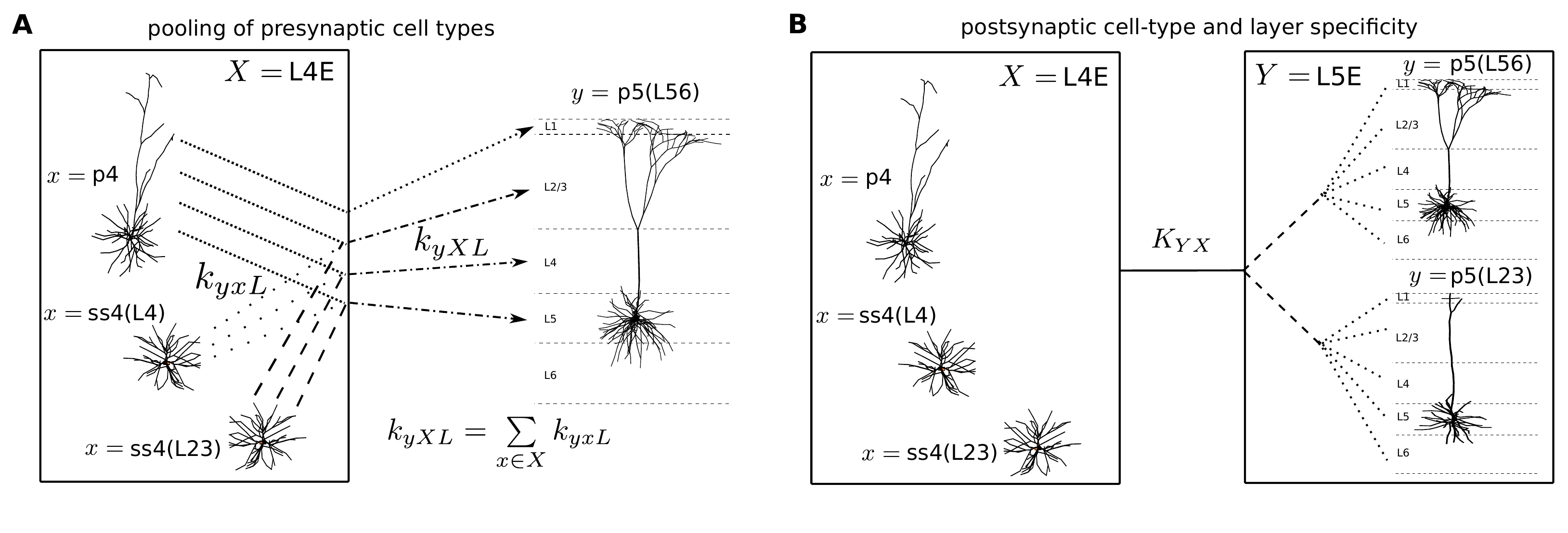}
\caption{
\textbf{Constructing spatial synaptic connectivity for the cortical microcircuit model.} 
\textbf{A)}~Illustration of pooling of presynaptic cell types.
Presynaptic populations $X$ in the point-neuron model (left box; here $X=$~L4E) consist of multiple cell 
types $x$ (here $x\in\{$p4, ss4(L4), ss4(L23)$\}$).
The layer-specific number of synapses $k_{yXL}$ (dash-dotted lines) formed between one cell of 
postsynaptic cell type $y$ (right part of panel A: morphology projected onto cortical layers 1--6; here $y=$~p5(L56)) 
and a presynaptic population $X$ is given by the sum of all individual cell-type resolved 
synapse counts $k_{yxL}$ (dotted or dash-dotted lines).
\textbf{B)}~Bi-directional cell- and layer-specific pooling and dispersing of synapses between pre- and postsynaptic cell types. 
Postsynaptic populations $Y$ (right box; here $Y=$~L5E) in the point-neuron model consist of multiple cell types 
$y$ (here $y\in\{$p5(L56), p5(L23)$\}$). A given presynaptic population
$X$ (left box; here $X=$~L4E) containing cell types $x$ (here $x\in\{$p4, ss4(L4), ss4(L23)$\}$)
forms cell-type and layer-specific connections within $Y$ (black connection tree).
For the number of synapses $K_{yXL}$ between population $X$ and cells of type $y$ in layer $L$ 
(right-most branching of connection tree) 
the synapse count $K_{YX}$ between all cells in $X$ and $Y$ 
can be obtained by pooling all synapses onto cell types $y\in~Y$
and input layers $L$. 
Conversely,  for a given total number of synapses $K_{YX}$ between all cells in 
$X$ and $Y$, the number of synapses $K_{yXL}$ onto a specific cell type $y$ and layer $L$
can, as described by \Fref{eq:KyXL2}, be obtained by calculating the 
cell-type and layer specificity of connections $\mathcal T_{yX}$ and 
$\mathcal L_{yXL}$ (see \Fref{fig:5}) from anatomical data (\Fref{tab:7}).
}
\label{fig:3}
\end{figure}

\begin{figure}[!h]
\centering
\includegraphics[width=\textwidth]{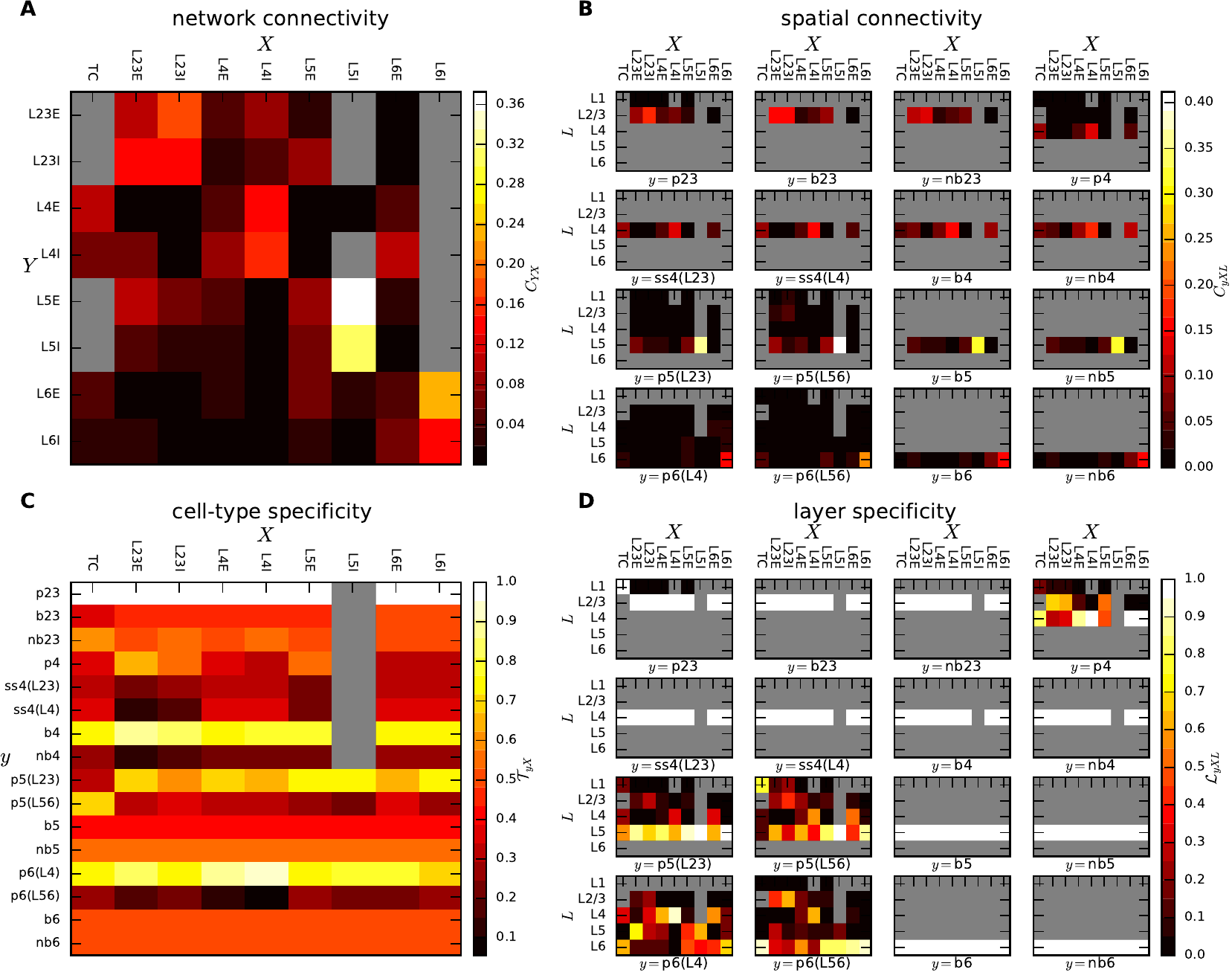}
\caption{
  \textbf{Connectivity of the cortical microcircuit model}. 
  \textbf{A}) Connection probability $C_{YX}$ between presynaptic population $X$ and postsynaptic population $Y$ of the cortical microcircuit model by \citet{Potjans2014} given in \Fref{tab:5}.
  Zero values are shown as gray here and in subsequent plots.
 \textbf{B})~Layer- and cell-type specific connectivity map $C_{yXL}$, where $X$, $y$ and $L$ denote presynaptic populations, 
  postsynaptic cell types and the synapse location (layer), respectively. This map is computed from the connectivity of the
  point-neuron network (panel A) and cell-type (panel C) and layer specificity (panel D) of connections.
  \textbf{C}) Cell-type specificity $\mathcal{T}_{yX}$ of connections quantified as
  the fraction of synapses between pre- and postsynaptic populations $X$ and $Y$ formed with a specific postsynaptic cell type $y$.
  \textbf{D}) Layer specificity $\mathcal{L}_{yXL}$ of connections denoting the fraction of synapses 
  between population $X$ and cell type $y$ formed in a particular layer $L$. 
  Both $\mathcal{T}_{yX}$ and $\mathcal{L}_{yXL}$ in panels C and D respectively are calculated from anatomical data \citep{Binzegger2004,Izhikevich2008}, cf.~\Fref{tab:8}. 
  }
\label{fig:5}
\end{figure}

For our example point-neuron network model, the cortical microcircuit model by \citet{Potjans2014}, 
the connectivity is to a large extent based on
anatomical data from cat visual cortex \citep{Binzegger2004,Izhikevich2008} (cf.~\Fref{tab:8}).
From \Fref{tab:8} we obtain
(i) the number $N_y$ of neurons belonging to cell type $y$,
(ii) the average total number $k_{yL}$ of synapses on all compartments in layer $L$ (input layer) of a single postsynaptic neuron of type $y$, 
and (3) the fraction $p_{yxL}$ of the $k_{yL}$ synapses formed with
presynaptic neurons of cell type $x$. 
The quantity 
\begin{equation}
 k_{yxL} = p_{yxL} \, k_{yL} 
 \label{eq:kyxL}
\end{equation}
defines the number of synapses between all presynaptic cells of type $x$ and a single postsynaptic cell of type $y$ in input layer $L$ 
(cf. network connectivity in \citet{Izhikevich2008}).
The number of synapses between all neurons in $x$ and all neurons in $y$, irrespective of the input layer $L$,
is given by 
\begin{equation}
 K_{yx} = N_y  \sum_L k_{yxL}~.
 \label{eq:Kyx}
\end{equation}
The number $K_{yx}$ of connections in combination with a chosen connectivity model (e.g., random graphs with binomially distributed \citep{Erdos59},
fixed in-/out-degree \citep{Newman2003} or random graphs with defined higher-order statistics \citep{Song2005, Zhao2011}) 
is sufficient for setting up the point-neuron network.
Assuming independently drawn synapses (allowing multiple connections between neurons),
the probability $C_{yx}$  of at least one
connection between a neuron of type $x$ and a neuron of type $y$ 
can be obtained from $K_{yx}$ as \citep{Potjans2014}
\begin{equation}
 C_{yx}=1-\left(1-\frac{1}{N_xN_y}\right)^{K_{yx}}~.
 \label{eq:Cyx}
\end{equation}
In our case, 
the point-neuron microcircuit model consists of excitatory and inhibitory populations $X,Y$ 
(see \Fref{tab:1}-\ref{tab:2}) pooling different pre- and postsynaptic cell types $x\in X$ and $y \in Y$ 
(cf. \Fref{fig:4}). 
Given a single multicompartment model neuron of type $y$ we compute the number $k_{yXL}$ of incoming connections 
(in-degree)
from cell types $x$ in each presynaptic population $X$ in a given layer $L$ by pooling all connections as illustrated in \Fref{fig:3}A as
\begin{equation}
k_{yXL} = \sum_{x\in X} k_{yxL}~.
\label{eq:kyXL}
\end{equation}
The total number of connections onto postsynaptic cells $y$ from cells in $X$ is then
\begin{equation}
K_{yXL} = N_y   k_{yXL}~.
\label{eq:KyXL}
\end{equation}
The layer-specific connection probability $C_{yXL}$ (\Fref{fig:5}B) can be derived 
from \Fref{eq:KyXL} analogously to \Fref{eq:Cyx}
for a presynaptic population size $N_X$ (here, $N_X=\sum_{x \in X} N_x$).
In order to obtain the connectivity within the point-neuron network, i.e., between populations $X,Y$,
we also need to pool over all synapses of input layers $L$ and cell types $y$ within the postsynaptic population $Y$
(dashed/dotted lines in \Fref{fig:3}B). Thus 
\begin{equation}
K_{YX} = \sum_{y \in Y} K_{yX} = \sum_{y \in Y} \sum_{L} K_{yXL}~,
\label{eq:KYX}
\end{equation}
which yields the connectivity of the simplified network structure $C_{YX}$ (cf. \Fref{eq:Cyx}, \Fref{fig:5}A).

\subsubsection{From pooled to specific network connectivity}
\label{sec:2.3.2}

In case of an already existing point-neuron network model such as introduced in \Fref{sec:2.1}, 
the reverse task of creating a spatial connectivity $C_{yXL}$ 
from a given point-neuron network connectivity $C_{YX}$ is necessary.
This inverse procedure compared to pooling over cell types and input layers
entails introducing the cell-type specificity 
\begin{equation}
  \mathcal T_{yX} = \frac{K_{yX}}{K_{YX}}~,
  \label{eq:TyX}
\end{equation}
which describes the fraction of synapses between populations $X$ and $Y$ that are formed with a specific postsynaptic cell type $y$
(\Fref{fig:5}C), 
and the layer specificity of connections
\begin{equation}
  \mathcal L_{yXL} = \frac{K_{yXL}}{K_{yX}},
  \label{eq:LyXL}
\end{equation}
denoting the fraction of synapses between population $X$ and all cells of cell type $y$ formed in a particular layer $L$
(\Fref{fig:5}D). The product $\mathcal T_{yX}  \mathcal L_{yXL}$ defines the probability 
of a synapse between populations $X$ and $Y$ 
formed with a specific postsynaptic cell type $y$ in a particular layer $L$ (\Fref{fig:3}B). 
Thus, if $K_{YX}$ is given, 
the total number of connections in layer $L$ onto postsynaptic cells $y$ from cells in $X$ is 
\begin{equation}
K_{yXL} = K_{YX} \mathcal T_{yX}  \mathcal L_{yXL}~.
 \label{eq:KyXL2}
\end{equation}
If $K_{YX}$ is constructed from
the same data as $\mathcal T_{yX}$ and $\mathcal L_{yXL}$, 
Equations \ref{eq:TyX}-\ref{eq:KyXL2} are fully consistent.
However, $K_{YX}$ can also be computed from any given 
point-neuron network connectivity $C_{YX}$.
This is particularly relevant for the network connectivity $C_{YX}$ (\Fref{fig:5}A) of \citet{Potjans2014} that
includes additional data sets for which
spatial information on synapse locations is not available. 
Here the number of synapses $k_{yXL}$ from population $X$ established in layer $L$ on each multicompartment model neuron of type $y$ is obtained from (9) as $k_{yXL}=K_{yXL}/N_y$.

\subsection{Forward modeling of extracellular potentials}
\label{sec:2.4}

The LFP signal reflects transmembrane currents weighted according to the distance from the source to the measurement location 
\citep{Einevoll2013},  and here we compute the LFP from the model neurons using a now well established
forward-modeling scheme combining multicompartment neuron modeling and electrostatic 
(volume-conduction) theory \citep{Holt1999, Gold2006, Pettersen2008, Linden2010, Linden2011, Reimann2013, Linden2014, Tomsett2014}. 
Each morphology was
spatially discretized into compartments using the \texttt{d\_lambda} rule \citep{Hines2001} with electrotonic length constants computed at $f=100$~Hz.
In this forward modeling scheme, 
localized synaptic activation of a morphologically detailed neuron results in spatially
distributed transmembrane currents across the neuronal membrane as calculated using standard cable theory, see e.g.,~\citet{DeSchutter2009}.
The extracellular potentials, including the LFP, are in turn given as a weighted sum of  transmembrane currents as described by volume conduction 
theory~\citep{Holt1999,Einevoll2013}. 

The cable-equation description is summarized in box D in \Fref{tab:4}.
\Fref{eq:imem} relates  synaptic input currents $I_{jin}$ onto compartment $n$ in a neuron $j$ from presynaptic neurons $i$, the membrane voltages $V_{\text{m}jn}$, 
and transmembrane currents $I_{\text{m}jn}$ and is derived from the assumption (Kirchoff's law) of current balance in the intracellular node of the equivalent electrical circuit of a cylindrical compartment $n$ with $m$ neighboring compartments.
We use the standard convention that a positive membrane current is a positive current from the intracellular to the extracellular space across the membrane. 
$I_{\text{m}jn}$ is assumed to be homogeneously distributed across the outer surface of the cylindrical compartment, 
and the calculation assumes that the electrical potential on the outside boundary of the membrane is zero at all times. 
Hence, there are no mutual interactions 
(i.e., ephaptic coupling \citep{Anastassiou2011})
between the extracellular potential estimated using volume conduction theory, 
the transmembrane currents, 
and intracellular potentials.

The associated extracellular potential resulting from the transmembrane currents is calculated based on volume conduction theory \citep{Nunez2006,Einevoll2013}.  
In the present application where the signal frequencies are well below 1,000~Hz, 
this calculation is simplified by applying the quasistatic approximation to Maxwell's equations, 
i.e., terms with time derivatives of the electrical and magnetic fields are omitted, cf. \citet[p. 426]{Hamalainen1993}.
Further, we assume the extracellular medium to be linear, isotropic, homogeneous and ohmic~\citep{Pettersen2012,Einevoll2013} 
and represented by a scalar extracellular conductivity $\sigma_\text{e}$.  

Given a time-varying point current source with magnitude $I(t)$ at position $\mathbf{r'}$, 
the scalar extracellular potential $\phi(\mathbf{r}, t)$ at position $\mathbf{r}$ and time $t$ is then given by \citep{Nunez2006, Linden2014}
\begin{equation}
 \phi(\mathbf{r},t) = \frac{I(t)}{4\pi\sigma_\text{e}|\mathbf{r} - \mathbf{r'}|}~.
\label{eq:extracellular1}
\end{equation}
Contributions to the extracellular potential from multiple current sources, i.e., 
transmembrane currents of all individual compartments $n$ from all cells $j$ in a population of $N$ cells 
sum linearly. 
In accordance with the assumed homogeneous current distribution along each cylindrical compartment, the \emph{line-source} approximation is used for dendritic compartments~\citep{Holt1999}. The line-source forward-modeling formula is
obtained by integrating \Fref{eq:extracellular1} along the cylindrical axis of each compartment $n$, and summing the contributions from all $n_\text{comp}$ compartments~\citep{Holt1999, Pettersen2008, Linden2014}:
\begin{equation}
\phi(\mathbf{r}, t) = \sum^N_{j=1} \sum^{n_\text{comp}}_{n=1} \frac{I_{\text{m}jn}(t)}{4 \pi \sigma_\text{e}}\int \frac{1}{|\mathbf{r} - \mathbf{r}_{jn}|}d\mathbf{r}_{jn}~.
\label{eq:extracellular2}
\end{equation}
Presently, we approximate the thick soma compartments as spherical current sources, 
and thus combine the point-source equation (\Fref{eq:extracellular1}) with the line-source formula (\Fref{eq:extracellular2}) for dendrite compartments, 
obtaining \citep{Linden2014}:
\begin{align}
\phi(\mathbf{r}, t) 
	&= \sum^N_{j=1} \frac{1}{4 \pi \sigma_\text{e}} 
		\left(\frac{I_{\text{m}j,\text{soma}}(t)}{|\mathbf{r} - \mathbf{r}_{j, \text{soma}}|} 
			+ \sum^{n_\text{comp}}_{n=2} \int \frac{I_{\text{m}jn}(t)}{|\mathbf{r} - \mathbf{r}_{jn}|}d\mathbf{r}_{jn} \right) \nonumber \\
	&= \sum^N_{j=1} \frac{1}{4 \pi \sigma_\text{e}} 
		\left( \frac{I_{\text{m}j,\text{soma}}(t)}{|\mathbf{r} - \mathbf{r}_{j,\text{soma}}|}
			+ \sum_{n=2}^{n_\text{comp}}  \frac{I_{\text{m}jn}(t)}{\Delta s_{jn}} 
				\ln \left| \frac{\sqrt{h_{jn}^2+r_{\perp jn}^2}-h_{jn}}{\sqrt{l_{jn}^2+r_{\perp jn}^2}-l_{jn}} \right| \right) ~.
\label{eq:extracellular3}
\end{align}
Here, $\Delta s_{jn}$ denotes compartment length, 
$r_{\perp jn}$ the perpendicular distance from the electrode point contact to the axis of the line compartment, 
$h_{jn}$ the longitudinal distance measured from the start of the compartment, 
and $l_{jn} = \Delta s_{jn}+h_{jn}$ the longitudinal distance from the other end of the compartment.
If the distance between electrode contacts and dendritic current sources becomes 
smaller than the radius of the dendritic segment,
an unphysical singularity in our extracellular potential may occur. 
In these cases singularities are avoided by setting $|{\mathbf r}-{\mathbf r}_{j,\text{soma}}|$ or $r_{\perp jn}$ equal to the compartment radius.
Electrode contacts of real recording devices have finite spatial extent and are not point contacts as assumed above. 
However, the recorded signal can be well approximated as the mean of the potential averaged across the uninsulated surface~\citep{Robinson1968,Nelson2008,Nelson2010,Ness2015}, 
at least for current sources positioned further away than an electrode radius or so~\citep{Ness2015}. 
Here we employed the \emph{disc-electrode} approximation to the potential~\citep{Camunas-Mesa2013, Linden2014, Ness2015}: 
\begin{equation}
\phi_{\text{disc}}(\mathbf{u},t)  = \frac{1}{A_S} \iint_{S} \phi(\mathbf{u},t) \,d^2 r \approx \frac{1}{m} \sum_{h=1}^m \phi({\bf u}_h,t) ~.
\label{eq:extracellular4}
\end{equation}
We further considered circular electrode contacts with a radius of 
$r_{\text{contact}}$=7.5~$\mu$m, 
and we averaged the point-contact potential in \Fref{eq:extracellular3} over $m=50$ random locations $\mathbf{u}_h$ across the contact surface $S$,  $A_S$ being the surface area.
The chosen locations were distributed with uniform probability on circular discs representing each contact surface, with 
surface vectors oriented perpendicular to the electrode axis~\citep{Linden2014}.
Calculations of extracellular potentials were facilitated by \texttt{LFPy} (\url{http://LFPy.github.io}) \citep{Linden2014}, in which \texttt{NEURON} simulation software is used for calculations of transmembrane currents (i.e., solving \Fref{eq:imem}) \citep{Hines2001, Hines2009, Carnevale2006}.

\subsection{Data analysis and software}
\label{sec:2.5}

\subsubsection{Model measurements}
\label{sec:2.5.1}

The main simulation output of the hybrid scheme consists of spike trains of each neuron in the point-neuron network, 
`ground-truth'  current-source density (CSD) and local field potentials (LFP) of each neuron in the morphologically detailed postsynaptic model populations (see \Fref{tab:3} and \ref{tab:4}). 
Here the term 'ground-truth' refers to the fact that the CSD is computed from transmembrane currents rather than estimated from the LFP.

As the transmembrane current of each compartment (\Fref{eq:imem}) is known at each simulation time step, 
we follow the procedure of \citet{Pettersen2008} to compute the `ground-truth' CSD in addition to the LFP.
From $N$ model neurons with $n_\text{comp}$ compartments 
having membrane currents  \(I_{\text{m}jn}(t)\) and lengths \(\Delta s_{jn}\), 
we calculate the CSD signals $\rho(\mathbf{r},t)$ inside cylinder elements $V_\rho(\mathbf{r})$ around each electrode contact as:
\begin{equation}
\rho(\mathbf{r},t) = \frac{1}{\pi r^2 h_\text{elec}}\sum_{j=1}^N \sum_{n=1}^{n_\text{comp}} I_{\text{m}jn}(t)\frac{\Delta s_{jn, \text{inside}}(\mathbf{r})}{\Delta s_{jn}}~.
\label{eq:csd}
\end{equation}
$\Delta s_{jn, \text{inside}}(\mathbf{r})$ denotes the length of the line source  contained within $V_\rho(\mathbf{r})$.  
In contrast to \citet{Pettersen2008}, we do not apply a spatial filter to the CSD.
The volumes have radii equal to the population radius $r$ and heights equal to  
the electrode separation $h_\text{elec}$ (cf. \Fref{tab:6}).

In the present example application, 
the extracellular potential is computed at locations corresponding to a laminar 
multi-electrode array with 16~recording electrodes 
with an inter-electrode distance of $h_\text{elec}=100~\mu$m, positioned at the cylindrical axis of the model column
with the topmost contact at the pial surface (cf. \Fref{fig:1}D, see \Fref{tab:6} for details). 
Each electrode contact is set to have a radius of $7.5~\mu$m (cf. \Fref{eq:extracellular4} in \Fref{sec:2.4}).
In the network we also record membrane voltages and input currents from a subset of cells in each of the eight cortical populations (see \Fref{tab:9}).

We ran our simulations for a total duration of $T=5{,}200$~ms using a temporal resolution of $dt=0.1$~ms (cf. \Fref{tab:5} and \ref{tab:11}).
However, LFP and CSD signals were resampled prior to file storage to 
a temporal resolution of $dt_{\psi}=$1~ms by (i) applying a 4th-order Chebyshev type I  
low-pass filter with critical frequency $f_\text{c}=400$~Hz and 0.05 dB ripple in the passband 
using a forward-backward linear filter operation, 
and (ii) then selecting every 10th time sample. 

\subsubsection{Post-processing and data analysis}
\label{sec:2.5.2}

As the contributions to the CSD and LFP of the different cells sum linearly 
(cf. \Fref{eq:extracellular3}), 
we compute population-resolved signals as the sum over contributions from
all cells in a population,
and the full compound signals as the sum over all population signals (see \Fref{tab:9}).
In \Fref{sec:downscaling} we derive a
rescaled 'low-density predictor' $\phi^{\gamma \xi}(\mathbf{r},t)$ of the LFP
from random subsets of neurons in all populations. 
Thereby, we make a downscaled LFP-generating model setup 
with the same column volume, but with neuron density reduced to a factor 
$\gamma \in (0,1)$ of the original density, while preserving the in-degrees,
i.e., the number of synaptic connections onto individual neurons.
The LFP from the downscaled
setup is multiplied by an overall scaling factor $\xi$ chosen to roughly preserve the LFP from the full-scale model.
For analysis and plotting, the initial 200~ms of results after simulation onset
was removed, and the signal mean was subtracted from LFP and CSD traces 
emulating DC filtering during experimental data acquisition.
The contribution from each population to the overall LFP signal was assessed by computing and comparing the signal variances 
(see \Fref{tab:10}).

Cross-correlations between single-cell LFPs $\phi_i(\mathbf{r},t)$ were quantified by analyzing the power spectrum of the compound extracellular signals.
Given that the compound LFP/CSD is a linear superposition of 
single-cell LFP/CSD contributions, 
the power spectrum of the compound signal $P_{\phi}(\mathbf{r},f)$ can be obtained as the sum of all single-cell
power spectra $P_{\phi_i}(\mathbf{r},f)$ and all pairwise cross-spectra $C_{\phi_i \phi_j}(\mathbf{r},f)$, 
or equivalently from the average single-cell
power spectrum $\overline{P_{\phi}}(\mathbf{r},f)$, 
the average pairwise cross-spectrum $\overline{C_{\phi}}(\mathbf{r},f)$, and the total cell count $N$, as shown in \Fref{tab:10}. 
Note that $\overline{C_{\phi}}(\mathbf{r},f)$ and hence the average pairwise single-cell LFP coherence $\overline{\kappa_{\phi}}(\mathbf{r},f)$ (\Fref{eq:coherence}) are real, 
while $C_{\phi_i\phi_j}(\mathbf{r},f)$ is complex. 
Note that this definition allows for negative values of $\overline{\kappa_{\phi}}(\mathbf{r},f)$.
In the sum over all $i$ and $j$ in \Fref{eq:mean-cross-spectrum}, 
the imaginary parts of $C_{\phi_i\phi_j}(\mathbf{r},f)$ and $C_{\phi_j\phi_i}(\mathbf{r},f)$ cancel because $C_{\phi_i\phi_j}(\mathbf{r},f)=C_{\phi_j\phi_i}(\mathbf{r},f)^*$ (${}^*$ denotes the complex conjugate).
The power spectrum $P_{\phi^{\gamma \xi}}(\mathbf{r},f)$ 
of the compound signals of the `downscaled' network (i.e., low-density LFP predictor, see above)
is given by the reduced cell count $\gamma N$, the average single-cell power spectrum $\overline{P_{\phi^{\gamma \xi}}}(\mathbf{r},f)$, and the average pairwise
single-cell cross-spectrum $\overline{C_{\phi^{\gamma \xi}}}(\mathbf{r},f)$ calculated from that subset of neurons (see \Fref{tab:10}).
These single-cell averages are the same as 
the respective single-cell averages of the full-scale 
model setup, apart from variability due to subsampling.

Throughout this paper, signal power spectra are estimated using Welch's average 
periodogram method \citep{Welch1967} 
(with the \texttt{matplotlib.mlab.psd} implementation in \texttt{Python}, see \Fref{tab:11}).
Temporal cross-correlations are quantified as the zero time-lag correlation coefficient (\Fref{eq:cc} in \Fref{tab:10}).
Spike-triggered average LFP (staLFP) signals are computed as the
cross-covariance between the time-resolved population spike rate $\nu_X(t)$ and the compound LFP $\phi(\mathbf{r}, t)$, divided by the total number of spikes (i.e., $\int_0^T \nu_X(t) dt$).

\subsubsection{The \texttt{hybridLFPy} \texttt{Python} package}
\label{sec:2.5.3}

To facilitate usage of the hybrid scheme by other users,
a novel \texttt{Python} software package, \texttt{hybridLFPy},
has been made publicly available under the General Public License version 3 (GPLv3, \url{http://www.gnu.org/licenses/gpl-3.0.html}) 
on GitHub (\url{http://github.com/INM-6/hybridLFPy}).
Compatibility with a host of different machine architectures and operating systems ($\ast$nix, OSX, Windows) is ensured with the freely available, 
object-oriented programming language \texttt{Python} (\url{http://www.python.org}).
\texttt{Python} adds tremendous flexibility in terms of interfacing a large number of packages and libraries for, e.g., 
performing numerical analysis and data visualization, 
such as \texttt{numpy} (\url{http://www.numpy.org}) 
and \texttt{matplotlib} (\url{http://www.matplotlib.org}), 
while several other neural simulation softwares also come with their own \texttt{Python} interfaces, 
such as \texttt{NEST} (\url{http://www.nest-initiative.org}) \citep{Eppler2008} and \texttt{NEURON} 
(\url{http://www.neuron.yale.edu}) \citep{Hines2009}.

The source code release of \texttt{hybridLFPy} provides a set of classes implementing the hybrid scheme, 
as well as example network simulation codes implemented with \texttt{NEST} \citep{eppler_2015_32969} 
(at present a simplified two-population network \citep{Brunel2000} and the full cortical microcircuit model of \citet{Potjans2014} adapted from public codes (see \Fref{sec:2.1})).
The class \texttt{hybridLFPy.CachedNetwork} uses an efficient \texttt{sqlite3} database implementation for reading in all point-neuron network spike events and interfacing network spike events with the main simulation in which the LFP and CSD are calculated.
The class \texttt{hybridLFPy.Population} defines 
populations of multicompartment model neurons representing each cell type, 
assigns synapse locations across the laminae,
selects spike trains for each synapse location from the appropriate presynaptic population,
and calculates the LFP and CSD. 
The single-cell calculations are handled using \texttt{LFPy}
(\url{http://LFPy.github.io})
\citep{Linden2014} which builds on \texttt{NEURON}~\citep{Carnevale2006, Hines2009}.
As there are no mutual interactions between the multicompartment model neurons in the calculation of LFPs, 
these calculations remain embarrassingly parallel operations. 
Finally,
the class \texttt{hybridLFPy.PostProcess} constructs the full compound signals in terms of LFPs and CSDs created by multiple instances of the \texttt{Population} class, 
and performs the main analysis steps as described in \Fref{sec:2.5.2}. 
Reproducible simulation and data analysis 
are assured by tracking code revisions using \texttt{git} 
and by fixing random number generation seeds and the number of parallel processes.
Further documentation and information on installing and using \texttt{hybridLFPy} is provided online, see \url{http://github.com/INM-6/hybridLFPy}.

The present implementation of \texttt{hybridLFPy} and corresponding simulations were made possible by
\texttt{Open MPI} (v.1.6.2),
\texttt{HDF5} (v.1.8.13),
\texttt{sqlite3} (v.3.6.20),
\texttt{Python} (v.2.7.3) with modules 
\texttt{Cython} (v.0.23dev),
\texttt{NeuroTools} (v.0.2.0dev),
\texttt{SpikeSort} (v.0.13),
\texttt{h5py} (v.2.5.0a0),
\texttt{ipython} (v.0.13),
\texttt{matplotlib} (v.1.5.x),
\texttt{mpi4py} (v.1.3),
\texttt{numpy} (v.1.10.0.dev-c63e1f4),
\texttt{pysqlite} (v.2.6.3) and 
\texttt{scipy} (v.0.17.0.dev0-357a1a0). 
Point-neuron network simulations were performed using \texttt{NEST} (v.2.8.0 ff71a29), 
and 
simulations of multicompartment model neurons using \texttt{NEURON} (v.7.4 1186:541994f8f27f) through \texttt{LFPy} 
(dev. v.3761c4). 
All software was compiled using \texttt{GCC} (v.4.4.6).
Simulations were performed in parallel (256 threads) on the Stallo high-performance computing facilities 
(NOTUR, the Norwegian Metacenter for Computational Science) consisting of 2.6 GHz Intel E5-2670 CPUs running the Rocks Cluster Distribution (Linux) operating system (v.6.0).


\section{Results: LFP generated by a cortical microcircuit}
\label{sec:results}

To illustrate the application of the hybrid scheme for predictions of LFPs from point-neuron networks, we here present
results for a modified version of the point-neuron network model of \citet{Potjans2014} with $\sim 78{,}000$  
neurons mimicking a 1 mm$^2$ patch of cat primary visual cortex (see \Fref{sec:2.1}). 
The microcircuit model has realistic cell density and deliberately neglects many biological 
details on the single-cell level, focussing on the effect of the connectivity on the local network dynamics in such circuits.
Despite this simplicity, the model displayed firing rates across populations in agreement with experimental observations \citet{Potjans2014}, as well as propagation of spiking activity across layers. Likewise, the microcircuit model in conjunction with simplified, passive multicompartment populations is used here to study the effect of the (spatial) connectivity on the laminar pattern of spontaneous and stimulus-evoked CSD and LFP signals. For this, 
CSD and LFP signals for a laminar multielectrode recording at different cortical depths are computed.
The large network size of the model is further used to illustrate the effect of correlations and neuron density on CSD and LFP predictions.

\subsection{Spontaneous vs. stimulus-evoked LFP}
\label{sec:3.1}

\begin{figure}[!htbp]
\centering
\includegraphics[width=\textwidth]{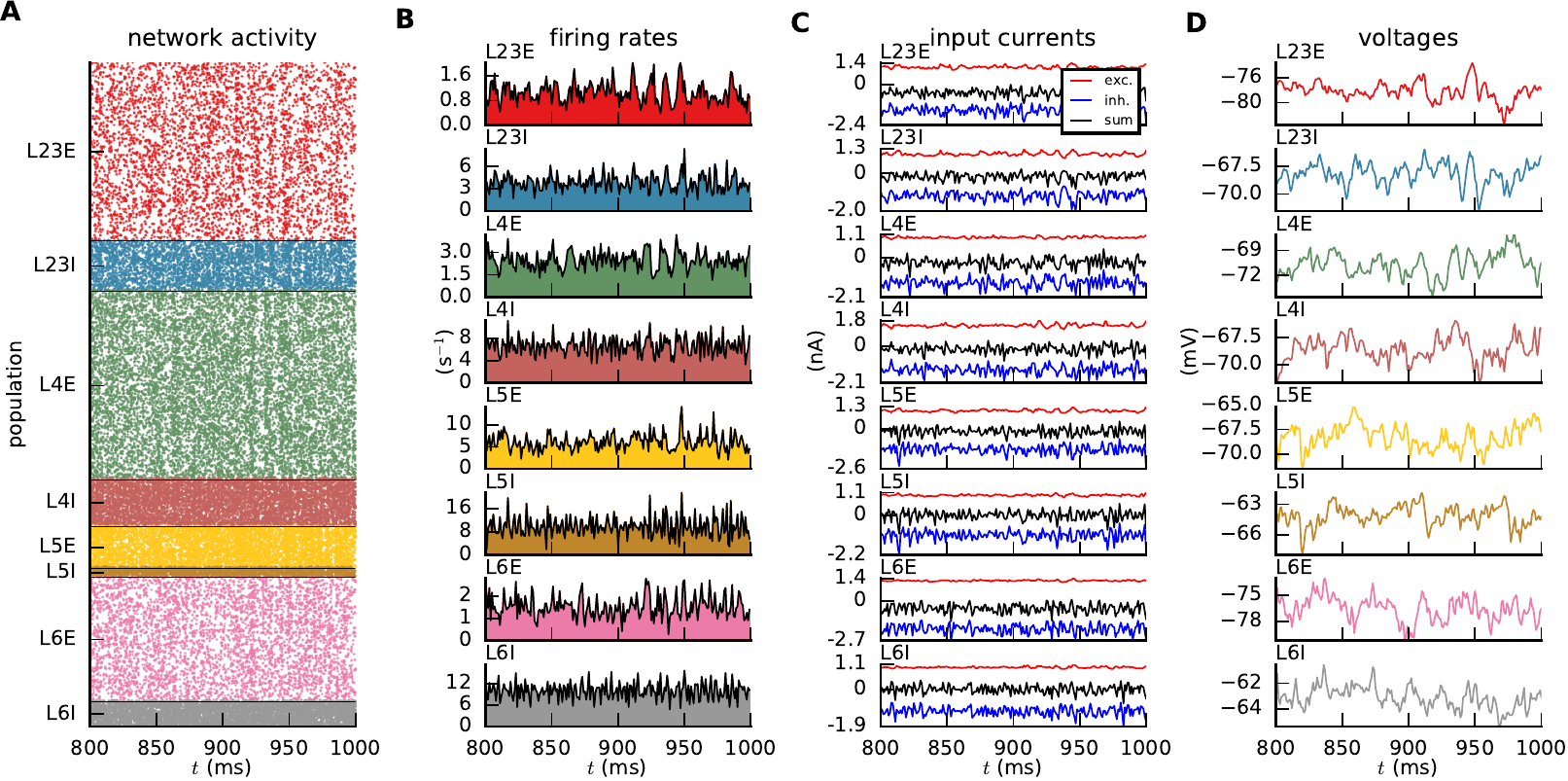}
\includegraphics[width=0.75\textwidth]{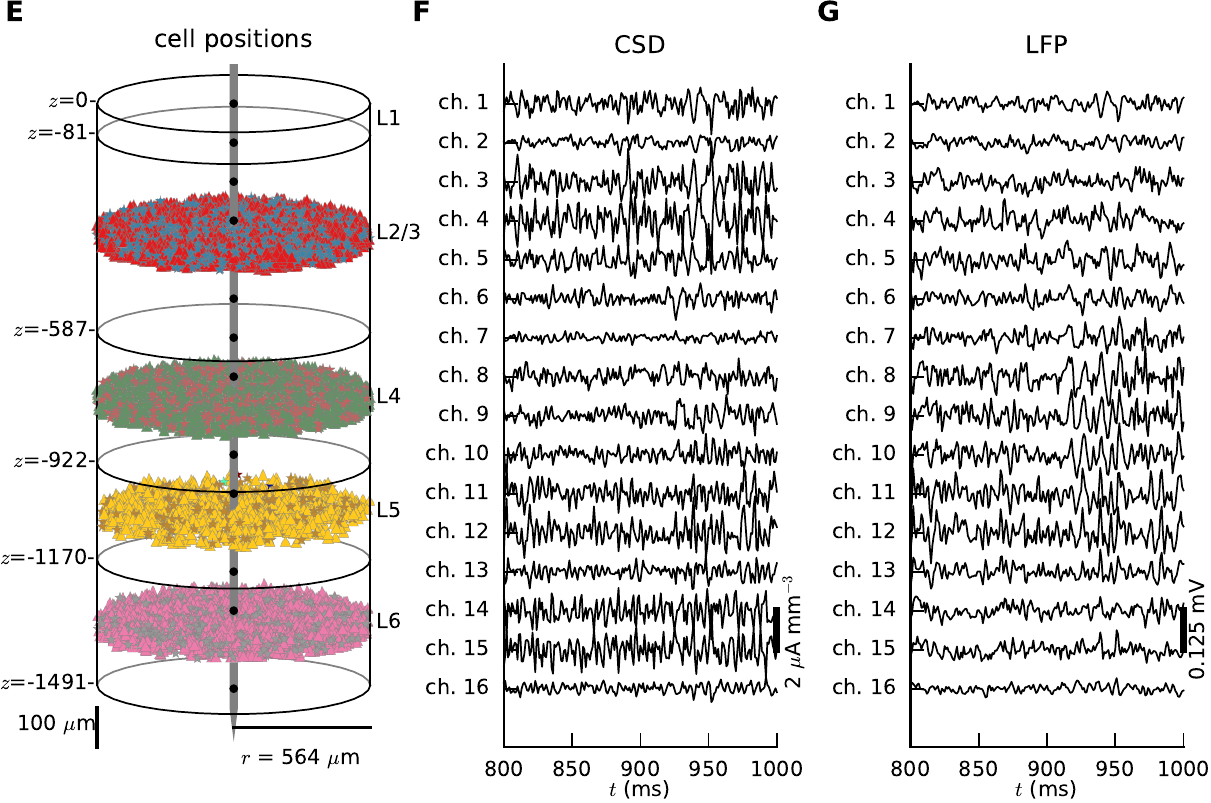}
\caption{
\textbf{Overview of output signals obtained from application of the hybrid scheme to a cortical microcircuit (spontaneous activity).}
\emph{Point-neuron network:}
\textbf{A})~Spiking activity. Each dot represents the spike time of a point neuron
(color coding as in \Fref{fig:1}).
\textbf{B})~Population-averaged firing rates for each population.
\textbf{C})~Population-averaged somatic input currents (red: excitatory, blue: inhibitory, black: total).
\textbf{D})~Population-averaged somatic voltages. 
Averaged somatic input currents and voltages are obtained from $100$ neurons in each population.
\emph{Multicompartment model neurons:}
\textbf{E})~Somas of excitatory (triangles) and inhibitory (stars) multicompartment cells and layer boundaries (gray/black ellipses). Illustration of a laminar
electrode (gray) with 16 recording channels (black circles).
\textbf{F})~Depth-resolved current-source density (CSD) obtained from
summed transmembrane currents in cylindrical volumes centered at each
contact.
\textbf{G})~Depth-resolved local field potential (LFP) calculated at each electrode contact 
from transmembrane currents of all neurons in the column. 
Channel 1 is at pial surface, channel 2 at 100~$\mu$m depth, etc. 
}
\label{fig:6}
\end{figure}

We first consider the LFP generated by spontaneous network activity.
The output of our hybrid scheme covers various scales and measurement modalities, 
from spikes of each neuron (\Fref{fig:6}A), 
population-averaged firing rates (\Fref{fig:6}B), 
excitatory, inhibitory, and total synaptic input currents (\Fref{fig:6}C), and
membrane voltages (\Fref{fig:6}D),
to the compound CSD and LFP stemming from all populations of different cell types (\Fref{fig:6}E-G). 
For spontaneous activity (cf. \Fref{sec:2.1}, \Fref{tab:1}),
i.e., no modulated thalamic input, we observe asynchronous irregular 
spiking in all populations (\Fref{fig:6}A) and firing rates 
similar to the original model \citep{Potjans2014} (\Fref{fig:6}B). In particular, 
layer 2/3 exhibits low firing rates, and generally inhibitory neurons fire with higher rates than excitatory neurons of the respective layers. 
The network is in a balanced regime \citep{Brunel2000}, 
reflected by the substantial cancellation of population-averaged excitatory and inhibitory input currents (\Fref{fig:6}C). 
The population-averaged membrane potential fluctuates below the fixed firing threshold $\theta=-50$~mV down to $\sim-80$ mV (\Fref{fig:6}D). 

The corresponding compound CSD and LFP signals with contributions from all cortical populations are shown in \Fref{fig:6}F~and~G, respectively. 
As expected for spontaneous cortical network activity, the LFP signal amplitudes are small, 
$\simeq$0.1~mV  (intriguingly close to what has been seen in experiments, see, e.g., Fig.~1 in \citet{Maier2010}  for macaque visual cortex, 
Fig.~7 in \citet{Hagen2015} for mouse visual cortex), 
and exhibit strong across-channel covariance 
in line with experimental observations, e.g., \citep{Einevoll2007,Riehle2013, Hagen2015}.
In the model these correlations stem from dendritic cable properties and volume conduction effects \citep{Pettersen2008,Linden2011,Leski2013}.
The correlations across channels are, as expected, generally less visible in the more localized CSD signal where volume conduction effects are absent ~\citep{Nicholson1975, Pettersen2006, Pettersen2008}.

\begin{figure}[h]
\centering
\includegraphics[width=\textwidth]{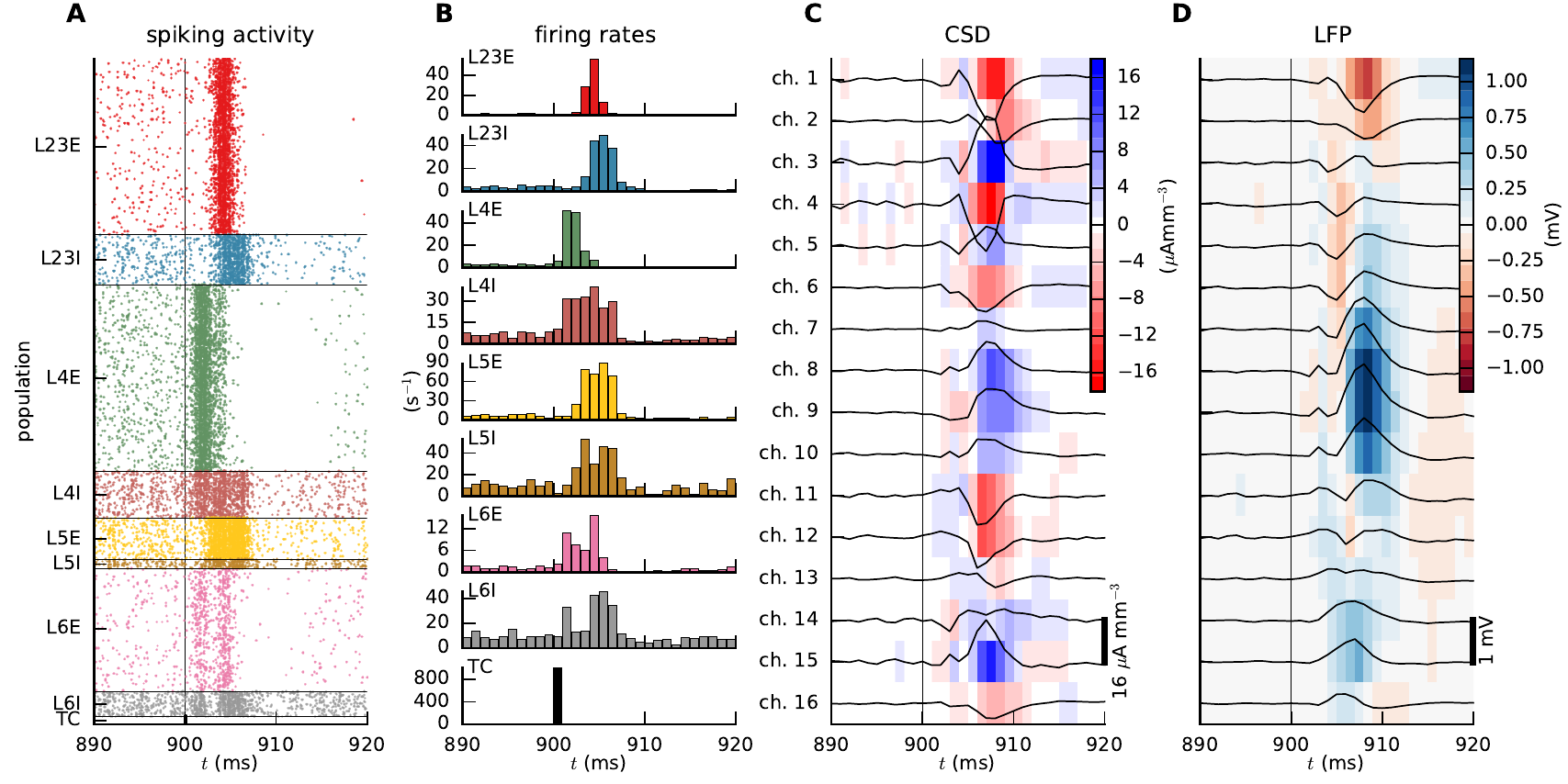}
\caption{
\textbf{Network activity following transient activation of thalamocortical afferents}.
\textbf{A})~Raster plot of spiking activity before and after $\delta$-shaped thalamic stimulus 
presented at $t=900$~ms (vertical black line in panels A, C and D). 
\textbf{B})~Population-averaged firing rate histogram for each population (color coding as in panel A).
\textbf{C})~Depth-resolved compound current-source density (CSD) of all populations (shown both in color and by the black traces). 
\textbf{D})~Depth-resolved compound local field potential (LFP, shown both in color and by the black traces) at each electrode channel as generated by all populations.  
Channel 1 is at pial surface, channel 2 at 100~$\mu$m depth, etc.
}
\label{fig:7}
\end{figure}

LFP and associated CSD studies have commonly been used to investigate stimulus-evoked responses in sensory cortices, see, e.g.,  
\citet{Mitzdorf1979,Mitzdorf1985,Di1990,Schroeder1998,Swadlow2002,Einevoll2007,Sakata2009,Szymanski2009,Maier2010,Jin2011,Szymanski2011}, 
as well as \citet{Einevoll2013} and references therein.
To model the situation with a sharp onset of a visual stimulus (or direct electrical stimulation of the thalamocortical pathway~\citep{Mitzdorf1979})
we drive the network with a short thalamic pulse mimicking a volley of incoming spikes onto primary visual cortex from 
the visual thalamus (lateral geniculate nucleus, LGN). The activation targets populations in layers 4 and 6 
(see \Fref{fig:1}A or \Fref{fig:5}A,B) 
and propagates in the network to populations in layers 2/3 and 5  (\Fref{fig:7}A,B). 
At the level of spiking activity, the results match the behavior of the original model 
\citep[Fig. 10A,B]{Potjans2014} and even agree qualitatively with experimental findings in rodents from stimulus-evoked activation of 
auditory cortex~\citep{Sakata2009} and somatosensory cortex~\citep{Armstrong-James1992,Einevoll2007,Reyes-Puerta2015}.
This points to a general biological plausibility of this generic network model, based largely on data from cat V1.

The corresponding CSD and LFP profiles across depth associated with this spiking activity are determined by the synapse locations and dendritic filtering by cell-type specific morphologies (see \Fref{fig:2}, \Fref{fig:5}D). 
For the CSD (\Fref{fig:7}C) a complex alternating spatiotemporal pattern of current sinks (negative CSD) and sources (positive CSD) is observed.
Due to volume conduction this detailed spatial pattern is largely smeared out in the LFP profile (\Fref{fig:7}D), which displays a strong positivity
 across the middle layer around $t$=910~ms. We also note the different spatiotemporal profiles of the spiking activity (\Fref{fig:7}A,B) compared
to CSD and LFP laminar profiles. Not only do the CSD and LFP signals typically fade 
out 5--10 ms later than the spiking, the spatial profiles are also very different. For example, the LFP signal is very weak between channels 11 and 12 (i.e., between 1,100 and 1,200~$\mu$m depth), even if the firing rate of layer 5 positioned between these channels is very high. We also note  that the predicted LFP magnitudes are an order of magnitude larger compared to the
LFP predicted for spontaneous activity, that is, $\sim$1~mV for stimulus-evoked versus $\sim$0.1~mV for spontaneous activity. 

Although the present example has not been tuned to address specific experiments, we nevertheless observe that the model predictions display
several features seen in experiments. For example, stimulus-evoked LFP amplitudes on the order of 1~mV are similar
to the maximal amplitudes ($\sim$1--3~mV) observed in cat V1 following electric stimulation of thalamocortical axons (optical radiation) 
\citep{Mitzdorf1985}, in visually evoked LFPs in monkey V1~\citep{Schroeder1998}, 
and in rat somatosensory (barrel) cortex following whisker 
flicks~\citep{Di1990,Einevoll2007,Reyes-Puerta2015}. 
Also some qualitative features of the spatiotemporal LFP and CSD patterns from the cat visual cortex experiments of \citet{Mitzdorf1985} can be recognized. One example is the early CSD sink around layer 4 (i.e., at channel 6 close to the boundary between layers 2/3 and 4), another is the large positivity in LFP extending through most of the layers following the initial response to the thalamic input volley.

The observed time courses of the LFP and CSD in our simulations with $\delta$-pulse thalamic activations are not as directly 
comparable with experimental stimulus-evoked activity,  
where the input is temporally filtered by several cell populations
before reaching thalamus on the way to
cortex. However, we note that very swift stimulus-evoked responses lasting not much longer than the $~\sim$10~ms response volleys seen in our simulations also are observed in the somatosensory system~\citep{Di1990,Einevoll2007,Reyes-Puerta2015}.

\subsection{Effect of network dynamics on LFP}
\label{sec:3.2}

The spiking activity in the microcircuit model is highly sensitive to modification of intrinsic model parameters and external input \citep{Bos2015}.
In general LFPs reflect synaptic input both from local and distant neurons~\citep{Herreras2015}, 
and also depend on network state \citep[see, e.g., ][]{Kelly2010, Gawne2010}. 
With the hybrid scheme,  we first illustrate as an example 
the dependence of the spontaneous LFP, i.e., the LFP without thalamic input, 
on local network dynamics as determined by intrinsic network parameters. In particular we 
compare our reference network model with the original model proposed by \citet{Potjans2014} 
where the strength of connections from L4I to L4E neurons is weaker 
and the synaptic weights are drawn from a Gaussian distribution. 
We next investigate the 
LFP when the network is stimulated by sinusoidally modulated thalamic input.

\begin{figure}[!htbp]
\centering
\includegraphics[width=\textwidth]{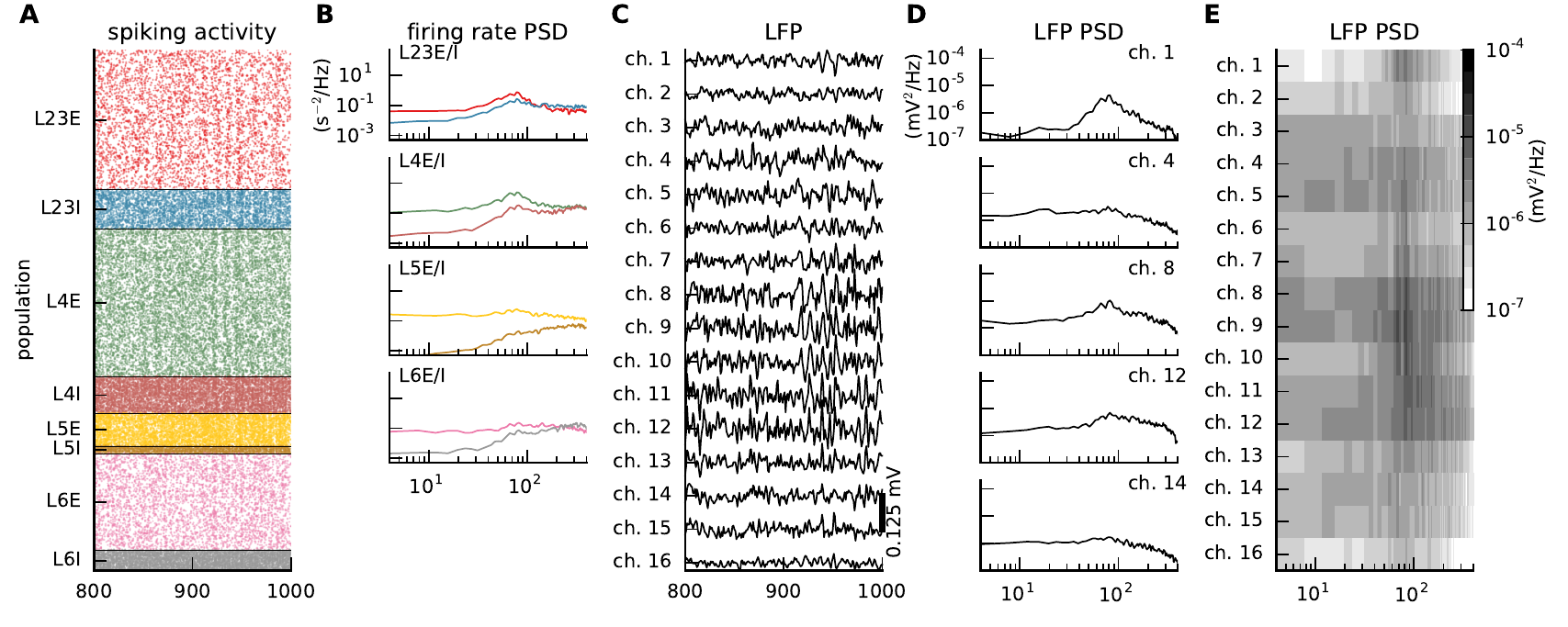}
\includegraphics[width=\textwidth]{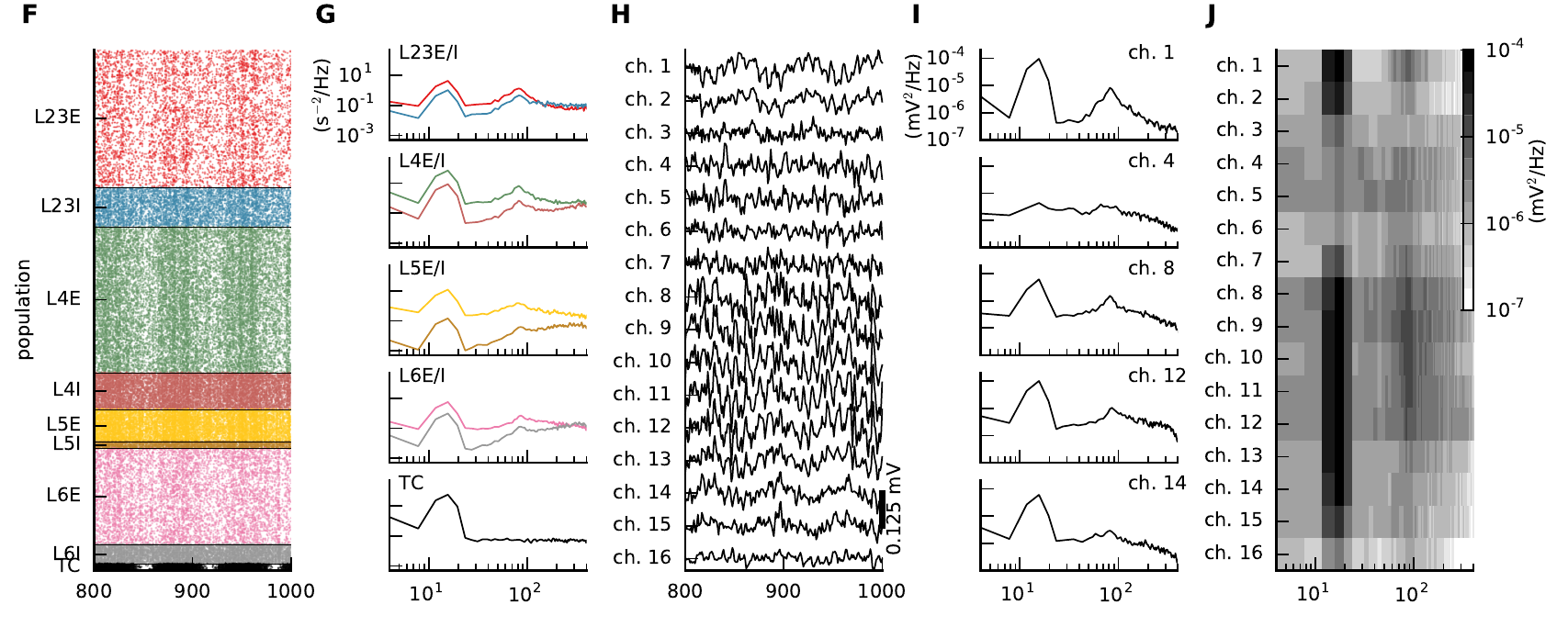}
\includegraphics[width=\textwidth]{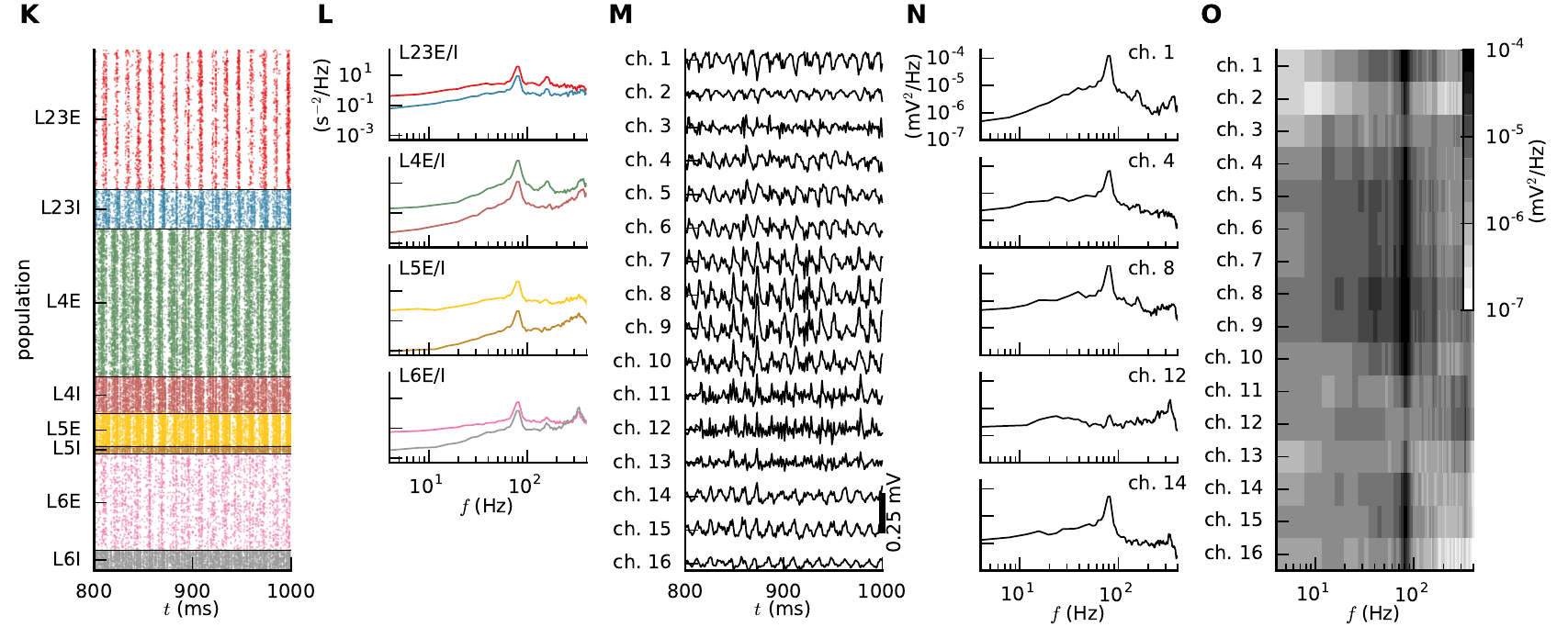}
\caption{
  \textbf{Effect of network dynamics on LFP.}
  Comparison of two different thalamic input scenarios and two different networks.
  {\bf Top:}~Reference network, spontaneous activity.
  {\bf Center:}~Reference network, oscillatory thalamic activation.
  {\bf Bottom:}~Original model by \citet{Potjans2014}, spontaneous activity.
  {\bf A,F,K})~Population-resolved spiking activity.
  {\bf B,G,L})~Population-averaged firing rate spectra.
  {\bf C,H,M})~Depth-resolved LFP.
  {\bf D,I,N})~LFP power spectra in layer 1 and at typical somatic depths of network populations.
  {\bf E,J,O})~LFP power spectra across all channels.
Channel 1 is at pial surface, channel 2 at 100~$\mu$m depth, etc.
}
\label{fig:8}
\end{figure}

For spontaneous activity (\Fref{fig:8}A-E), the spiking (\Fref{fig:8}A) 
is asynchronous irregular in all populations. The firing-rate power spectra (\Fref{fig:8}B)
vary from relatively flat to more band-pass-like with a maximum power around 80~Hz. 
The suppressed power 
at lower frequencies arises from active decorrelation due to inhibitory feedback \citep{Tetzlaff2012}. 
In combination with the low-pass filtering involved in the generation of LFPs from spiking activity~\citep{Linden2010,Leski2013},
the firing-rate spectra translate into an LFP power spectrum that, depending on recording  depth, has either low-pass or 
band-pass filter characteristics (\Fref{fig:8}D). The notably sharper attenuation of the LFP power spectra
compared to the firing-rate power spectra above $\gtrsim$~100~Hz is expectedly due to the intrinsic dendritic filtering
effect~\citep{Linden2010} (however, the frequencies $\gtrsim$~400~Hz are sharply attenuated due 
to our anti-aliasing filter, cf. \Fref{sec:2.5.1}). 
As the effect of this filtering depends on the position of the electrode compared to the neuronal morphology~\citep{Linden2010}, the result is a variable LFP power spectral density (PSD) 
profile across cortical depths, even in the absence of any structured external input
(\Fref{fig:8}E). 

There is experimental evidence 
that cortical microcircuits receive oscillating input 
at various frequencies from remote areas and subcortical structures \citep{Bastos2015, Kerkoerle2014, Ito2014}.
To illustrate the effect of an oscillatory external input in our model, 
we model thalamic input as independent realizations of non-stationary Poisson 
processes with a sinusoidal rate profile (see \Fref{tab:5}, \Fref{fig:8}F-J). 
The spiking activity of all populations as well as the CSD and LFP across depth are sensitive to thalamic input, 
showing that the network response goes beyond the layers receiving thalamic input, i.e., layers 4 and 6. 
The stimulus-evoked spiking activity follows the $15$~Hz modulated rate of the 
thalamic input (\Fref{fig:8}F). 
The 15~Hz oscillation is reflected in the firing-rate spectrum as a peak around 15~Hz (\Fref{fig:8}G) 
and is robustly transferred to the LFP (\Fref{fig:8}H-J). 
The LFP oscillation strength
varies with depth and is greatest in channels 1--2 and channels 8--14, 
while the oscillation is barely seen in channels 3--6.
Populations L4E/I and L6E/I  
receive thalamic inputs around the depths of channels 8--10 and channels 14--15,
and these channels are strongly affected by the stimulus. 
However, recurrent connections between populations, dendritic propagation of currents, and volume conduction  
produce strong LFP oscillations  also in other channels.

The LFP amplitudes are not only influenced by the temporal structure of the external input, but also by synaptic weights 
in two ways: first via their influence on the spiking dynamics in the point-network simulation and second
via the influence on the size of the synaptic currents setting up the transmembrane currents in the multicompartment models
in the LFP-computing step.
In order to illustrate the weight dependence, 
we compare the LFP under spontaneous activity in our reference network model 
(\Fref{fig:8}A-E) 
with the corresponding spontaneous LFP in the original model by \citet{Potjans2014} (\Fref{fig:8}K-O).
The lower 
inhibition from population L4I onto L4E and the narrow weight distribution 
compared to our model gives a higher degree of spike synchrony in all populations.
Although our model exhibits asynchronous irregular activity (\Fref{fig:8}A), the original model is closer to 
a synchronous irregular regime \citep{Brunel2000} (\Fref{fig:8}K), 
resulting in high-frequency oscillations 
around 80~Hz and in the 300-400~Hz band (\Fref{fig:8}L), 
both associated with delay loops in the multi-layered network \citep{Brunel2000}. 
The 80~Hz oscillation also appears in the LFP and its corresponding power spectra
(\Fref{fig:8}M-O), but the magnitude of the peak is not constant across depth.
The lowest magnitudes are located in the vicinity of layers 2/3 and 5 
(in channels 3--4 and channels 11--13).

\subsection{Contributions from individual populations to CSD and LFP}
\label{sec:3.4}

The direct interpretation of CSD and LFP signals in terms of the underlying activity of different populations or input pathways is inherently ambiguous and thus difficult:
For example, 
a CSD sink observed in cortical layer 2/3 can alternatively stem from excitatory synaptic inputs to the basal dendrites of layer 2/3 cells, 
similar inputs into the apical dendrites of layer 5 cells, or even return currents from appropriately placed inhibitory inputs onto the same cells~\citep{Linden2010,Einevoll2013}.
Several schemes for decomposition of CSD and LFP data into contributions from cortical populations have thus been proposed:
principal component analysis (PCA)~\citep{Di1990},  laminar population analysis~\citep{Einevoll2007}, and independent component analysis (ICA)~\citep{Leski2010,Makarov2010,Glabska2014, Herreras2015}. 
In our modeling world we have the benefit of having the contributions from the various populations, connections or different synapse types to the CSD and LFP signals directly
accessible. 

\begin{figure}[!htbp]
\centering
\includegraphics[width=\textwidth]{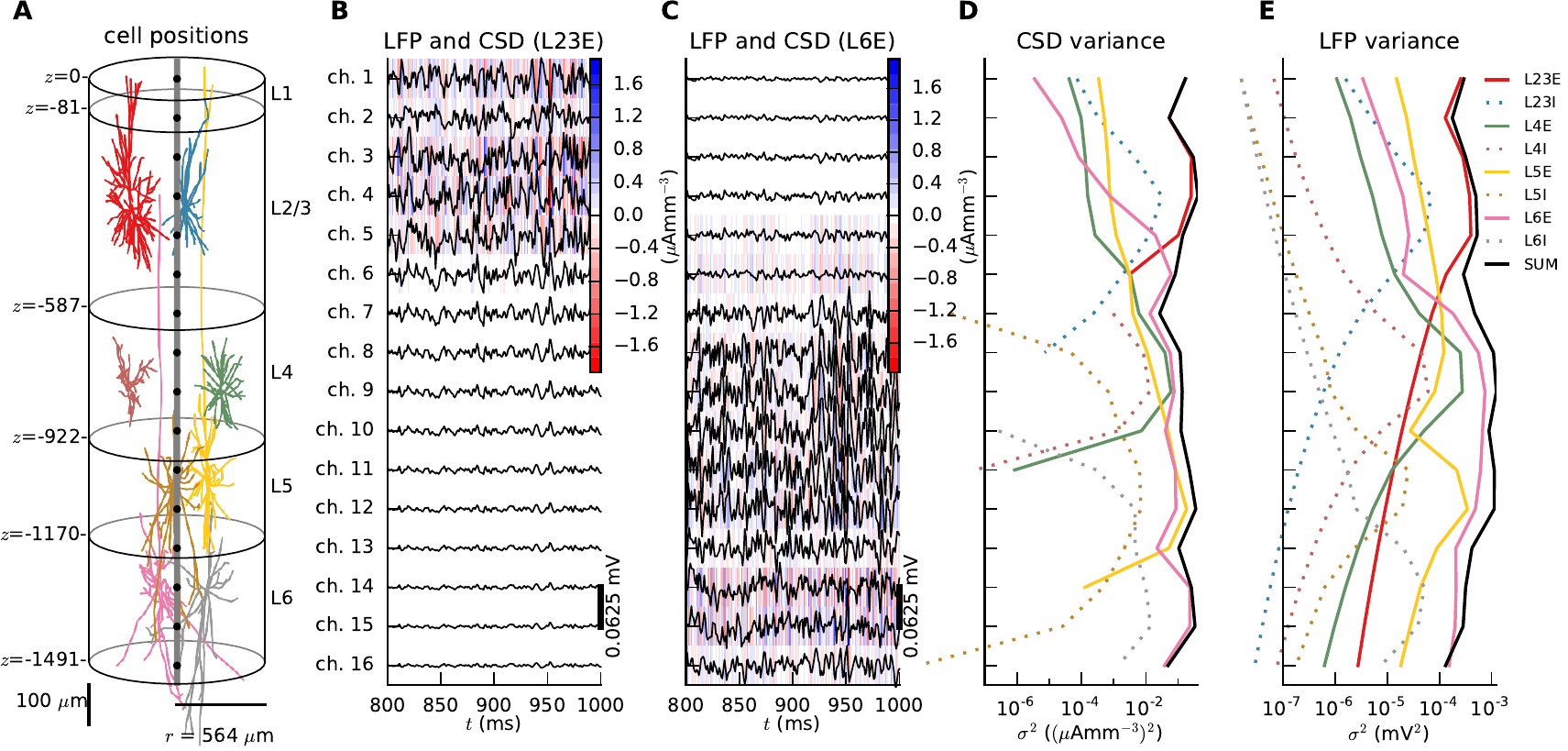}
\includegraphics[width=\textwidth]{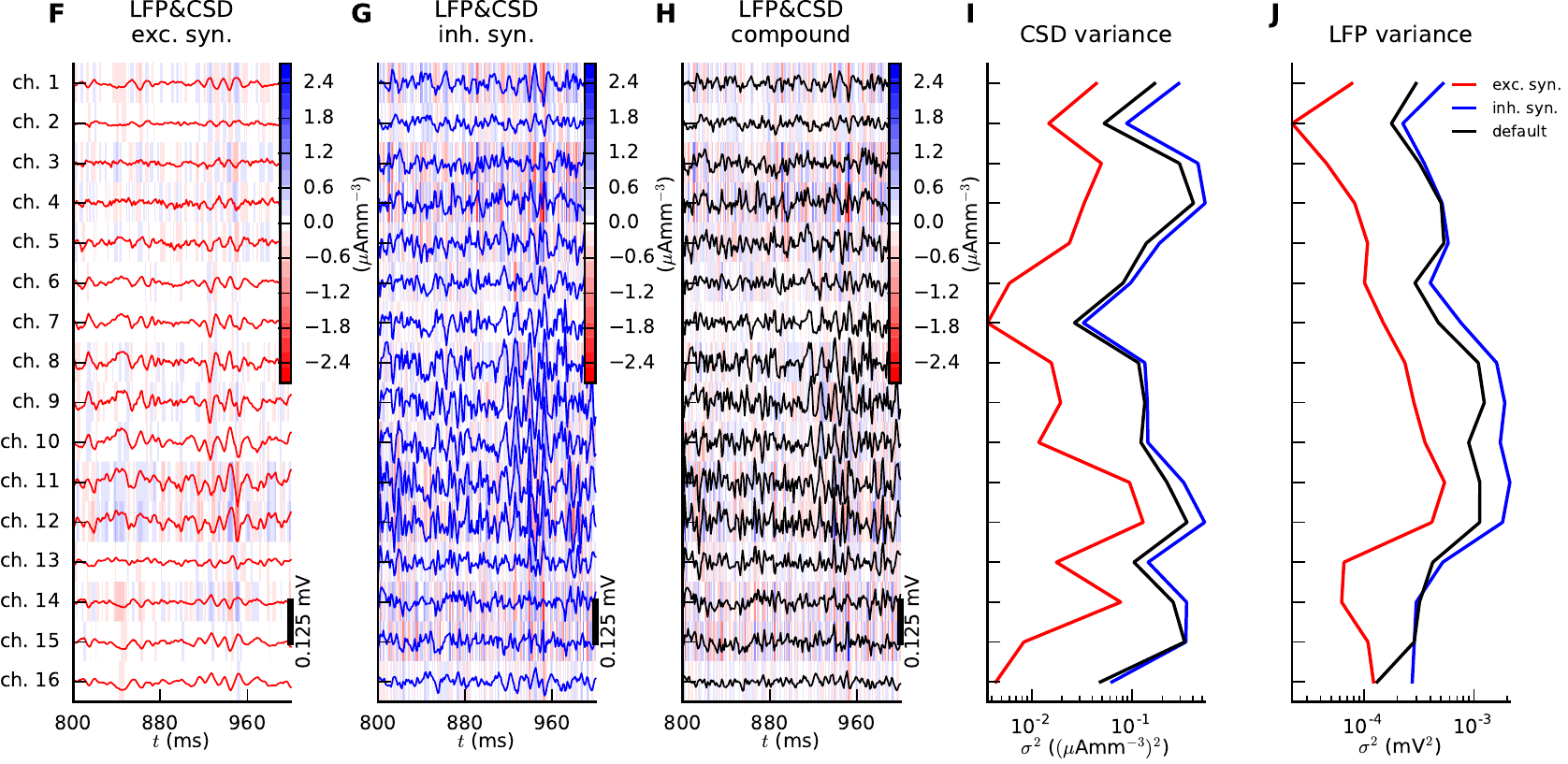}
\caption{{\bf Composition of CSD and LFP during spontaneous activity.}
{\bf A})~Representative morphologies of each population $Y$ illustrating dendritic extent.
{\bf B})~LFP (black traces) and CSD (color plot) produced by the superficial population L23E for  spontaneous activity in the reference network. 
{\bf C})~Similar to panel B for population L6E (summing over contributions of $y\in\{$p6(L4), p6(L56)$\}$). 
{\bf D})~CSD variance as function of depth for each individual subpopulation (colored) and for the full compound signal (black line).
{\bf E})~Same as in panel D, but for LFPs. 
{\bf F})~Compound LFP (red traces) and CSD (color plot) resulting from only excitatory input to the LFP-generating multicompartment model neurons.
{\bf G})~Conversely, LFP (blue traces) and CSD (color plot) resulting from only inhibitory input to the neurons.
{\bf H})~Full compound LFP (black traces) and CSD (color plot) resulting from both excitatory and inhibitory synaptic currents.
{\bf I})~Compound CSD variance as a function of depth with all synapses intact (black), or 
having only excitatory (red) or inhibitory synapse input (blue). 
{\bf J})~Same as in panel I, but for the LFP signal.  
}
\label{fig:9}
\end{figure}

Here, we focus on the LFP and CSD contributions from individual populations and different synapse types.
To quantify the contributions from each postsynaptic population, i.e., the CSD and LFP stemming from the transmembrane currents of
a population, we simply summed all single-cell CSD and LFP contributions from all neurons in the population. Results for spontaneous
network activity are shown in \Fref{fig:9}. 
As different cell types are assigned appropriate morphologies and cortical depths based on available anatomical data, 
a trivial consequence is that the neurons in each population make their main contributions to the CSD and LFP at depths spanned by their dendrites (\Fref{fig:9}B,C for post-synaptic populations L23E and L6E, respectively). For example, L6E neurons have 
a high density of afferent synapses on apical dendrites in layer 4,  and as a 
consequence,  the amplitude of CSDs and LFPs generated by population L6E  (\Fref{fig:9}C) is large in the vicinity of layer 4 and not just in layer 6. 

To quantify the relative contributions from the various populations we show in \Fref{fig:9}D,E the 
CSD and LFP variances across time 
(corresponding to power spectral densities summed over all non-zero frequencies) 
for all depths~\citep{Linden2011, Leski2013}.
For the compound CSD, only the L23E neurons contribute substantially in the superficial channels (ch. 1--4) (\Fref{fig:9}D). 
At deeper contacts the main contributing population is L6E.  L4E and L5E populations make sizable contributions only in the channels 
closest to their somatic location reflecting that the net associated return currents 
of their distributed synaptic inputs are largely restricted to somatic regions \citep{Linden2010}. 
The bulk of the variance of the CSD in layer 5 arises in about equal parts from the L6E and relatively sparse L5E populations.
The magnitude of the depth-resolved CSD variances of each layer's inhibitory population (L23I, L4I, ...) is consistently 
one order of magnitude or more smaller than that of the corresponding excitatory cell populations. Moreover, they span 
comparatively small depth ranges as determined by the maximum extent of the dendrites. 

The depth-resolved LFP variance (\Fref{fig:9}E) has similar features as the CSD variance, as expected from their common biophysical origin. However, volume conduction has some qualitative effects, such as the reduced relative contribution from the L6E population
in layer 2/3. As correlated synaptic inputs are known to amplify and increase the spread of the LFP generated from a cortical population~\citep{Linden2011,Leski2013} (see also \Fref{sec:correlations}), this may reflect that the synaptic inputs to the L23E population are more correlated than to the L6E population. 
Overall, we conclude that for the spontaneous activity in our network, the  L23E and L6E populations dominate the compound LFP and CSD
with only smaller contributions from the other excitatory populations.  The signal variances from the inhibitory populations are typically
much smaller than the contributions from the excitatory populations, suggesting that they can be safely neglected, 
in line with recent findings of \citet{Mazzoni2015}.

Although transmembrane currents in inhibitory neurons provide little of the observed CSD and LFP, the 
inhibitory synaptic inputs onto excitatory neurons provide a substantial contribution. In the present network,
inhibitory synaptic currents have a four-fold larger amplitude compared to most excitatory synapses (\Fref{tab:5}), 
inhibitory neurons have higher overall firing rates compared to excitatory neurons \citep{Potjans2014}, and 
inhibition specifically targets soma-proximal sections \citep{Markram2004}. 
Since our LFP-generating model is linear (passive cable formalism, linear synapse model), we can decompose the compound signal into contributions from each synapse type. For example, selective removal of either inhibitory or excitatory synaptic currents in the CSD and LFP modeling (\Fref{fig:9}F,G)
shows that the CSD and LFP signals (\Fref{fig:9}H) are dominated by inhibitory synaptic currents and their 
associated return currents (\Fref{fig:9}I,J) when the network operates in the spontaneous asynchronous irregular firing regime.
Further, at most depths the variance of the signal arising from inhibitory input exceeds the compound variance, 
implying that the inhibitory component is generally negatively correlated with the excitatory component (\Fref{fig:9}I,J). 
Visual inspection of \Fref{fig:9}F-H also reveals that the inhibitory dominance appears particularly strong at high frequencies, in accordance with the firing-rate PSDs in \Fref{fig:8}B showing less power of inhibitory spiking, and thus 
inhibitory synaptic input currents, at low frequencies.

\begin{figure}[!htbp]
\centering
\includegraphics[width=\textwidth]{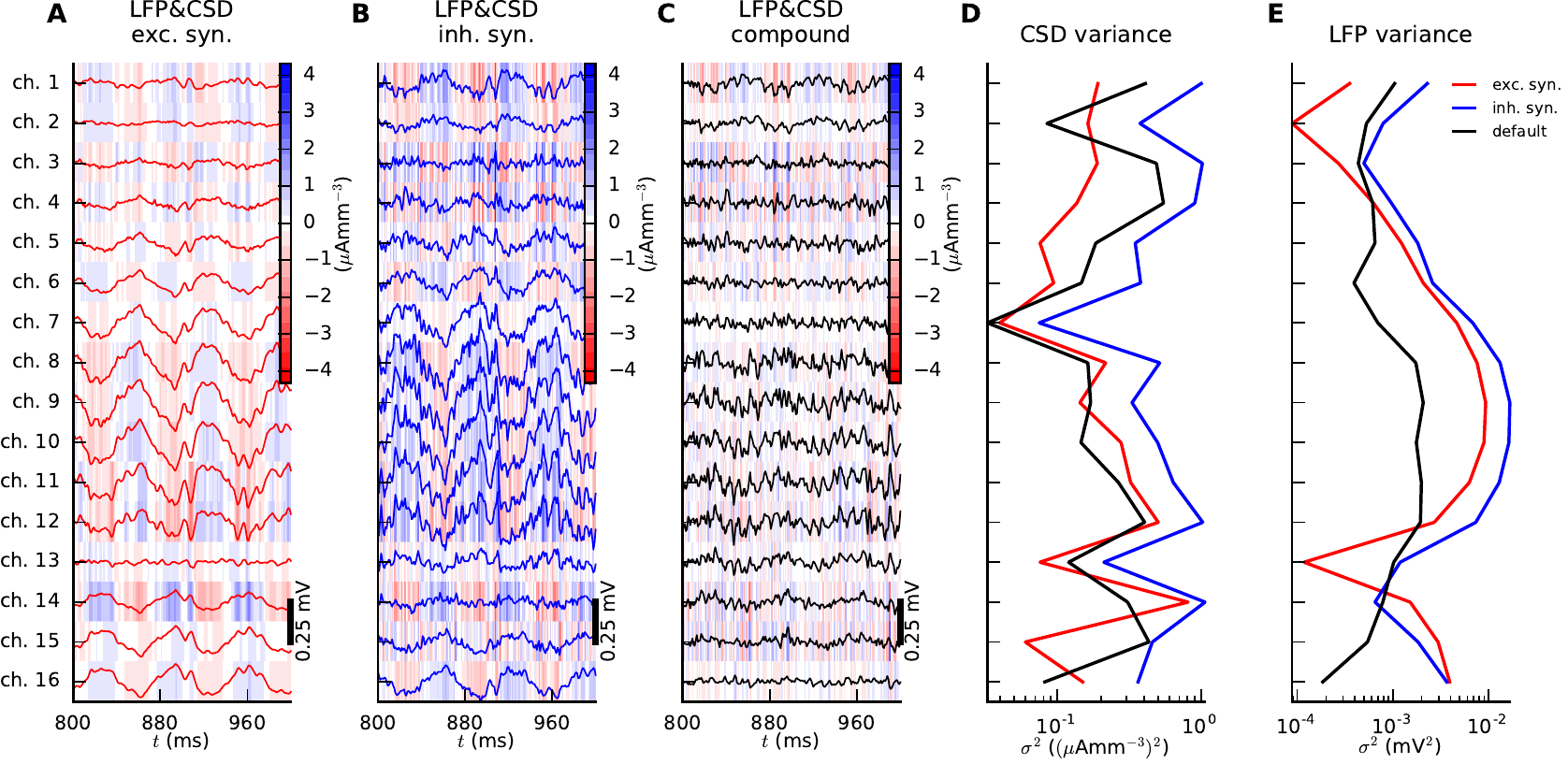}
\includegraphics[width=\textwidth]{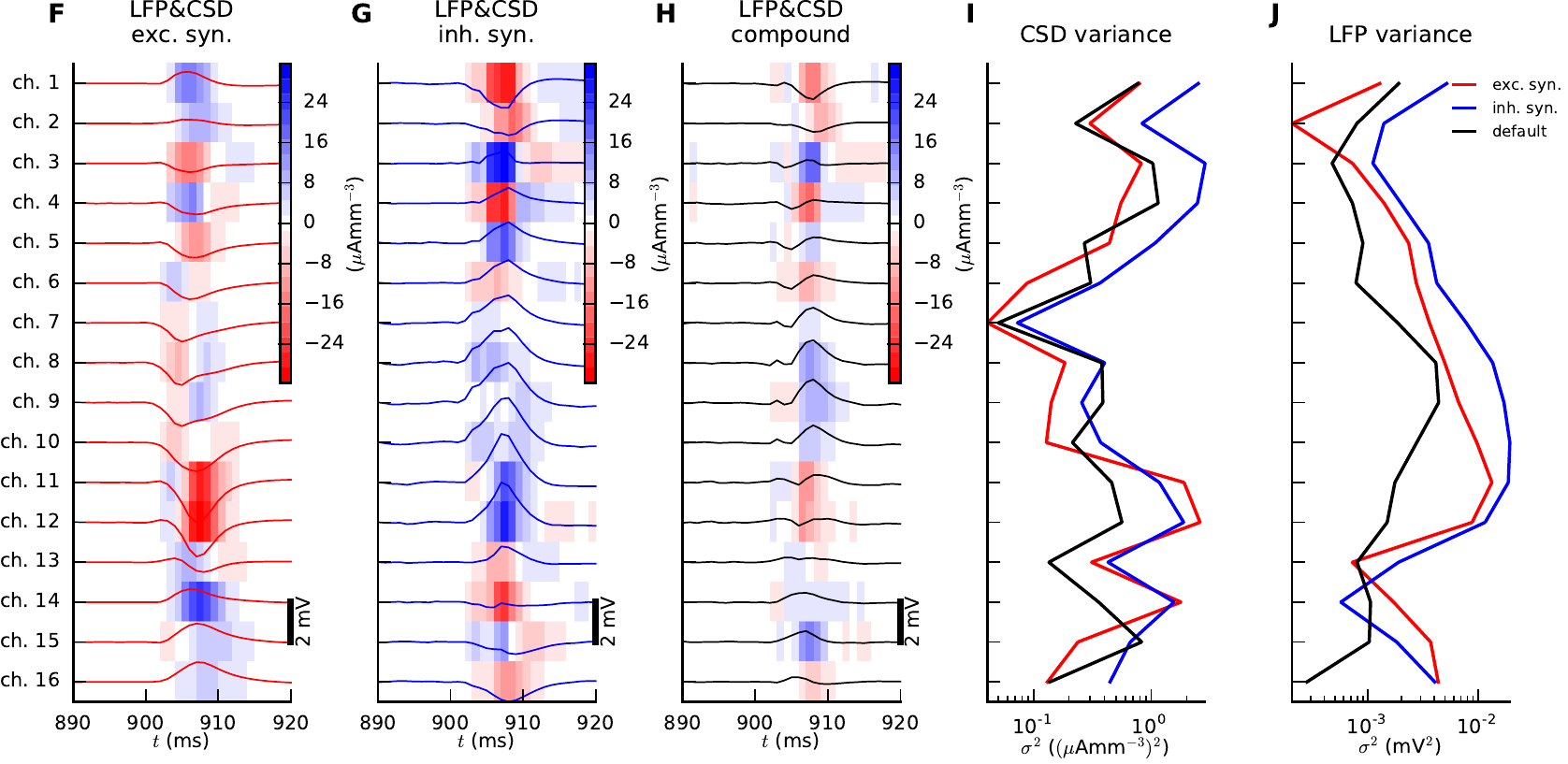}
\caption{\textbf{Decomposition of CSD and LFP into contributions due to excitatory and inhibitory inputs for thalamic activation.}
\textbf{A--E)}~Oscillatory thalamic activation ($f=15$~Hz). 
\textbf{F--J)}~Transient thalamic activations at $t=900+n \cdot 1000$~ms for $n = 0,1,2,3,4$.
Same row-wise figure arrangement as in \Fref{fig:9}F-J. 
}
\label{fig:10_2}
\end{figure}

The relative contribution from excitation and inhibition to the CSD and LFP depends on thalamic input and network state, however. 
For oscillatory thalamic input, the CSDs and LFPs from excitatory (\Fref{fig:10_2}A) or inhibitory synapses (\Fref{fig:10_2}B) alone show 
much stronger oscillations than the compound signals (\Fref{fig:10_2}C). 
This follows from the observations that the contributions to the CSD and LFP from excitatory and inhibitory synapses are anticorrelated, which, in turn, is a consequence of the dynamical balance between excitation and inhibition in asynchronous-irregular states of balanced random networks~\citep{Hertz10_427,Renart2010,Tetzlaff2012}.

For transient thalamic activation, the same cancellation of excitation and inhibition can be observed (\Fref{fig:10_2}F-H), although it is more pronounced at some depths (layers 5 and 6, i.e., channels 11-13) than at others, depending on whether excitation and inhibition are in phase or not (\Fref{fig:10_2}I,J). 
For such strong and transient thalamic input, the network activity is briefly imbalanced as inhibition cannot keep up with thalamic excitation on the short time scales.
Since thalamic input is most prominent in layer 4 (channels 7--10),  fewer cancellation effects are present here: 
in channel 9 in \Fref{fig:10_2}I the total CSD variance is, e.g., seen to be larger than the individual contributions from inhibitory and excitatory synapses.

\subsection{Effect of input correlations}
\label{sec:correlations}

Synaptic inputs to two neighboring cells are typically correlated because (i)  they receive, to some extent, inputs from the same presynaptic sources ('shared-input correlation'), and (ii) the spike trains of the presynaptic neurons may be correlated ('spike-train correlation'). 
The net synaptic-input correlation is determined by the interplay between these two contributions, shared-input correlations and spike-train correlations \citep{Renart2010,Tetzlaff2012}.
As the LFP is largely generated by synaptic inputs, synaptic-input correlations result in correlated single-cell LFP contributions $\phi_i(\mathbf{r},t)$ \citep[for details, see][]{Linden2011, Leski2013}.
As outlined in the following, these single-cell-LFP correlations play a dominating role for the spectrum of the compound LFP.

The power spectrum $P_{\phi}(\mathbf{r},f)$ (\Fref{eq:psd}) of the compound LFP 
$\phi(\mathbf{r},t) = \sum_{i=1}^N\phi_i(\mathbf{r},t)$ 
of a population of $N$ neurons is given by
\begin{equation}
 P_{\phi}(\mathbf{r},f) =  
     N \overline{P_{\phi}}(\mathbf{r},f) + N(N-1)\overline{P_{\phi}}(\mathbf{r},f)\overline{\kappa_{\phi}}(\mathbf{r},f)
    \label{eq:psd_compoundLFP}
\end{equation}
(see \Fref{sec:2.5.2} and \Fref{tab:10}).
Here $\overline{P_{\phi}}(\mathbf{r},f)$  
is the average single-cell LFP power spectrum (\Fref{eq:psd_mean_i}) and
$\overline{\kappa_{\phi}}(\mathbf{r},f)$  the average pairwise single-cell LFP coherence (\Fref{eq:coherence}), 
a measure for cross-correlations, across all cells.
Note that, while the first term in  \Fref{eq:psd_compoundLFP} scales linearly with the number of neurons $N$, the second term is proportional to $N(N-1)\approx N^2$ for large $N$.
Hence, for large $N$, even small cross-correlations may dominate the spectrum of the compound LFP. 
Here, we investigate this situation by calculating the power spectrum $P^{0}_{\phi}(\mathbf{r},f)$ 
of the compound LFP 
under the assumption of zero cross-correlation (where it simply reduces to a sum over single-cell spectra $P_{\phi_i}(\mathbf{r},f)$), 
and compare to the true spectrum $P_{\phi}(\mathbf{r},f)$.
The ratio between these quantities is given by
\begin{equation}
  \label{eq:P_ratio}
  \frac{P_{\phi}(\mathbf{r},f)}{P^{0}_{\phi}(\mathbf{r},f)} = 1+(N-1) \overline{\kappa_{\phi}}(\mathbf{r},f)\;.  
\end{equation}
With weak or no cross-correlations, i.e., 
$(N-1)\overline{\kappa_{\phi}}(\mathbf{r},f) \ll 1$, the ratio approaches unity, and the power of the compound LFP is essentially the sum of the power 
of the single-cell LFPs. For $N\overline{\kappa_{\phi}}(\mathbf{r},f) \gg 1$, i.e., in the correlation-dominated regime, this ratio is instead proportional to the number of neurons $N$. Note also 
that anti-correlated signals ($\overline{\kappa_{\phi}}(\mathbf{r},f)<0$) may lead to a ratio $P_{\phi}(\mathbf{r},f)/P^{0}_{\phi}(\mathbf{r},f) < 1$. 

\begin{figure}[!h]
\centering
\includegraphics[width=\textwidth]{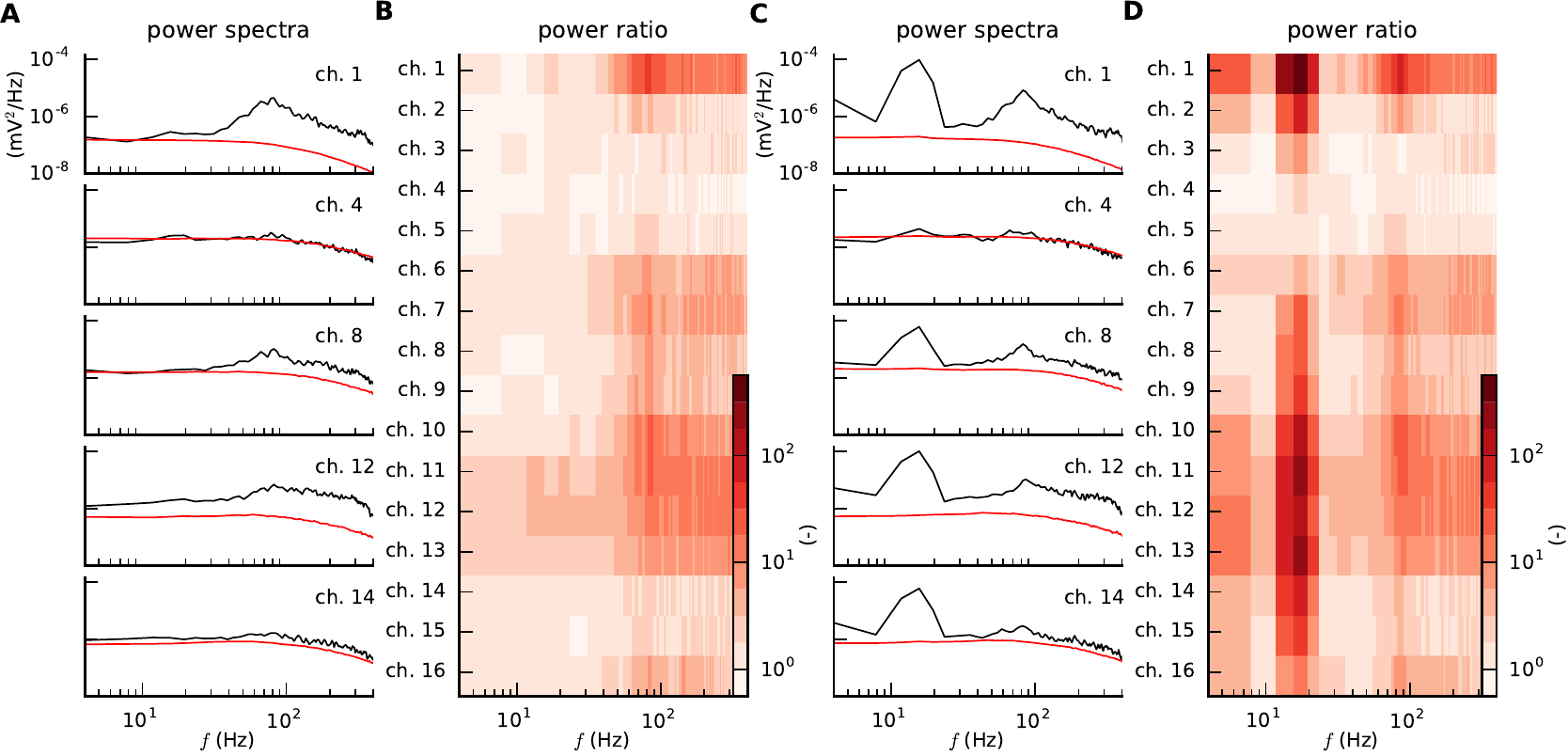}
\caption{
\textbf{Effect of single-cell-LFP cross-correlations on compound-LFP power spectra during spontaneous activity (A,B) and 
for oscillatory thalamic input (C,D).}
 \textbf{A,C)}~Compound-LFP power spectra $P_{\phi}(\mathbf{r},f)$ (black traces) and compound spectra 
$P^{0}_{\phi}(\mathbf{r},f)$ obtained when omitting cross-correlations between single-cell LFPs (red traces; 
see main text, \Fref{sec:correlations}, computed for 10\% of the cells and multiplied by a factor 10)
at recording channels corresponding to the centers of layers 1, 2/3, 4, 5 and 6.
\textbf{B,D)}~Depth and frequency-resolved ratio $P_{\phi}(\mathbf{r},f)/P^{0}_{\phi}(\mathbf{r},f)$ 
of LFP power spectra, cf.~\Fref{eq:P_ratio}.
}
\label{fig:11}
\end{figure}

In the example application, 
both for spontaneous (\Fref{fig:11}A,B) and for evoked activity (\Fref{fig:11}C,D) 
the compound power spectra $P_{\phi}(\mathbf{r},f)$ are systematically (across channels and frequencies) larger than 
$P^{0}_{\phi}(\mathbf{r},f)$, demonstrating the importance of cross-correlations in the present network.
Depending on the recording depth and frequency, the ratio
varies from $\sim 1$ to  $10^3$ (see \Fref{fig:11}B,D). 
For spontaneous activity (see \Fref{sec:3.2}, \Fref{fig:8}A-E), the largest effects of cross-correlations 
are typically found at higher frequencies (\Fref{fig:11}A,B). 
At low frequencies, 
cross-correlations are suppressed by inhibitory feedback (cf. \Fref{fig:8}B, \citet{Tetzlaff2012}). 
The thalamic sinusoidally modulated input to the network (\Fref{sec:3.2},  \Fref{fig:8}F-J) 
synchronizes single-cell CSDs and LFPs at the stimulus frequency and gives a large boost of the LFP power at this frequency (see peak in \Fref{fig:11}C,D). Close inspection reveals that there is also in fact a boost of the power at 
around 80 Hz, but much less so than around 15 Hz.
Note that the external activation hardly affects the single-cell spectra (see red curves in \Fref{fig:11}A,C).
LFP synchronization is thus mainly encoded in the phase of $\phi_i(\mathbf{r},t)$.

In conclusion, 
cross-correlations between single-cell LFP contributions play a pivotal role in 
shaping the compound LFP spectra (similar for CSD spectra, results not shown).
To account for the dominant features of the LFP (CSD) in such models, 
it is therefore essential to include the main factors determining the synaptic-input correlations, 
i.e.,~realistic correlations in presynaptic spike trains and shared-input structure.
The findings presented here for the cortical microcircuit model hold in general in the presence of correlated activity.
Only the details of the spectra depend on the specific underlying network dynamics.

\subsection{Network downscaling}
\label{sec:downscaling}

Due to the computational cost associated with modeling 
LFPs, it would be desirable to downsize the postsynaptic populations of multicompartment model neurons to a fraction $\gamma N$ ($\gamma\in\left(0,1\right)$~)
while leaving the point-neuron network at full size ($N$) 
and at the same time preserve the in-degrees of each postsynaptic cell. 
The power spectrum of the full-scale LFP can indeed be estimated from the population-averaged single-cell power spectra 
$\overline{P_{\phi^{\gamma}}}(\mathbf{r},f) \approx \overline{P_{\phi}}(\mathbf{r},f)$ and coherences $\overline{\kappa_{\phi^{\gamma}}}(\mathbf{r},f) \approx \overline{\kappa_{\phi}}(\mathbf{r},f)$ computed for downsized networks by means of \Fref{eq:psd_compoundLFP}.
These quantities are preserved except for deviations due to smaller sampling size $\gamma N$ (\Fref{eq:psd} in \Fref{tab:10}). 
However, due to lack of phase information in the power spectra, 
one cannot estimate the LFP time course.

\begin{figure}[!htbp]
\centering
\includegraphics[width=\textwidth]{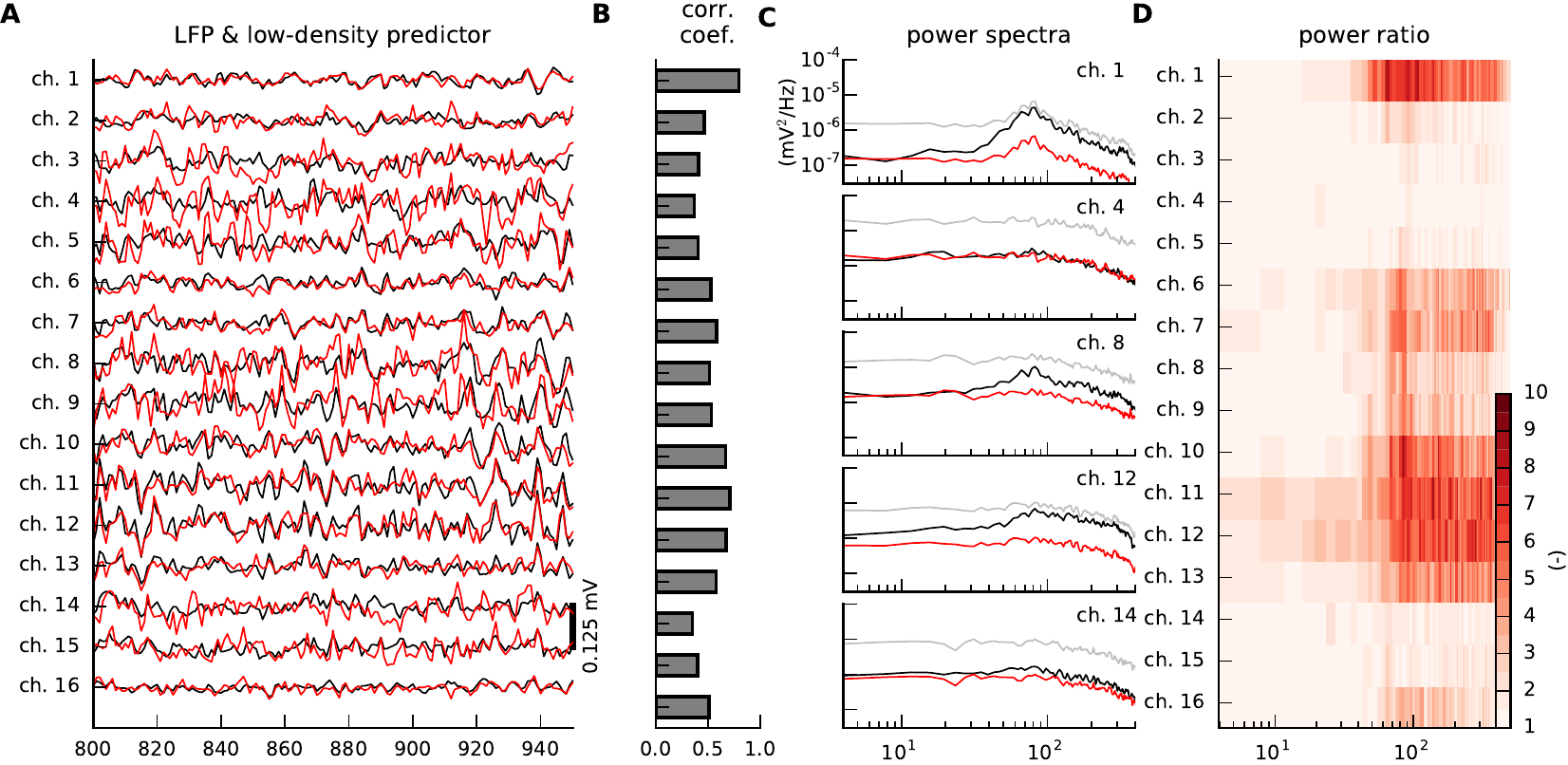}
\includegraphics[width=\textwidth]{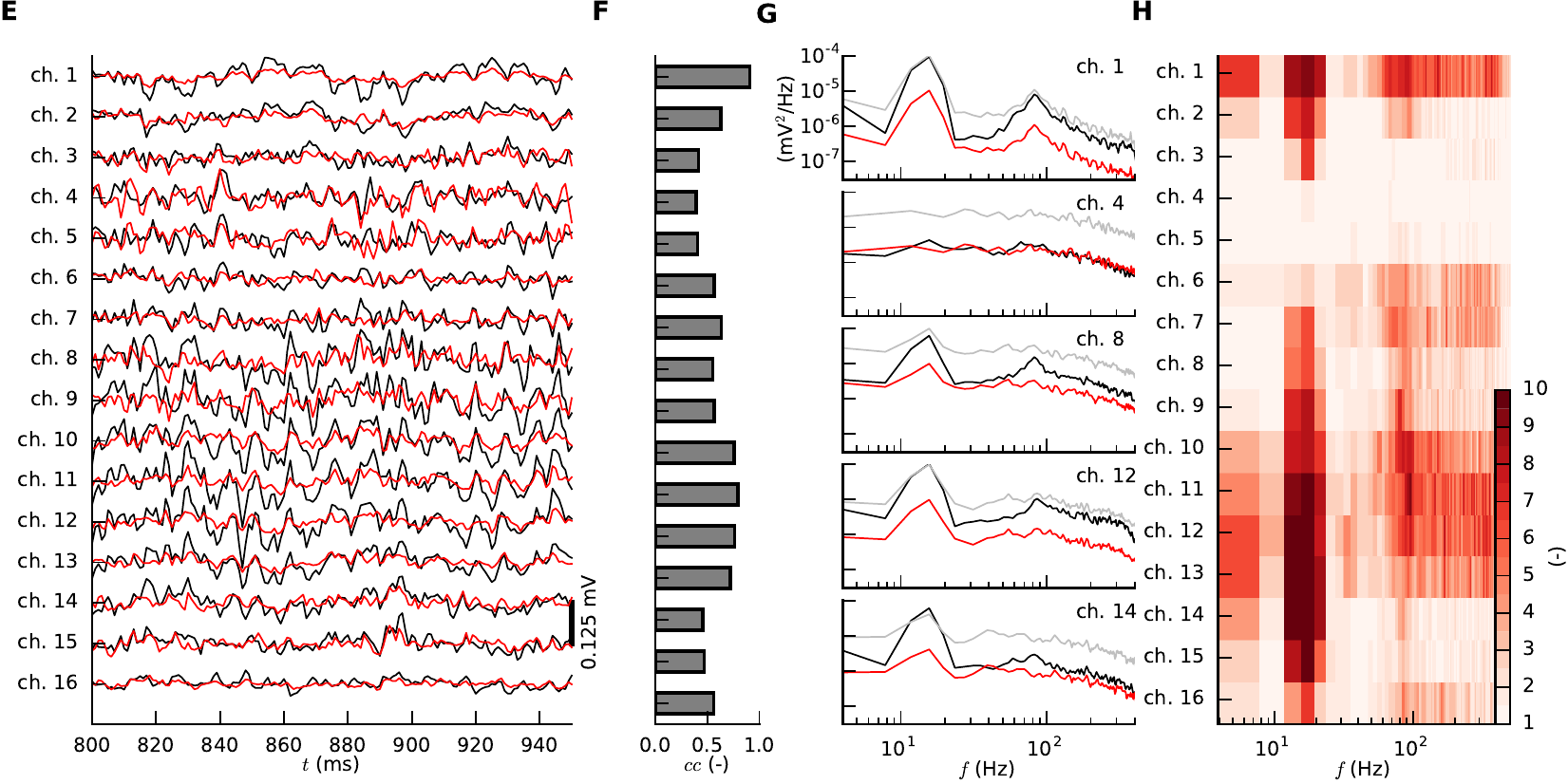}
\caption{\textbf{Prediction of LFPs from downsized networks.}
Top row: Spontaneous activity. Bottom row: Oscillatory thalamic activation.
{\bf A,E)}~Full-scale LFP traces $\phi(\mathbf{r},t)$ (black) and low-density predictors $\phi^{\gamma \xi}(\mathbf{r},t)$ (red) obtained from a fraction $\gamma=0.1$ of neurons in all populations and upscaling by a factor $\xi=\gamma^{-\frac{1}{2}}$. 
{\bf B,F})~Correlation coefficients between full-scale LFP and low-density predictor shown in panels A and E, respectively.
{\bf C,G})~Power spectra $P_{\phi}(\mathbf{r},f)$  and $P_{\phi^{\gamma \xi}}(\mathbf{r},f)$
of full-scale LFPs (black) and low-density predictors with $\gamma=0.1$ and $\xi=\gamma^{-\frac{1}{2}}$ (red) or $\xi=\gamma^{-1}$ (gray).
{\bf D,H}) Ratio $P_{\phi}(\mathbf{r},f)/P_{\phi^{\gamma \xi}}(\mathbf{r},f)$ between power spectra of full-scale LFP and low-density predictor with $\gamma=0.1$ and $\xi=\gamma^{-\frac{1}{2}}$ (cf.~ \Fref{eq:P_ratio2}). 
}
\label{fig:12}
\end{figure}

One could attempt to obtain a time-course estimate $\phi^{\gamma \xi}(\mathbf{r},t)$, i.e.,
a `low-density LFP prediction', of the full-scale signal $\phi(\mathbf{r},t)$
by upscaling single-cell LFPs $\phi_i(\mathbf{r},t)$ computed in the downsized setup 
by a scalar factor $\xi$ (cf. \Fref{eq:phi_gamma_xi}). 
Such a naive upscaling can grossly recover the amplitude of the full-scale LFP $\phi(\mathbf{r},t)$, 
but it still only partially reconstructs its detailed time course  (\Fref{fig:12}A,E).
Also, this approach does not generally give accurate power spectra as the two terms in \Fref{eq:psd_compoundLFP} 
scale differently with $\xi$:
The rescaling introduces a prefactor $\xi^2$ in the population-averaged single-cell power spectra 
$\overline{P_{\phi^{\gamma \xi}}}(\mathbf{r},f)\approx \xi^2 \overline{P_{\phi}}(\mathbf{r},f)$, while the coherences  $\overline{\kappa_{\phi^{\gamma \xi}}}(\mathbf{r},f) \approx \overline{\kappa_{\phi}}(\mathbf{r},f)$ 
are unchanged. 
Thus the compound spectra $P_{\phi}(\mathbf{r},f)$ and $P_{\phi^{\gamma \xi}}(\mathbf{r}, t)$ of the full-size LFP 
and the low-density LFP predictor, respectively, differ. Their ratio 
\begin{equation}
  \label{eq:P_ratio2}
  \frac{P_{\phi}(\mathbf{r},f)}{P_{\phi^{\gamma \xi}}(\mathbf{r},f)} 
  = \frac{1+(N-1) \overline{\kappa_{\phi}}(\mathbf{r},f)}
         {\gamma\xi^2+ \gamma\xi^2(\gamma N-1)  \overline{\kappa_{\phi}}(\mathbf{r},f)} 
  = \begin{cases} 1/(\gamma\xi^2) & \text{ for } \overline{\kappa_{\phi}}(\mathbf{r},f)=0 \\ 
    1/(\gamma^2\xi^{2}) & \text{ for } \overline{\kappa_{\phi}}(\mathbf{r},f)=1 \end{cases}
\end{equation}
demonstrates that in the general case there is no scaling factor $\xi$ which allows for the recovery of
the full-size compound LFP power, i.e., makes the ratio in \Fref{eq:P_ratio2} equal to one for all spatial positions $\mathbf{r}$ and frequencies $f$. 
This can only be done in the special case where $\overline{\kappa_{\phi}}(\mathbf{r},f$) is a constant $c$ ($0 \leq c \leq 1$).
Here the two extreme cases correspond to no correlation ($\overline{\kappa_{\phi}}(\mathbf{r},f)=0$ with $\xi=1/\sqrt{\gamma}$) and full correlation between all single-cell signals ($\overline{\kappa_{\phi}}(\mathbf{r},f)=1$ with $\xi=1/\gamma$).

The substantial scaling effects observed for our microcircuit model in the asynchronous state (\Fref{fig:12}A--D)
 suggest that correlations cannot be neglected even when modeling the LFP 
for spontaneous network activity. Choosing the scaling factor $\xi=1/\sqrt{\gamma}$ corresponding to 
 $\overline{\kappa_{\phi}}(\mathbf{r},f)=0$ (red lines in \Fref{fig:12}A,C)  leads to a severe underestimation of 
the full-size compound power spectrum (\Fref{fig:12}C,D).  Even though the correlation (i.e., Pearson's correlation coefficients)
between the full-size full-size LFP signals and low-density LFP predictions are quite high (\Fref{fig:12}B), the power ratios (\Fref{fig:12}D) reveal that the rescaled signals are systematically wrong in frequency bands where single-cell LFPs are most strongly correlated 
(i.e., the frequencies for which the compound spectra are much larger than the predictions when omitting cross-correlations  (\Fref{fig:11}A,B)).
Assuming the full-correlation scaling factor $\xi=1/\gamma$, on the other hand, typically overestimates 
the full-size compound power spectrum (cf. gray spectra in \Fref{fig:12}C), particularly at low frequencies. 

The results for the sinusoidally stimulated network (\Fref{fig:12}E--H) are quite similar to the spontaneous-activity results, 
except around the stimulation frequency 15~Hz where the modulated input leads to strongly correlated single-cell LFP
contributions and a strong boost of the compound LFP. 
The approximate downscaling procedure assuming the full-correlation scaling factor $\xi=1/\gamma$ thus essentially agrees 
with the full-size compound spectrum for 15~Hz (while giving a strong overestimation for frequencies other than $\sim$15 and $\sim$80 Hz).

We note in passing that in contrast to the power spectra, the computation of the \emph{spike-triggered averaged LFP} (staLFP)
\citep{Swadlow2002,Nauhaus2009,Jin2011,Denker2011} in downsized networks do not have a principled problem due to cross-correlations between single-cell LFPs.
As staLFPs are linearly dependent on the single-neuron LFP contributions, the only principled problem with downsizing is increased noise
in the estimates due to sampling over fewer postsynaptic neurons.

\subsection{LFP prediction from population firing rates}
\label{sec:3.3}

An important question in systems neuroscience is to what extent the dynamics of networks of thousands or millions of neurons
can be described by much simpler mathematical descriptions in terms of neural \emph{populations}~\citep{Deco2008,Blomquist2009}.
Likewise, we here ask the question of whether LFPs can be predicted from knowledge of the population firing 
rate~\citep{Einevoll2007,Moran2008,Einevoll2013}. The hybrid scheme 
is excellently suited for testing and development of simplified numerical schemes for LFP prediction as the ground truth, i.e., 
the LFP from the full network, is available as benchmarking data.

The use of current-based synapses and passive dendrites in the present application of the hybrid scheme, 
renders synaptic events independent of each other in the LFP prediction. 
This inherent linearity results in 
a unique spatio-temporal relation $H_{X}^{i}(\mathbf{r},\tau)$ 
for $\tau \in [-\infty, \infty]$ between a spike event of a point neuron $i$ in population $X$ and its contribution to the compound LFP $\phi(\mathrm{r},t)$ from all its postsynaptic multicompartment model neurons. 
In this scheme the link is causal, i.e., the spikes drive the LFP, so that $H_{X}^{i}(\mathbf{r},\tau)=0$ for $\tau<0$ (as in laminar population analysis (LPA)~\citep{Einevoll2007}). 
$H_{X}^{i}(\mathbf{r},\tau)$ encompasses connectivity, spike transmission delays and all postsynaptic responses including effects of synaptic input currents and passive return currents. 
With a linear, current-based model such as our example cortical column, 
it is in principle possible by linear superposition to fully reconstruct the compound LFP if $H_{X}^{i}(\mathbf{r},\tau)$ and the 
spike times $t^i_l$ are known for all neurons $i$ in each population of the network.
It is, however, in the case of large networks impractical to assess each $H_{X}^{i}(\mathbf{r},\tau)$, 
as the LFP response needs to be determined for every neuron separately.
In contrast, a large reduction in dimensionality can be achieved by determining the population-averaged 
LFP responses $\overline{H}_X(\mathbf{r},\tau)$ of a spike within each population $X$. We thereby ignore heterogeneity in kernels $H_{X}^{i}(\mathbf{r},\tau)$ due to the variability in the connections from neurons in population $X$.
An approximate compound LFP $\phi^\ast(\mathbf{r},t)$ based on population firing rates ~\citep{Einevoll2007} 
can be computed from these extracted population kernels by means of the 
convolution $\phi^\ast(\mathbf{r},t) = \sum_X \left(\nu_X \ast \overline{H}_X\right)(\mathbf{r}, t)$, where $\nu_X(t)$ are 
the instantaneous population firing rates. 

\begin{figure}[!htbp]
\centering
\includegraphics[width=\textwidth]{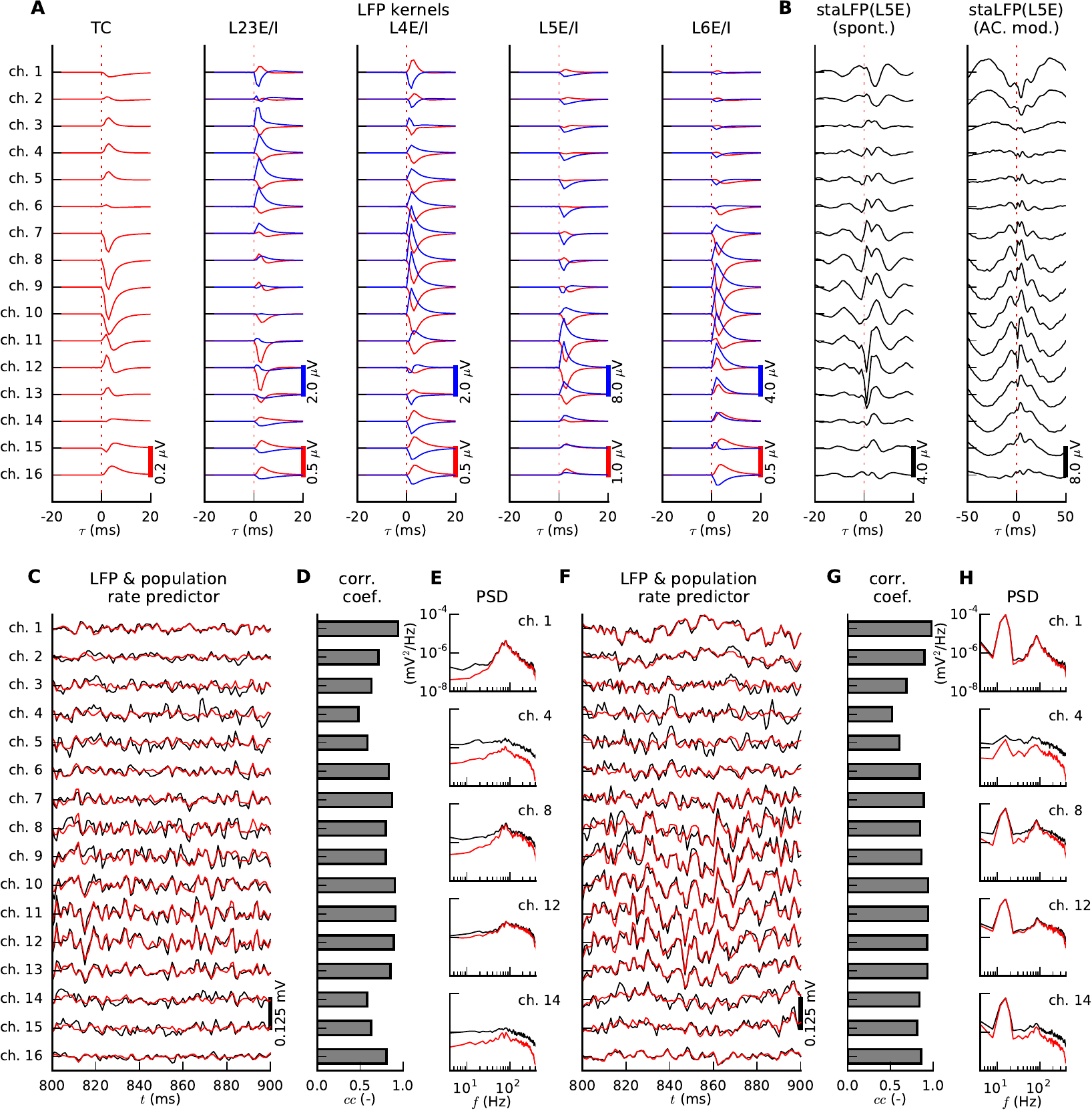}
\caption{ \textbf{Linear prediction of LFPs from population firing rates.}
{\bf A})~LFP responses $\overline{H}_X(\mathbf{r},\tau)$ (kernels) to simultaneous firing of all neurons in a single presynaptic population $X$ (see subpanel titles) at time $\tau=0~$ms, normalized by size $N_X$ of the presynaptic population (red/blue: responses to firing of excitatory/inhibitory presynaptic populations).
{\bf B})~LFPs triggered on spikes of L5E neurons during spontaneous activity (left) and oscillatory thalamic network activation (right), averaged across all L5E spikes ($T$=5~s simulation time).
{\bf C,F})~LFP traces of the full model (black) compared to predictions (red) obtained from superposition of linear convolutions of population firing rates $\nu_X$ with LFP kernels $\overline{H}_X(\mathbf{r},\tau)$ shown in A.
{\bf D,G})~Correlation coefficients between LFPs and population-rate predictors shown in C and F. 
{\bf E,H})~Power spectra of LFPs (black) and the population-rate predictors (red) for different recording channels.  
Panels C-E and F-H show results for spontaneous activity and oscillatory thalamic activation, respectively. 
}
\label{fig:13}
\end{figure}

Here we estimate the population LFP kernels by computing the response to synchronous activation of all neurons in a 
population (\Fref{fig:13}).
The spatio-temporal kernels $\overline{H}_X(\mathbf{r},\tau)$ are extracted from time slices $[t_X -20~\text{ms}, t_X+20~\text{ms}]$ 
of the compound LFP response $\phi(\mathbf{r},t) / N_X$, where $N_X$ is the number of neurons in a presynaptic population. 
The procedure results in unique kernels $\overline{H}_X(\mathbf{r},\tau)$ 
for each excitatory and inhibitory population in the network (\Fref{fig:13}A).

In the example application, the population kernels $\overline{H}_X(\mathbf{r},\tau)$ differ 
significantly between populations. 
Excitatory spike events result in prominent LFP negativities 
at depths where most connections are made, 
such as in layer 4 (channels 7--10) for thalamocortical connections 
($\overline{H}_\text{TC}(\mathbf{r},\tau)$, column 1 in \Fref{fig:13}A, cf. \Fref{fig:5}C).
In contrast,
spikes of inhibitory point neurons on average produce prominent LFP positivities in their corresponding layer, such as in layer 2/3 
(channels 3--6) for population L23I 
(column 2 in \Fref{fig:13}A). 
In all cases, the signatures of opposite-sign return currents and also other, 
weakly connected populations are seen across depth. 

As seen in \Fref{fig:13}C,F the population-rate predictions are in good qualitative 
agreement with the ground-truth LFP for both spontaneous and sinusoidally modulated network activity 
with correlation coefficients ($cc$) between 
0.48 and 0.94 for spontaneous activity (\Fref{fig:13}D) and 0.52 and 0.98 for thalamically evoked oscillations (\Fref{fig:13}G).
Overall, the population-rate predictions appear to be best for the lower frequencies (\Fref{fig:13}E,H), while the inherent variability in the 
individual point-neuron kernels $H_{X}^{i}(\mathbf{r},\tau)$ (which is not accounted for in the population approximation) has a larger
effect on the higher frequencies. This can be understood on biophysical grounds, as the lower frequencies are expected
to mainly reflect the gross 
anatomical features of the postsynaptic populations and their presynaptic connections patterns 
where the individual variability plays a lesser 
role~\citep{Linden2010,Pettersen2012}.
The correlation coefficients and power spectra of the population-rate prediction thus show that the population-rate LFP predictor is more accurate than the low-density 
LFP predictors (\Fref{fig:12}C,F) in case of substantial downscaling.

Although the spike-triggered average LFP (staLFP) \citep{Swadlow2002,Nauhaus2009,Jin2011,Denker2011}, 
calculated as the cross-covariance between the population spike rate $\nu_X(t)$ and the compound LFP $\phi(\mathbf{r}, t)$ 
divided by the total number of spikes (i.e., $\int_0^T \nu_X(t) dt$), 
is related to our LFP population kernels, 
it measures very different aspects  of cortical dynamics. 
The population kernels $\overline{H}_X(\mathbf{r},\tau)$ are causal and  
independent of effects of spike-train correlations. 
The staLFP, on the other hand, is non-causal and strongly depends on spike-train correlations, and thus also network state ~\citep{Einevoll2013}. 
The staLFP is thus not only very different from $\overline{H}_X(\mathbf{r},\tau)$, it also varies strongly between the spontaneous and sinusoidally modulated network state  (see example for L5E neurons in \Fref{fig:13}B).

\section{Discussion}
\label{sec:discussion}

We have here described a hybrid modeling scheme for computing the local field potential (LFP) incorporating both large-scale neural network dynamics 
and the biophysics underlying LFP generation on the single-neuron level. 
The hybrid modeling scheme was illustrated with a full-scale network model of a cortical column in early sensory cortex \citep{Potjans2014}, and
the impact of individual populations, network dynamics and cell density on the mesoscopic LFP signal was investigated.

\subsection{The hybrid LFP modeling scheme}

The hybrid scheme combines the simplicity and efficiency of point-neuron network models with 
the biophysics-based modeling of LFP by means of multicompartment model neurons with detailed dendritic morphologies.
The neuronal network dynamics are governed by the point-neuron network model independent of LFP predictions.
The spikes of the point-neuron network are distributed to the synapses of the multicompartment model neurons 
with realistic cell-type and layer-specific connectivity.
Synapse activation results in spatially distributed transmembrane currents, 
which are mapped to an LFP signal according to well-established volume-conduction theory.

A main motivation for developing the hybrid LFP modeling scheme was to obtain the ability to compute LFPs for 
a key class of network models that are amenable to mathematical analysis and can provide intuitive understanding of emerging network dynamics, 
namely point-neuron models. 
Similar to networks of anatomically and biophysically detailed neuron models, point-neuron networks can generate realistic spiking activity.
In addition, the hybrid modeling scheme brings a substantial computational advantage:
With present-day computing and software technologies, point-neuron networks with $\sim$100,000 neurons can be modeled with laptop computers, 
and networks comprising millions of point neurons can be routinely simulated on high-performance compute facilities~\citep{Helias2012, Kunkel2014}.
Until now, the largest simulation of LFPs based on networks of
multicompartmental neuron models with reconstructed morphologies, in contrast, comprised about 12,000 neurons and was done on 
a Blue Gene/P supercomputer with 4096 CPUs~\citep{Reimann2013}. 
The linearity of electromagnetic theory allows for the implementation of the LFP hybrid modeling scheme as an
``embarrassingly'' parallel operation \citep{Foster1995}.
Therefore, the results for the cortical microcircuit application with $\sim$78,000 neurons were obtained with only 256 CPUs, and 
could even be acquired with much smaller computing architectures. 

A full implementation of the hybrid scheme is provided by the freely available \texttt{Python} module \texttt{hybridLFPy} (\url{http://github.com/INM-6/hybridLFPy}).
Our model  implementation in \texttt{hybridLFPy} relies on the publicly available \texttt{NEURON} software as a simulation backend through \texttt{Python} with \texttt{LFPy} \citep{Linden2014} for calculating single-cell LFP contributions.
This ensures flexibility and compatibility with a large library of existing neuron models, with or without active channels 
and with morphologies of arbitrary levels of detail, obtained from ModelDB \citep{Hines2004}, NeuroMorpho.org \citep{Ascoli2007} or other resources.
While we did use \texttt{NEST} \citep{eppler_2015_32969} for simulating our reference network, 
the \texttt{hybridLFPy} module can be used in combination with any other neural-network simulation software.

The present hybrid LFP scheme involves several assumptions with respect to 
(i) the generation of realistic spiking activity, 
(ii) forward modeling of extracellular potentials, and 
(iii) the combined use of point-neuron networks and multicompartment modeling.
In the following we review the main assumptions and discuss potential extensions:

(i) \emph{Spike-train generation by point-neuron networks:}
Although highly simplified, 
single-compartment models of individual neurons (point-neuron models)
can mimic realistic spiking for a variety of cell types \citep{Izhikevich2008,Kobayashi2009,Yamauchi2011} 
and can make accurate predictions of single-cell firing responses under in-vivo like conditions \citep{Jolivet2008, Gerstner2009}. 
Moreover,  networks of point neurons can reproduce a number of activity features observed \emph{in vivo}, such as
spike-train irregularity \citep{Softky1993,Vreeswijk1996, Amit1997, Shadlen1998},
membrane-potential fluctuations \citep{Destexhe1999}, asynchronous
firing \citep{Ecker2010,Renart2010,Ostojic2014}, 
correlations in neural activity \citep{Gentet2010,Okun2008,Helias2013}, self-sustained activity \citep{Ohbayashi2003a,Kriener2014} 
and realistic firing rates across laminar cortical populations \citep{Potjans2014}. Note that the hybrid LFP modeling scheme is not necessarily restricted to point-neuron networks as generators of spiking activity. In principle, they could, for example, be replaced by statistical models of spike generation \citep{Linden2011,Leski2013}, 
or even experimentally measured spiking activity.

(ii) \emph{Biophysical forward modeling of LFPs:}
The biophysical forward model described by \Fref{eq:extracellular1}, implemented in \texttt{LFPy} \citep{Linden2014}, 
underlies the presently used computational scheme for LFPs of point-neuron networks. 
This forward model is based on well-established volume conductor theory~\citep{Rall1968,Holt1999} 
and assumes an \emph{infinite}, \emph{isotropic} (same in all directions), 
\emph{homogeneous} (same in all positions) and \emph{ohmic} (frequency-independent)
extracellular medium represented by a scalar conductivity $\sigma_\text{e}$.
However, one could generalize the forward model in a straightforward manner to account for anisotropy \citep{Nicholson1975, Logothetis2007, Goto2010}, 
or jumps in conductivities at tissue interfaces~\citep{Pettersen2006,Gold2006,Hagen2015,Ness2015}.
For even more complicated geometrical spatial variations of the conductivity, 
the forward modeling problem can always be solved by means of Finite Element Modeling (FEM)~\citep{Ness2015}.
Recent experiments have only found a small frequency dependence of the extracellular conductivity $\sigma_\text{e}$ at LFP frequencies 
($f\lesssim$~500~Hz, \citet{Logothetis2007,Wagner2014}), but see \citet{Gabriel1996,Gabriel2009,Bedard2009}. In any case the forward model could still be applied with frequency-dependent conductivity by means of Fourier decomposition where each frequency component of the LFP signal is considered separately. 
For more information on possible generalizations of the biophysical forward-modeling scheme, see \citet{Pettersen2012}.
Finally, we assumed the so-called disc-electrode approximation and averaged the computed LFP signal across the electrode 
surface~\citep{Moulin2008,Linden2014,Ness2015}. 
Although electrode impedance will affect the measurement, it appears that confounding effects from this can easily be avoided with present-day LFP recording techniques~\citep{Nelson2010}, hence few compelling reasons exist to incorporate additional temporal filters.

(iii) \emph{Combined use of point-neuron and multicompartmental models:}
The key approximation in the hybrid LFP scheme comes from the combined use of  point-neuron (single-compartment) and multicompartmental neuron models. 
The multicompartment neurons are mutually unconnected, have no outgoing (efferent) connections, 
and are solely used to compute the LFP.
Further, due to dendritic filtering,
the somatic postsynaptic potentials in the multicompartment model neurons are not identical to those of their point-neuron counterparts.
This inconsistency could, at least partially, be resolved by 
adjusting the amplitudes and temporal shapes of the synaptic currents in either the multicompartment neurons or 
the point neurons \citep{Koch1985,Wybo2013, Wybo2015}.

\subsection{Applications of the hybrid LFP scheme}

The hybrid scheme is not limited to the example point-neuron network model and the particular multicompartment neuron models chosen here.
It can be applied to networks
(i) of arbitrary topology (graph structure, distance dependencies, dimensions), 
(ii) with any number of populations,
(iii) with arbitrarily complex point-neuron (e.g. LIF, Izhikevich, MAT, Hodgkin-Huxley) and synapse dynamics (e.g., current-based, conductance-based, static, plastic), and
(iv) any level of biophysical detail in multicompartment neuron models (e.g., morphologies, active channels).

For illustration, we used the hybrid scheme to compute LFPs along a virtual laminar multielectrode from activity in a multilayered spiking point-neuron network, 
modeling signal processing in a patch of primary visual cortex. The network consisted of $\sim$78,000 neurons organized in four layers, each with an excitatory and an inhibitory population, representing a cortical patch of $\sim$1~mm$^2$~\citep{Potjans2014}. 
Altogether 16 different cell types and 10 different morphologically reconstructed neurons were used in the LFP calculation. 
This cortical microcircuit model is well-suited for the illustration of the hybrid scheme due to its
(i) minimum level of detail in single-neuron dynamics of both point neurons (LIF) and multicompartment neurons (passive membranes),
(ii) realistic neuron density allowing investigation of the effects of correlations and scaling of network size, and 
(iii) its spatial organization of multiple populations across cortical layers which yields cancellation effects 
not captured by LFP proxies such as in \citet{Mazzoni2015}.

Even though the example application was based on a generic network model biased towards cat visual cortex
and not tuned to address specific experiments, its spiking activity nevertheless matched experimental findings \citep{Potjans2014}.
We even observed the predicted LFP to be in qualitative accordance with LFP measurements in primary sensory cortices 
from a variety of animal species and sensory modalities in terms of 
(i)~LFP amplitude, for both spontaneous ($\simeq$0.1~mV~\citep{Maier2010, Hagen2015}) and stimulated activity
($\sim$1--3~mV)~\citep{Mitzdorf1979,Mitzdorf1985,Castro-Alamancos1996,Schroeder1998,Di1990,Einevoll2007}, 
and (ii)~stimulus-evoked spatiotemporal LFP and CSD patterns~\citep{Mitzdorf1985,Einevoll2007,Reyes-Puerta2015}. 
This supports the overall biological plausibility of the hybrid LFP scheme.

For the present example the LFP was dominated by synaptic inputs, and their associated return currents, on excitatory neurons, 
in particular onto pyramidal cells in layers 2/3 and 6. Further, contributions from inhibitory synaptic inputs typically 
dominated the contributions from the excitatory inputs, 
particularly for LFPs stemming from spontaneous network activity. 
Although the main point of employing the present example was to illustrate the use of the hybrid LFP scheme and not to  make  predictions for specific neural systems, 
we note in passing that a dominance of inhibitory synaptic inputs appears to be in agreement with LFPs generated in the CA3 region of hippocampus as observed in an \emph{in vitro}  setting~\citep{Bazelot2010}.
In accordance with a previous study~\citep{Linden2011} we found 
that correlations in synaptic input play a major role in determining the CSD and LFP stemming  from the network activity, 
for both spontaneous and stimulus-evoked activity. 
We further showed that due to inevitable correlations between synaptic input currents
the main features of the LFP can only be correctly predicted by a full-scale model.
As our ambition is to compute LFPs also for extended point-neuron networks with millions of neurons covering, for example,
entire cortical areas, we finally demonstrated 
how the hybrid modeling scheme can  predict LFPs from population firing rates rather than from spikes of individual neurons~\citep{Einevoll2007}.

The present microcircuit model application 
involving simplified, passive multicompartment populations 
was used here to study the effect of the spatial connectivity on the laminar pattern of spontaneous and stimulus-evoked CSD and LFP signals. 
The application thus represented a minimal approach incorporating spatial features 
in LFP predictions of multilaminar point-neuron networks. 
However, several of the simplifying model assumptions 
made in the present example application can straightforwardly be generalized. 
In particular, such generalizations concern 
(i) the synaptic connectivity between point neurons and the equivalent multicompartment neurons,
(ii) the absence of active conductances,
(iii) the positioning of the cells,
(iv) the reconstructed morphologies, 
(v) the representation of external inputs, and 
(vi) the fact that the model only encompassed the 
local circuitry of a $\sim$1~mm$^2$ patch of cortex.

(i) \emph{Synaptic connectivity:} 
While the population-specific connection probabilities, delay distributions, synapse time constant and mean synaptic weights were identical 
for the connections in the point-neuron network and those between point neurons and 
LFP-generating multicompartment neurons, the exact realizations of the two types of connectivities were different.
In contrast to the point-neuron network, each cell of a particular type had a fixed in-degree, i.e., a fixed number of synaptic inputs, 
and a fixed synaptic current amplitude 
in the LFP modeling step. 
We positioned the synapses randomly with the prescribed layer specificity, i.e., without clustering onto specific dendrites.
The use of the hybrid LFP scheme, however, is not restricted to these or any other specific assumptions about the synaptic connectivity patterns. 
One could, for example, gather all point-neuron network connections and corresponding weights and delays 
for use with the LFP-generating multicompartment neuron populations. 
However,  for large networks this would require additional computing and memory 
resources as the number of recorded connection weights and delays grow proportionally to $N^2$ in a network of $N$ neurons
with fixed connection probabilities.

(ii) \emph{Active conductances:} 
In the present study we neither included the active channels underlying spike generation, nor 
active dendritic conductances in the multicompartment neuron models~\citep{Remme2011}.
Experiments suggest that the contribution to the LFP from the former is small 
in stimulus-evoked recordings from sensory cortex, at least for the low frequencies of the LFP~\citep{Pettersen2008}, 
but see \citet{Ray2011a}.
In any case the contribution of the spikes to the extracellular potentials, including the LFP, 
could be included in the present scheme by 
giving each spike produced in the point-neuron network simulations a cell-type-specific spatiotemporal signature in the computation of the extracellular potential 
(e.g., as calculated in \citet{Holt1999,Hagen2015}).
Substantial effects of active dendritic conductances on the LFP were observed in a two-layered model by \citet{Reimann2013}, but this 
should be further explored. Recent modeling 
results~\citep{Ness2015b} suggest that for the purposes of LFP prediction, active dendritic conductances can, 
at least for subthreshold potentials, be effectively described by means of "quasi-active linearized" 
theory~\citep{Sabah1969,Koch1984a,Remme2011}. This simplifies the LFP modeling substantially and makes the computation similar 
in complexity to the present case with purely passive membranes.

(iii) \emph{Soma positioning:}
For model conciseness, 
the somas of all neurons belonging to a specific cortical populations were set to have the same cortical depth, 
cf.~\Fref{fig:6}E. By instead assuming a biologically more plausible, 
distribution of soma depths, the CSD profiles 
are expected to be spatially smoothed compared to the present profiles, e.g., \Fref{fig:6}F. Also the LFP profiles will be affected, 
but to a lesser degree since the LFP profiles already are 
spatially smoothed due to volume conduction effects.

(iv) \emph{Reconstructed morphologies:}
We further chose to rely on a small number of highly detailed reconstructed dendritic morphologies from experimental preparations, 
and partly reuse morphologies across neural populations. Obviously, larger sets of distinct reconstructed dendritic morphologies
can be used in future studies as they become available. The effect of dendritic morphologies on generated LFP can be further 
assessed by use of stylized~\citep{Tomsett2014,Glabska2014} or artificially grown morphologies resembling real 
neurons~\citep{Cuntz2010,Cuntz2011,Torben-Nielsen2014,Mazzoni2015}. 

(v) \emph{External input:}
In our point-neuron network modified from \citet{Potjans2014}, we assumed the depolarizing input from surrounding cortex and remote areas to
be represented by deterministic, fixed-amplitude input currents (DC inputs) rather than independent Poisson spike trains. We thus avoided
the generation and storage  of $O(10^5)$ high-rate uncorrelated Poisson spike trains and distributing the corresponding spike events onto the multicompartment model neurons, but this can be introduced in future applications.

(vi) \emph{Scale:}
Our network model represents only an isolated cortical column under $\sim$1~mm$^2$ of pial surface \citep{Potjans2014}, 
but work is underway to extend the network to a larger scale, e.g., 
by incorporating additional cortical areas~\citep{Schmidt2015} or extending the network size in the lateral directions~\citep{Senk2015}.
In addition to allowing for predictions of LFPs at several lateral positions as measured by multishank electrodes, the anticipated outcome is also an altered network dynamics and consequently altered LFPs. 
The present columnar model lacks structured input from other parts of cortex which corresponds to about 50\% of all excitatory synapses
(see \citet{Potjans2014} and references therein). 
For example, 
accounting for different cortical areas and their interactions would allow for the detailed investigation of pathway-specific LFP contributions 
as recently reviewed in hippocampus by \citet{Herreras2015}.

\subsection{Effects of correlations and network size}
Synaptic inputs to neurons are typically correlated, and as shown in this article and in earlier studies~\citep{Linden2011,Leski2013},
these correlations have a major impact on the properties of the generated LFP (and corresponding CSD). 
There are different contributions to these input correlations,
shared presynaptic neurons and 
correlations in the presynaptic spiking activity, see, e.g., \citep{Renart2010,Tetzlaff2012}, and
correct LFP predictions require that both effects are properly taken into account. 
The generation of presynaptic spike trains with
realistic correlation structure 
requires networks of realistic size~\citep{Albada2015}. 
The contribution from shared presynaptic input 
 can only be accounted for by using realistic statistics of inputs to the LFP-generating multicompartment model neurons, 
i.e., realistic synaptic connection probabilities resulting in realistic statistics of shared inputs.
If these two requirements are fulfilled, the properties of single-cell LFPs as well as the correlations between pairs of single-cell LFPs, will be
correctly accounted for.

Given synaptic inputs with realistic statistics from sufficiently large point-neuron networks,
do we actually need to represent the full population of LFP generating neurons in order to predict a realistic compound LFP?
Or can we alternatively get a good estimate of the compound population-LFP signal from a downscaled population of neuronal LFP generators if we know the correct single-cell and pairwise LFP statistics, 
thereby reducing computational costs?
As shown in this article (\Fref{sec:correlations}), the answer is negative: 
In the presence of (even tiny) synaptic-input correlations, a realistic compound LFP can only be generated by  
multicompartmental neuron populations with realistic size and cell densities.

The hybrid modeling scheme allows one to account for both, i.e., realistic sizes of networks to generate spike trains with correct correlation structure and realistic sizes and cell densities of the LFP-generating multicompartmental neuron populations. This can be achieved since 
point-neuron networks can be simulated very efficiently, and the multicompartmental neurons are independent and can be simulated serially 
(or in an embarrassingly parallel manner). 

\subsection{Outlook}

While we here have focused on the computation of the LFP based on output from spiking point-neuron networks,
a similar hybrid approach could be used when the network dynamics is rather modeled in terms of firing rates or even neural
fields~\citep{Deco2008}.  For our example case of a four-layered cortical network with an excitatory and an inhibitory population in each layer, the present scheme could be adapted directly by replacing the set of spike trains for each population with the corresponding 
population firing rates~\citep{Schuecker2015} in the LFP-generating step. 
However, the feasibility and prediction accuracy of such a scheme would have to be investigated in detail.

Another natural development would be to consider other measurement modalities. The present LFP scheme already incorporates the prediction of 
ECoG (electrocorticography) signals, i.e., the electrical signals recorded at the cortical surface, 
although the LFP forward-modeling scheme may have to be adjusted  to account for the discontinuity in electrical conductivity at the cortical surface~\citep{Nunez2006}. 
An extension to EEG (electroencephalography) and MEG (magnetoencephalography) would in principle also be straightforward as the key variable linking single-neuron activity and the measured signal is the single-neuron current dipole moment~\citep{Hamalainen1993, Nunez2006}.
This dipole moment can be computed from multicompartmental neuron models when the transmembrane or axial currents are known~\citep{Linden2010,Ahlfors2015}. 
Given the magnitude and orientation of the current dipole moments for all contributing neurons, the EEG and MEG signal can be computed by a linear superposition 
of single-cell contributions (given an appropriate extracellular volume conductor model for the EEG signal).
Another measurement that could be modeled is voltage-sensitive dye imaging (VSDi) where 
the signal largely reflects average membrane potentials of dendrites close to the cortical surface~\citep{Chemla2010}. 
The spatial profile of the weights in the averaging procedure of the VSDi forward-model will be
determined by the spatial distribution of dye and the propagation of light in the neural tissue~\citep{Chemla2010a,Tian2011}.

Even though the LFP has been measured for more than half a century, the interpretation of the recorded data has so far largely been 
qualitative~\citep{Einevoll2013}. The biophysical origin of the signal on the single-cell level
appears well understood
\citep{Rall1968,Holt1999,Gold2006,Linden2010,Buzsaki2012,Pettersen2012,Einevoll2013}, 
and several modeling studies have explored the link between neuron and network 
activity~\citep{Pettersen2008,Linden2010,Linden2011,Leski2013,Reimann2013,Tomsett2014,Glabska2014}.
However, the computation of LFPs from network activity has until now been too cumbersome and computer-intensive to allow for
practical exploration of the links between different types of network dynamics and the resulting LFP. Thus a validation of network models against
measured LFP data has essentially been absent.  With the present hybrid LFP scheme, accompanied by the release of the simulation tool \texttt{hybridLFPy},
we believe that a significant step has been taken 
towards the goal of making combined modeling and measurement of the LFP signal a practical research tool for probing neural circuit activity.


\section*{Acknowledgments}

The research leading to these results has received funding from 
the European Union Seventh Framework Programme (FP7/2007-2013) under grant agreement 604102 (Human Brain Project, HBP) and grant agreement 269912 (BrainScaleS), 
the Helmholtz Association through the Helmholtz Portfolio Theme ``Supercomputing and Modeling for the Human Brain'' (SMHB), 
Juelich Aachen Research Alliance (JARA), the Danish Council for Independent Research and FP7 Marie Curie Actions COFUND (grant id: DFF-1330--00226)
and the Research Council of Norway (NFR, through ISP, NOTUR -NN4661K). 
In addition, we would thank Zolt\'{a}n F. Kisv\'{a}rday and Armen Stepanyants for useful discussions in the early phase of this study and providing us with experimentally reconstructed morphologies of cat visual cortex neurons. 

\section*{Statement on competing interests}
The authors have no financial or non-financial competing interests.


\printbibliography

\clearpage


\begin{table}[!htbp]
\small
  \begin{tabularx}{\linewidth}{|p{0.15\linewidth}|X|}
    \hline\modelhdr{2}{A}{Model summary}\\\hline
    \bf Structure & Multi-layered excitatory-inhibitory (E-I) network \\\hline
    \bf Populations & 8 cortical in 4 layers, 1 thalamic (TC) \\\hline 
    \bf Connectivity & Random, independent, population-specific, fixed number of connections \\\hline
    \bf External input & Cortico-cortical: constant current with population-specific strength \\\hline
    \bf Neuron model & Cortex: leaky integrate-and-fire (LIF); TC: point process \\\hline
    \bf Synapse model & Exponential postsynaptic currents, static weights, population-specific weight distributions \\\hline
    \bf Measurements & Spike activity, input currents, membrane potential of each neuron \\\hline
  \end{tabularx}\vspace{\tabspace}
   \begin{tabularx}{\linewidth}{|p{0.15\linewidth}|X|}
     \hline\modelhdr{2}{B}{Network model}\\\hline
     \bf Connectivity & 
     Connection probability $C_{YX}$ ($X,Y \in \{\text{L2/3, L4, L5, L6}\} \times \{\text{E,I}\} \cup \text{TC}$, $C_{YX}=0$ for $Y=\text{TC}$) \\
     &-~Fixed number of synapses $K_{YX}$ between populations $X$ and $Y$  \\
     &-~Binomial in-/outdegrees\\
      \bf Input & Cortico-cortical direct current $I_Y^\text{ext}$ \\\hline
   \end{tabularx}\vspace{\tabspace}
  \begin{tabularx}{\linewidth}{|p{0.15\linewidth}|X|}
    \hline\modelhdr{2}{C}{Neuron model}\\\hline
    \multicolumn{2}{|c|}{ \bf \cellcolor[gray]{0.9}  Cortex}  \\\hline
    \bf Type & Leaky integrate-and-fire neuron\\\hline 
    \bf Description & 
    Dynamics of membrane potential $V_i(t)$ (neuron $i\in[1,N]$):
    \begin{itemize}
      \setlength{\itemsep}{0pt}
    \item[-] Spike emission at times $t^i_l$ with $V_i(t^i_l)\ge\theta$
    \item[-] Subthreshold dynamics:
    \item[]
      $\tau_\text{m}\dot{V_i}=-V_i+ R_\text{m} I_i(t)$ \quad if
      $\forall l:\,t\notin(t^i_l,t^i_l+\tau_\text{ref}]$
    \item[-] Reset + refractoriness: 
      $V_i(t)= V_\text{reset}$ \quad if $\forall l:\,t\in(t^i_l,t^i_l+\tau_\text{ref}]$
    \end{itemize}\\    
    & Exact integration with temporal resolution $dt$ \citep{Rotter99a} \\
    & Uniform distribution of membrane potentials at $t=0$ \\
    \hline
    \multicolumn{2}{|c|}{ \bf \cellcolor[gray]{0.9}  Thalamus}  \\\hline
    \bf Type &- DC current for constant background input \\
     &- Nonstationary Poisson process for modulation \\\hline 
    \bf Description & DC current included in external DC input \\
    & \\
    & Types of thalamic input modulation: \\
    &- Spontaneous activity: no modulation in activation of thalamic neurons\\
    &- Thalamic pulses: fixed-interval coherent activation of all $N_\text{TC}$ thalamic neurons\\
    &- AC modulation: Poisson spike trains with sinusoidally modulated rate profile (discretized with time resolution $dt$):\\
    & \,\hfill
    \begin{equation}
      \nu_\text{th}(t) 
      = \overline\nu_\text{TC} + \Delta\nu_\text{TC} \sin{(2\pi t f_\text{TC})}
      \label{eq:ac_modulation}
    \end{equation}\hfill\,
   \\\hline
  \end{tabularx}\vspace{\tabspace}
  \caption{Description of point-neuron network for the cortical microcircuit model (continued in \Fref{tab:2}) following the guidelines of \citet{Nordlie2009}.
  }
  \label{tab:1}
\end{table}


\begin{table}[!htbp]
\small
  \begin{tabularx}{\linewidth}{|p{0.15\linewidth}|X|}
    \hline\modelhdr{2}{D}{Synapse model}\\\hline
    \bf Type  & Exponential postsynaptic currents, static weights \\\hline
    \bf Description &
    Input current of neuron $j$ of synapses formed with presynaptic neurons $i$: \\
    & \,\hfill  \begin{math}
      I_j(t) = \sum_i J_{ji}\sum_l \exp{(-(t-t^i_l-d_{i})/\tau_\text{s})}~ H(t-t^i_l-d_{i}) + I_{j}^\text{ext}
    \end{math}\hfill\, \\
    & \begin{itemize}
       \item[-] Static synaptic weights $J_{ji}=\text{sgn}(X) \left|J_{YX}\right|$ ($i \in X$, $j \in Y$);
         $\text{sgn}(X)=1$~for \(X~\in~\{\text{L2/3E, L4E, L5E, L6E, TC}\}\), $-1$ otherwise
      \item[-]Absolute weights $\left|J_{YX}\right|$ drawn from lognormal distribution
         \begin{equation}
           p(\left|J_{YX}\right|)=\frac{1}{\sqrt{2\pi}\sigma_{YX}  \left|J_{YX}\right|}\exp{\left(-\frac{(\ln{\left|J_{YX}\right|}-\mu_{YX})^2}{2 \sigma_{YX}^2}\right)}
         \end{equation}
         or normal distribution 
         \begin{equation}
           p(\left|J_{YX}\right|)=\frac{1}{\sqrt{2\pi}\sigma_{YX}}\exp{\left(-\frac{(\left|J_{YX}\right|-\mu_{YX})^2}{2 \sigma_{YX}^2}\right)}
         \end{equation}
         with $\mu_{YX}=g_{YX} J$ and $\sigma_{YX}=\sigma_{J,\text{rel}} \, \mu_{YX}$
       \item[-] Delays $d_{i}=d_{X}$ ($i \in X$) drawn from (left-clipped) Gaussian distribution 
         \begin{equation}
           p(d_{X})=\frac{1}{\sqrt{2\pi}\sigma_{X}}\exp{\left(-\frac{(d_{X}-\mu_{X})^2}{2 \sigma_{X}^2}\right)}
         \end{equation}
         with mean $\mu_{X}=d_\text{E},d_\text{I}$ for $X$ exc., inh., stdev  $\sigma_{X}=\sigma_{d,\text{rel}} \,  \mu_{X}$ and $d_X \in [dt,\infty)$
       \item[-] $ H(t)= 1 \text{ for } t \geq{0}~, \text{ and } 0 \text{ elsewhere.}$
       \item[-] External DC input $I_{j}^\text{ext} = I_Y^\text{ext}= I^\text{ext} k^\text{ext}_Y$ ($j \in Y$) 
       \end{itemize}
     \\\hline 
  \end{tabularx}\vspace{\tabspace}
  \caption{Description of point-neuron network for the cortical microcircuit model (continuation of \Fref{tab:1}).}
  \label{tab:2}
\end{table}


\begin{table}[!htbp]
\small
  \begin{tabularx}{\linewidth}{|p{0.15\linewidth}|X|}
    \hline\modelhdr{2}{A}{Model summary}\\\hline
    \bf Topology & Cortical column under $1\,\text{mm}^2$ of cortical surface \\\hline
    \bf Populations & 8 excitatory and 8 inhibitory cell types \\\hline 
    \bf Input & Spiking activity of thalamic and cortical populations as modeled by point-neuron network
    \\\hline
    \bf Neuron model & Multicompartment, passive cable formalism \\\hline
    \bf Synapse model & Exponential postsynaptic current, static weights \\\hline
    \bf Measurements & Current source density (CSD), local field potential (LFP) \\\hline
  \end{tabularx}\vspace{\tabspace}
  \begin{tabularx}{\linewidth}{|p{0.15\linewidth}|X|}
     \hline\modelhdr{2}{B}{Topology}\\\hline
     \bf Type &  Cylindrical volume with layer-specific distribution of cell types and synapses\\\hline
     \bf Description & Cylinder radius $r$ \\
      & Laminar, defining upper/lower boundaries of layers 1, 2/3, 4, 5, 6 \\
     \hline
   \end{tabularx}\vspace{\tabspace}
   \begin{tabularx}{\linewidth}{|p{0.15\linewidth}|X|}
      \hline\modelhdr{2}{C}{Populations}\\\hline
      \bf Type & Each cell type $y$ assigned to population $Y$, $y \in Y$ \\\hline
      \bf Description & Populations $Y \in \{\text{L2/3, L4, L5, L6}\} \times \{\text{E,I}\}$ (population size $N_Y$, cell types $y \in Y$) \\
       & (e.g.,~$\text{L4E}=\{\text{p4},~\text{ss4(L23)},~\text{ss4(L4)}\}$, cf.~\Fref{fig:2}).\\
       & Cell types $y$:\\
       & -~Size $N_y = F_y  \sum_{Y} N_Y$, $F_y$ is the occurrence of cell type $y$ in the full model\\
       & -~Morphology $M_y$  \\
       & -~Extrapolated according to spatial connectivity data (\Fref{tab:7}) \\
       & Somatic placement, population $Y$: \\
       & -~Random soma placement in cylindrical volumes with radius $r$, thickness $h$ \\
       & -~Volumes centered between boundaries of layers 2/3--6 \\
 	\hline 
      \multicolumn{2}{|c|}{ \bf \cellcolor[gray]{0.9} Morphologies}  \\\hline
      \bf Type & 3D histological reconstructions from slice preparations
      (see \citet{Jacobs2009, DeSchutter2009}) of cat visual and somatosensory cortices \\\hline
      \bf Description 
      & One morphology $M_y$ per cell type: \\
      & -~Excitatory and inhibitory cells in layers 2/3--6 \\
      & -~For all cells $j \in y$: $M_j = M_y$ \\
      & -~For some cell types $y,y'$: $M_y = M_{y'}$ (limited availability) \\
      & Orientations: \\
      & -~Pyramidal cells: apical dendrites oriented along depth axis with random depth-axis rotation \\
      & -~Interneurons, stellate cells: random rotation around all axes \\
      & Corrections: \\
      & -~Apical dendrites of pyramidal cells elongated to accommodate spatial connectivity \\
      & -~Axons removed if present \\
      & Reconstructed morphologies (cf. \Fref{fig:4}): \\
      & -~Cat visual cortex \citep{Kisvarday1992, Mainen1996, Contreras1997, Stepanyants2008} \\
      & -~Cat somatosensory cortex from NeuroMorpho.org \citep{Contreras1997, Ascoli2007}. \\
      \hline
      \end{tabularx}\vspace{\tabspace}
      \caption{Description of multicompartment-neuron populations for the cortical microcircuit model (continued in \Fref{tab:4}).
      }
   \label{tab:3}
\end{table}


\begin{table}[!htbp]
\small
  \begin{tabularx}{\linewidth}{|p{0.15\linewidth}|X|}
    \hline\modelhdr{2}{D}{Neuron models}\\\hline
    \bf Type & Passive, multicompartment, reconstructed morphologies \\\hline
    \bf Description
    & Compartment $n$ membrane potential $V_{\text{m}jn}$ of cell $j$ having length~$l_{jn}$, diameter~$d_{jn}$ and surface area~$A_{jn}$:
    \begin{eqnarray}
    C_{\text{m}jn} \frac{dV_{\text{m}jn}}{dt} &=& \sum_{k=1}^m I_{\text{a}jkn} - G_{\text{L}jn}(V_{\text{m}jn} - E_\text{L}) - \sum_i I_{jin} ~, \label{eq:cable} \\
    C_{\text{m}jn} &=& c_\text{m} A_{jn} ~, \\
    I_{\text{a}jkn} &=& G_{\text{a}jkn}\left(V_{\text{m}jk} - V_{\text{m}jn}\right) ~, \\
	G_{\text{a}jkn} &=& \pi(d_{jk}^2 + d_{jn}^2)  / 4r_\text{a}(l_{jk} + l_{jn})~,\\
    G_{\text{L}jn} &=& A_{jn} / r_\text{m} ~, \\
    I_{\text{m}jn} &=& C_{\text{m}jn} \frac{dV_{\text{m}jn}}{dt} + G_{\text{L}jn}(V_{\text{m}jn} - E_\text{L}) - \sum_i I_{jin}~. \label{eq:imem}
    \end{eqnarray}
    $C_{\text{m}jn}$~is compartment capacitance,
    $G_{\text{L}jn}$~its passive leak conductance, 
    $E_\text{L}$~the passive leak reversal potential, 
    $I_{\text{a}jkn}$~axial current between compartment $n$ and neighboring compartment $k$ (out of $m$ compartments), 
    $G_{\text{a}jkn}$~axial conductance between $n$ and $k$,
    $I_{jin}$~synaptic currents, 
    and $I_{\text{m}jn}$~transmembrane current of compartment $n$.
    For specific parameter values, see \Fref{tab:6}.
    Membrane potentials and transmembrane currents are computed using \texttt{NEURON} through \texttt{LFPy} \citep{Carnevale2006, Linden2014}, 
    assuming the extracellular potential to be zero everywhere on the outside of the neuron, that is, an infinite extracellular conductivity. \\
    \hline 
\end{tabularx}\vspace{\tabspace}
  \begin{tabularx}{\linewidth}{|p{0.15\linewidth}|X|}
    \hline\modelhdr{2}{E}{Synapse model}\\\hline
    \bf Type  & Exponential postsynaptic current, static weights \\\hline
    \bf Description 
    & Neuron $j$ input current of synapse formed with presynaptic neuron $i$:
    \begin{eqnarray}
      I_{ji}(t) &=& I_{ji,\text{max}} \sum_l \exp{(-(t-t^i_l-d_i)/\tau_\text{s})} ~ H(t-t^i_l-d_i) ~, \\
      I_{ji,\text{max}} &=& C_\text{m}\,\mu_{YX} ~ \text{of point-neuron network,}\\
      H(t) &=& 1 \text{ for } t \geq{0}~, \text{ and } 0 \text{ elsewhere.}
    \end{eqnarray}    
      -~Static synaptic weights $J_{ji} = \mu_{YX}$  ($j \in Y$, $i \in X$) (see \Fref{tab:2})  \\
    & -~Delays $d_{i}$ from Gaussian distribution with mean $d_{X}$ ($i \in X$), relative standard deviation $\sigma_{d,\text{rel}}$ \\
    & -~Synapse activation times: network spike trains plus delay \\
    & -~No cortico-cortical connections: $I^\text{ext} = 0$ (cf. \Fref{tab:5})  \\
    \hline 
  \end{tabularx}\vspace{\tabspace}
      \begin{tabularx}{\linewidth}{|p{0.15\linewidth}|X|}
      \hline\modelhdr{2}{F}{Input}\\\hline
      \bf Type &  Spike times $t^i_l$ of spiking neuron network (including thalamic input spikes), no cortico-cortical input \\\hline
      \bf Description & Synapse placement, postsynaptic cell $j \in y,~y \in Y$ (\Fref{sec:2.3}): \\
      & -~Number of synapses from presynaptic population $X$ in layer $L$: $k_{yXL}$ (\Fref{eq:KyXL2}) \\
      & -~Compartment specificity of connections: $A_{jn} / \sum_{n \in L}A_{jn}$, compartment $n \in L$ \\
      & -~Synapse locations within layers are chosen randomly among dendritic compartments only \\
      \hline
   \end{tabularx}\vspace{\tabspace}
  \begin{tabularx}{\linewidth}{|p{0.15\linewidth}|X|}
     \hline\modelhdr{2}{G}{Measurements}\\\hline
     \bf Type & Local field potential (LFP) and current source density (CSD) \\
     \hline
     \bf Description & Laminar multielectrode, see parameter values in \Fref{tab:6}: \\
     & -~Axis perpendicular to pial surface \\
     & -~$n_\text{contacts}$: number of contacts \\
     & -~$h_\text{contacts}$: intercontact distance \\
     & -~$r_\text{contact}$: contact surface radius \\
     \hline
   \end{tabularx}\vspace{\tabspace}
   \caption{Description of multicompartment-neuron populations for the cortical microcircuit model (continuation of \Fref{tab:3}).}
   \label{tab:4}
\end{table}


\begin{table}[!htbp]
\small
  \tabcolsep=0.11cm
  \begin{tabularx}{\linewidth}{|p{0.10\linewidth}|p{0.20\linewidth}|X|}
      \hline
      \modelhdr{3}{~A}{Global simulation parameters} \\
      \hline
      \bf Symbol & \bf Value & \bf Description \\
      \hline
      $T$ & 5,200 ms & simulation duration \\
      $dt$ & 0.1 ms & temporal resolution \\
      \hline 
  \end{tabularx}\vspace{\tabspace}
  \tabcolsep=0.11cm
  \begin{tabularx}{\linewidth}{|p{0.1\linewidth}|ccccccccc|X|}
    \modelhdr{11}{~B}{Point-neuron network} \\
    \parameterhdr{11}{}{Populations and external input}  \\
  \hline 
    \bf Symbol & \multicolumn{9}{l|}{ \bf Value} & \bf Description \\ \hline
    $X$ & L23E & L23I & L4E & L4I & L5E & L5I & L6E & L6I & TC & Name\\
    $N_X$ & 20,683 & 5,834 & 21,915 & 5,479 & 4,850 & 1,065 & 14,395 & 2,948 & 902 & Size \\
    $k_X^\text{ext}$ & 1,600 & 1,500 & 2,100 & 1,900 & 2,000 & 1,900 & 2,900 & 2,100 &  & Ext. in-degree per neuron\\
    $I^{\text{ext}}$ & $\tau_\text{syn}\nu_\text{bg}J$,  & $\nu_\text{bg}=$ &8 Hz&&&&&&& DC ampl. per ext. input \\ \hline
 \end{tabularx}\vspace{\tabspace}
 \begin{tabularx}{\linewidth}{|p{0.1\linewidth}|XXXXXXXXXX|}
 	\hline
	\parameterhdr{11}{}{Connectivity} \\
    \hline 
    $C_{YX}$ 	& \multicolumn{10}{c|}{from $X$} \\ \hline
    			& & L23E & L23I & L4E & L4I & L5E & L5I & L6E & L6I & TC \\
    to $Y$ 		& L23E & 0.101 & 0.169 & 0.044 & 0.082 & 0.032 & 0.0 & 0.008 & 0.0 & 0.0 \\
    			& L23I & 0.135 & 0.137 & 0.032 & 0.052 & 0.075 & 0.0 & 0.004 & 0.0 & 0.0 \\
    			& L4E & 0.008 & 0.006 & 0.050 & 0.135 & 0.007 & 0.0003 & 0.045 & 0.0 & 0.0983 \\
    			& L4I & 0.069 & 0.003 & 0.079 & 0.160 & 0.003 & 0.0 & 0.106 & 0.0 & 0.0619 \\
    			& L5E & 0.100 & 0.062 & 0.051 & 0.006 & 0.083 & 0.373 & 0.020 & 0.0 & 0.0 \\
    			& L5I & 0.055 & 0.027 & 0.026 & 0.002 & 0.060 & 0.316 & 0.009 & 0.0 & 0.0 \\
    			& L6E & 0.016 & 0.007 & 0.021 & 0.017 & 0.057 & 0.020 & 0.040 & 0.225 & 0.0512 \\
    			& L6I & 0.036 & 0.001 & 0.003 & 0.001 & 0.028 & 0.008 & 0.066 & 0.144 & 0.0196 \\ \hline
  \end{tabularx}\vspace{\tabspace}
  \begin{tabularx}{\linewidth}{|p{0.1\linewidth}|p{0.20\linewidth}|X|}
    \hline
    \parameterhdr{3}{}{Connection parameters} \\ \hline
    \bf Symbol & \bf Value  & \bf Description \\ \hline
    $J$ & $87.81$ pA  & Reference synaptic strength. All synapse weights are measured in units of $J$. \\
    $\sigma_{J,\text{rel}} $ &  & Relative width of synaptic strength distribution\\
    & 3 & -~for lognormal distribution \\
    & 0.1  & -~for Gaussian distribution \\
    $g_{YX}$ & & Relative synaptic strength: \\
    & 1  & $X \in \{\text{TC, L23E, L4E, L5E, L6E}\}, $\\
    & $-4$ & $X \in \{\text{L23I, L4I, L5I, L6I}\}$, except for\\
    & 2 & $(X,Y)= (\text{L4E, L23E})$ \\
    & $-4.5$ & $(X,Y)= (\text{L4I, L4E})$ \\
    $d_\text{E}$ & 1.5 ms & Mean excitatory spike transmission delay  \\
    $d_\text{I}$ & 0.75 ms & Mean inhibitory spike transmission delay \\ 
    $\sigma_{d,\text{rel}}$ & 0.5 & Relative width (stdev/mean) of transmission delay distributions \\ 
    \hline
    \parameterhdr{3}{}{Neuron model} \\ 
    \hline
    \bf Symbol & \bf Value & \bf Description \\ \hline
    $R_\text{m}$ & 40 M$\Omega$ & Membrane resistance \\
    $C_\text{m}$ & 250 pF &  Membrane capacitance \\
    $\tau_\text{m}$ & $R_\text{m}C_\text{m}$ (10 ms) & Membrane time constant \\
    $E_\text{L}$ & $-65$ mV & Resting potential \\
    $\theta$  & $-50$ mV &  Fixed firing threshold \\
    $V_\text{m}(t=0)$ & $[-65, -50]$ mV & Uniformly distributed initial membrane potential \\
    $V_\text{reset}$ & $E_\text{L}$ & Reset potential \\
    $\tau_\text{ref}$ & 2 ms & Absolute refractory period \\
    $\tau_\text{syn}$ & 0.5 ms & Postsynaptic current time constant \\
    \hline
    \parameterhdr{3}{}{Thalamocortical input} \\ 
    \hline
    \bf Symbol & \bf Value & \bf Description \\ \hline
    $\overline\nu_\text{TC}$&  30 s$^{-1}$& Mean firing rate per thalamocortical neuron\\
    $ \Delta\nu_\text{TC}$ &  30 s$^{-1}$ & Firing-rate modulation amplitude per thalamocortical neuron \\
    $f_\text{TC}$ & 15 Hz& Frequency of sinusoidal firing-rate  modulation \\ \hline
  \end{tabularx}\vspace{\tabspace}
  \caption{Parameters of the cortical microcircuit model.}
  \label{tab:5}
\end{table}


\begin{table}[!htbp]
\small
  \begin{tabularx}{\linewidth}{|p{0.1\linewidth}|p{0.20\linewidth}|X|}
      \modelhdr{3}{}{Multicompartment model neurons} \\
      \bf Symbol & \bf Value & \bf Description \\
      \hline
      $c_\text{m}$ & 1.0 $\mu$Fcm$^{-2}$ & Membrane capacity \\
      $r_\text{m}$ &  $\tau_\text{m} / c_\text{m}$ & Membrane resistivity \\
      $r_\text{a}$ & 150 $\Omega$cm &  Axial resistivity \\
      $E_\text{L}$ & $E_\text{L}$ & Passive leak reversal potential \\
      $V_\text{init}$ & $E_\text{L}$ & Membrane potential at $t=0$ ms \\
      $\lambda_f$ & 100 Hz & Frequency of AC length constant \\
      $\lambda_d$ & 0.1 & Factor for \texttt{d\_lambda} rule \citep{Hines2001} \\
      $\sigma_\text{e}$	& 0.3 Sm$^{-1}$ & Extracellular conductivity \\
      $r$	& $\sqrt{1,000^2/\pi} ~\mu$m & Population radius \\
      $h$	& $50 ~\mu$m &	Soma layer thickness \\
      $n_\text{contact}$ & 16 	& Number of electrode contacts \\
      $h_\text{elec}$ & 100 $\mu$m 	& Laminar-electrode intercontact distance \\      
      $r_\text{contact}$ & 7.5 $\mu$m 	& Electrode contact-point radius \\
      \hline 
  \end{tabularx}\vspace{\tabspace}
  \caption{Parameters of the multicompartment model neuron populations and calculations of extracellular potentials. 
   Values for $\tau_\text{m}$ and $E_\text{L}$ are inherited from network parameters in \Fref{tab:5}.
  }
  \label{tab:6}
\end{table}


\begin{table}[!htbp]
\small
  \begin{tabularx}{\linewidth}{|X|X|X|X|X|}
    \hline\modelhdr{5}{}{Morphology files}\\\hline
    \bf Cell type $y$ & \bf Morphology $M_y$ & \bf File & \bf Source & \bf Online source \\\hline   
    p23 & p23 & oi24rpy1.hoc & \citep{Kisvarday1992} & \#NMO\_00851 (\#NMO\_10045) \\\hline  
    b23 & i23 & oi38lbc1.hoc & \citep{Stepanyants2008} & - \\
    nb23 & i23 & oi38lbc1.hoc & \citep{Stepanyants2008} & - \\\hline   
    p4 & p4 & oi53rpy1.hoc & \citep{Kisvarday1992} & \#NMO\_00855 (\#NMO\_10040) \\   
    ss4(L23) & ss4 & j7\_L4ste.hoc & \citep{Mainen1996} & \#MDB\_2488, \#NMO\_00905 \\   
    ss4(L4) & ss4 & j7\_L4ste.hoc & \citep{Mainen1996} & \#MDB\_2488, \#NMO\_00905 \\\hline   
    b4 & i4 & oi26rbc1.hoc & \citep{Stepanyants2008} & - \\
    nb4 & i4 & oi26rbc1.hoc & \citep{Stepanyants2008} & - \\\hline   
    p5(L23) & p5v1 & oi15rpy4.hoc & \citep{Kisvarday1992} & \#NMO\_00850 (\#NMO\_10046) \\  
    p5(L56) & p5v2 &  j4a.hoc & \citep{Mainen1996} & \#MDB\_2488\\\hline   
    b5 & i5 & oi15rbc1.hoc & \citep{Stepanyants2008} & -  \\  
    nb5 & i5  & oi15rbc1.hoc & \citep{Stepanyants2008} & -  \\\hline   
    p6(L4) & p6 & 51-2a.CN.hoc & \citep{Contreras1997} & \#NMO\_00879 \\  
    p6(L56) & p5v1 & oi15rpy4.hoc &\citep{Kisvarday1992} & \#NMO\_00850 (\#NMO\_10046) \\\hline   
    b6 & i5 & oi15rbc1.hoc  & \citep{Stepanyants2008} & -  \\   
    nb6 & i5 & oi15rbc1.hoc & \citep{Stepanyants2008} & -  \\\hline   
  \end{tabularx}
  \vspace{\tabspace}
  \caption{Morphology types and file names used for each cell type in the model (p - pyramidal cell, ss - spiny stellate, i - interneuron). Online source numbers on the form \#NMO\_$\ast$ refer to NeuroMorpho.org identifiers, the form \#MDB\_$\ast$ refers to ModelDB identifiers.}
  \label{tab:7}
\end{table}


\begin{table}[!htbp]
\small
\centering
\rotatebox{90}{\parbox{0.98\textheight}{
\tabcolsep=0.05cm
\begin{tabular}{|cc|c|c|c|c|cc|ccc|cc|cc|cc|cc|cc|cc|}
\hline
 \multicolumn{2}{c|}{\color{white}\cellcolor[gray]{0.0} Postsyn.} & \color{white}\cellcolor[gray]{0.0}Layer & \color{white}\cellcolor[gray]{0.0}Occurr.  & \color{white}\cellcolor[gray]{0.0}Num.& \multicolumn{18}{c|}{\color{white}\cellcolor[gray]{0.0}Presynaptic populations $X$ and cell types $x$} \\ 
\color{white}\cellcolor[gray]{0.0} pop. & \color{white}\cellcolor[gray]{0.0}cell type  & \color{white}\cellcolor[gray]{0.0} &\color{white}\cellcolor[gray]{0.0}  & \color{white}\cellcolor[gray]{0.0} syn. &\multicolumn{1}{c|}{L23E} &\multicolumn{2}{c|}{L23I} &\multicolumn{3}{c|}{L4E} &\multicolumn{2}{c|}{L4I} &\multicolumn{2}{c|}{L5E} &\multicolumn{2}{c|}{L5I} &\multicolumn{2}{c|}{L6E} &\multicolumn{2}{c|}{L6I} &\multicolumn{2}{c|}{TC}  \\
&  & &  & & p23 & b23 & nb23 & ss4(L4) & ss4(L23) & p4 & b4 & nb4 & p5(L23) & p5(L56) & b5 & nb5 & p6(L4) & p6(L56) & b6 & nb6 & TCs & TCn \\ \hline
$Y$& $y$ &$L$ & $F_y$ (\%) & $k_{yL}$ &\multicolumn{18}{c|}{ $p_{yxL}$ (\%)} \\\hline
 L23E& p23 & 2/3 & 26.7 & 5,800 & 59.9 & 9.1 & 4.4 & 0.6 & 6.9 & 7.7 & . & 0.8 & 7.4 & . & . & . & 2.3 & . & . & 0.8 & . & . \\  
&  & 1 &  & 1,306 & 6.3 & 0.1 & 1.1 & . & . & 0.1 & . & . & 0.1 & . & . & . & . & . & . & . & .  & 4.1 \\  \hline
 L23I& b23 & 2/3 & 3.2 & 3,854 & 51.6 & 10.6 & 3.4 & 0.5 & 5.8 & 6.6 & . & 0.8 & 6.3 & . & . & . & 2.1 & . & . & 0.7& .  & 0.5 \\ 
& nb23 &2/3 & 4.3 & 3,307 & 48.6 & 11.4 & 3.3 & 0.5 & 5.5 & 6.2 & . & 0.8 & 5.9 & . & . & . & 1.8 & . & . & 0.6& .& 0.7 \\  \hline
 L4E& ss4(L4) &4 & 9.4 & 5,792  & 2.7 & 0.2 & 0.6 & 11.9 & 3.7 & 4.1 & 7.1 & 2 & 0.8 & 0.1 & . & . & 32.7 & . & . & 5.8& 1.7&1.3 \\  \hline
& ss4(L23) &4& 9.4 & 4,989 & 5.6 & 0.4 & 0.8 & 11.3 & 3.8 & 4.3 & 7.2 & 2.1 & 1.1 & 0.1 & . & . & 31.1 & . & . & 5.5& 1.7&1.3 \\ 
& p4 & 4 &9.4 & 5,031 & 4.3 & 0.2 & 0.6 & 11.5 & 3.6 & 4.2 & 7.2 & 2.1 & 1.2 & 0.1 & . & . & 31.4 & 0.1 & . & 5.9& 1.7& 1.3 \\  
&     & 2/3 & & 866  & 63.1 & 5.1 & 4.1 & 0.6 & 7.2 & 8.1 & . & 0.6 & 7.8 & . & . & . & 2.5 & . & . & 0.8&. &. \\  
&     & 1 & & 806  & 6.3 & 0.1 & 1.1 & . & . & 0.1 & . & . & 0.1 & . & . & . & . & . & . & .&. &4.1 \\  \hline
 L4I& b4 &4  & 5.5 & 3,230 & 5.8 & 0.5 & 0.8 & 11 & 3.8 & 4.2 & 8.4 & 2.4 & 1.1 & . & . & . & 30.3 & . & . & 5.4&1.6 &1.2 \\ 
& nb4 &4 & 1.5 & 3,688 & 2.7 & 0.2 & 0.6 & 11.7 & 3.6 & 4 & 8.2 & 2.3 & 0.8 & 0.1 & . & . & 32.2 & . & . & 5.7& 1.7&1.3 \\  \hline
 L5E& p5(L23) & 5 & 4.8 & 4,316 & 45.9 & 1.8 & 0.3 & 3.3 & 2 & 7.5 & . & 0.9 & 11.7 & 1 & 0.8 & 1.1 & 2.3 & 2.1 & . &11.5&0.1 &0.4 \\
&   & 4 & & 283  & 2.8 & 0.1 & 0.7 & 12.2 & 3.8 & 4.2 & 5.2 & 1.5 & 0.8 & 0.1 & . & . & 33.7 & . & . & 5.9& 1.8&1.4 \\  
&   & 2/3 & & 412 & 63.1 & 5.1 & 4.1 & 0.6 & 7.2 & 8.1 & . & 0.6 & 7.8 & . & . & . & 2.5 & . & . & 0.8& .&. \\ 
&   & 1 &  & 185  & 6.3 & 0.1 & 1.1 & . & . & 0.1 & . & . & 0.1 & . & . & . & . & . & . & .&. &4.1 \\ 
& p5(L56) & 5 & 1.3 & 5,101 & 44.3 & 1.7 & 0.2 & 3.2 & 2 & 7.3 & . & 0.8 & 11.3 & 1.2 & 0.8 & 1.1 & 2.3 & 2.5 & 0.3 & 11.3&0.2 &0.5 \\
&   & 4 & & 949 & 2.8 & 0.1 & 0.7 & 12.2 & 3.8 & 4.2 & 5.2 & 1.5 & 0.8 & 0.1 & . & . & 33.7 & . & . & 5.9& 1.8&1.4 \\  
&   & 2/3 & & 1,367 & 63.1 & 5.1 & 4.1 & 0.6 & 7.2 & 8.1 & . & 0.6 & 7.8 & . & . & . & 2.5 & . & . & 0.8& .&. \\ 
&   & 1 & & 5,658 & 6.3 & 0.1 & 1.1 & . & . & 0.1 & . & . & 0.1 & . & . & . & .& . & . & .& .& 4.1 \\ \hline
 L5I& b5 & 5& 0.6 & 2,981 & 45.5 & 2.3 & 0.2 & 3.3 & 2 & 7.5 & . & 1.1 & 11.6 & 1 & 0.9 & 1.3 & 2.3 & 2 & . & 11.4&0.1 &0.4 \\
& nb5 & 5& 0.8 & 2,981 & 45.5 & 2.3 & 0.2 & 3.3 & 2 & 7.5 & . & 1.1 & 11.6 & 1 & 0.9 & 1.3 & 2.3 & 2 & . & 11.4& 0.1&0.4 \\  \hline
 L6E& p6(L4) & 6 & 14.0 & 3,261 & 2.5 & 0.1 & 0.1 & 0.7 & 0.9 & 1.3 & . & 0.1 & 0.1 & 4.9 & . & 0.3 & 1.2 & 13.2 & 7.7 & 7.7&0.6 &2.9 \\ 
&  & 5 & & 1,066 & 46.8 & 0.8 & 0.3 & 3.4 & 2.1 & 7.7 & . & 0.6 & 11.9 & 1 & 0.6 & 0.8 & 2.3 & 2.1 & . & 11.7& 0.1&0.4 \\  
&  & 4 & & 1,915 & 2.8 & 0.1 & 0.7 & 12.2 & 3.8 & 4.2 & 5.2 & 1.5 & 0.8 & 0.1 & . & . & 33.7 & . & . & 5.9& 1.8&1.4 \\  
&  & 2/3 & & 121 & 63.1 & 5.1 & 4.1 & 0.6 & 7.2 & 8.1 & . & 0.6 & 7.8 & . & . & . & 2.5 & . & . & 0.8& .&. \\  
& p6(L56) & 6 & 4.6 & 5,573  & 2.5 & 0.1 & 0.1 & 0.7 & 0.9 & 1.3 & . & 0.1 & 0.1 & 4.9 & . & 0.3 & 1.2 & 13.2 & 7.8 & 7.8&0.6 &2.9 \\
&& 5 & & 257  & 46.8 & 0.8 & 0.3 & 3.4 & 2.1 & 7.7 & . & 0.6 & 11.9 & 1 & 0.6 & 0.8 & 2.3 & 2.1 & . & 11.7& 0.1&0.4 \\
& & 4 & & 243 & 2.8 & 0.1 & 0.7 & 12.2 & 3.8 & 4.2 & 5.2 & 1.5 & 0.8 & 0.1 & . & . & 33.7 & . & . & 5.9& 1.8&1.4 \\  
& & 2/3 & & 286 & 63.1 & 5.1 & 4.1 & 0.6 & 7.2 & 8.1 & . & 0.6 & 7.8 & . & . & . & 2.5 & . & . & 0.8& .&. \\  
&  & 1 & & 62 & 6.3 & 0.1 & 1.1 & . & . & 0.1 & . & . & 0.1 & . & . & . & . & . & . & .& .&4.1 \\  \hline
 L6I&  b6 &6  & 2.0 & 3,230 & 2.5 & 0.1 & 0.1 & 0.7 & 0.9 & 1.3 & . & 0.1 & 0.1 & 4.9 & . & 0.4 & 1.2 & 13.2 & 7.7 & 7.7&0.6 &2.9 \\ 
& nb6 &6  & 2.0 & 3,230  & 2.5 & 0.1 & 0.1 & 0.7 & 0.9 & 1.3 & . & 0.1 & 0.1 & 4.9 & . & 0.4 & 1.2 & 13.2 & 7.7 & 7.7& 0.6&2.9 \\  \hline
\end{tabular}
}}
\caption{
Spatial distribution of cell types and synapses of the cortical microcircuit model adapted from \citet{Binzegger2004, Izhikevich2008}. 
Note that occurrences are renormalized between cell types within layers 2/3, 4, 5 and 6 so that $\sum_y F_y = 100 \%$.
}
\label{tab:8}
\end{table}


\begin{table}[!htbp]
\small
  \begin{tabularx}{\linewidth}{|p{0.1\linewidth}|p{0.45\linewidth}|X|}
    \hline\modelhdr{3}{}{Measurements and derived signals}\\\hline
    \parameterhdr{3}{}{Measurements}  \\\hline  
    \bf Symbol &  \bf Description & \bf Number of recorded units \\\hline
    $I^{\text{ex}}_i(t)$ & Excitatory synaptic input current of neuron $i$ & $100$ per population $X$ \\
    $I^{\text{in}}_i(t)$ & Inhibitory synaptic input current of neuron $i$ & $100$ per population $X$ \\
    $V_i(t)$ & Somatic voltage of neuron $i$  & $100$ per population $X$ \\
    $s_i(t)$ & Spike train of neuron $i$ & $N_X$ neurons  \\
    $\phi_i(\mathbf{r},t)$ & Single-cell LFP generated by neuron $i$ & $N_X$ neurons  \\
    $\rho_i(\mathbf{r},t)$  & Single-cell CSD generated by neuron $i$ & $N_X$ neurons  \\\hline 
    \parameterhdr{3}{}{Derived signals}  \\\hline  
    \bf Symbol & \bf Definition & \bf Description \\\hline 
    $I_i(t)$ & $I^\text{ex}_i(t) + I^\text{in}_i(t)$   & Total synaptic input current of neuron $i$ \\
    $\overline{I}^{\text{ex}}_X(t)$ & $\frac{1}{n_\text{av}}\sum \limits_{i\in X}^{n_\text{av}} I^\text{ex}_i(t)$ & Average excitatory synaptic input current of population $X$ ($n_\text{av}=100$) \\
    $\overline{I}^{\text{in}}_X(t)$ &  $\frac{1}{n_\text{av}}\sum \limits_{i\in X}^{n_\text{av}}  I^\text{in}_i(t)$ & Average inhibitory synaptic input current of population $X$ ($n_\text{av}=100$) \\
    $\overline{I}_X(t)$ & $\frac{1}{n_\text{av}}\sum \limits_{i\in X}^{n_\text{av}}  I_i(t)$ & Average total synaptic input current of population $X$ ($n_\text{av}=100$) \\
    $\overline{V}_X(t)$ & $\frac{1}{n_\text{av}}\sum \limits_{i\in X}^{n_\text{av}}  V_i(t)$ & Average membrane voltage of population $X$ ($n_\text{av}=100$) \\
    $\nu_X(t)$ & $\frac{n^\text{s}_{X}(t)}{t_\text{bin}}$ & Instantaneous population ($X$) firing rate, with $n^\text{s}_{X}(t)$ 
    being the number of spikes in $[t,t+t_{\text{bin}})$ of all cells in population $X$, $t_{\text{bin}}=1$ ms \\

    $\overline{\nu}_X(t)$ & $\frac{\nu_X(t)}{N_X}$ & Average instantaneous firing rate of population $X$ \\

    $\phi_X(\mathbf{r},t)$ & $\sum_{i\in X} \phi_i(\mathbf{r},t)$ (cf. \Fref{eq:extracellular4} \& \Fref{sec:2.4})  & Population LFP of population $X$  \\
    $\rho_X(\mathbf{r},t)$ &  $\sum_{i\in X} \rho_i(\mathbf{r},t)$ (cf. \Fref{eq:csd} \& \Fref{sec:2.5.1}) & Population CSD of population $X$  \\
    $\phi(\mathbf{r},t)$ & $\sum_{X} \phi_X(\mathbf{r},t)$   & Compound LFP of all cells  \\
    $\rho(\mathbf{r},t)$ &  $\sum_{X} \rho_X(\mathbf{r},t)$   & Compound CSD of all cells  \\\hline 
    \parameterhdr{3}{}{Rescaled signals  }  \\\hline 
    $\phi^{\gamma}(\mathbf{r},t)$ & 
    $\sum_{X}\sum_{i\in X^\prime \subset X} \phi_i(\mathbf{r},t)$
    & Compound LFP signal from subset of neurons $i \in X^\prime \subset X$ 
      with $N_{X^\prime}=\gamma  N_X$, $\gamma \in [0,1]$ \\
    $\phi^{\gamma \xi}(\mathbf{r},t)$ & 
    $\xi  \phi^{\gamma}$$(\mathbf{r},t)~\refstepcounter{equation}(\theequation)\label{eq:phi_gamma_xi}$
    & Compound LFP signal from subset of neurons $i \in X^\prime \subset X$ 
      with $N_{X^\prime}=\gamma  N_X$, $\gamma \in [0,1]$, rescaled by factor $\xi$ \\\hline
  \end{tabularx}
  \caption{Measurements and derived signals obtained from the hybrid scheme for LFP simulations.
  }
  \label{tab:9}
\end{table}


\begin{table}[!htbp]
\small
  \begin{tabularx}{\linewidth}{|p{0.15\linewidth}|p{0.45\linewidth}|X|}
    \hline\modelhdr{3}{}{Data analysis}\\\hline
    \bf Symbol & \bf Definition & \bf Description \\\hline 
    $\psi,\psi^\prime$ & $\psi,\psi^\prime \in \{\phi_i,\phi_X,\phi,\rho_i,\rho_X,\rho,\nu_X\}$  & Signal (LFPs, CSDs, firing rates) \\
    $\mu_{\psi}(\mathbf{r})$ & $\frac{dt}{T}\sum_{h=1}^{T/dt} \psi(\mathbf{r},h\,dt) $ & Temporal mean of signal $\psi(\mathbf{r},t)$\\
    $\mathrm{cov}_{\psi\psi^\prime}(\mathbf{r})$ & $\frac{dt}{T}\sum_{h=1}^{T/dt} \psi(\mathbf{r},h\,dt)\psi^\prime(\mathbf{r},h\,dt) - \mu_{\psi}(\mathbf{r})\mu_{\psi^\prime}(\mathbf{r})$& Temporal covariance of signals $\psi(\mathbf{r},t),\psi^\prime(\mathbf{r},t)$ \\
    $\sigma^2_{\psi}(\mathbf{r})$ & $\mathrm{cov}_{\psi\psi}(\mathbf{r})$& Temporal variance of signal $\psi(\mathbf{r},t)$ \\
    $cc_{\psi\psi^\prime}(\mathbf{r})$ & 
$ \mathrm{cov}_{\psi\psi^\prime}(\mathbf{r})/\sqrt{\sigma^2_{\psi}(\mathbf{r}) \sigma^2_{\psi^\prime}(\mathbf{r})}\hfill\refstepcounter{equation}(\theequation)\label{eq:cc}$
& Zero time-lag correlation coefficient of signals  $\psi(\mathbf{r},t),\psi^\prime(\mathbf{r},t)$  \\
    
    $C_{\psi\psi^\prime}(\mathbf{r},f)$ & $\mathcal{F}[\psi](\mathbf{r},f)^* \mathcal{F}[\psi^\prime](\mathbf{r},f)$ (implemented using Welch's method) 
    & Pairwise cross-spectral density of signals $\psi(\mathbf{r},t),\psi^\prime(\mathbf{r},t)$, $\mathcal{F}[\psi]$: Fourier transform of $\psi$  \\
    $P_{\psi}(\mathbf{r},f)$ & $C_{\psi\psi}(\mathbf{r},f)$ 
    & Power spectral density (PSD) of signal $\psi(\mathbf{r},t)$ \\
    $\overline{P_{\phi}}(\mathbf{r},f)$  & $\frac{1}{N}\sum \limits_{i=1}^N P_{\phi_i}(\mathbf{r},f)\hfill\refstepcounter{equation}(\theequation)\label{eq:psd_mean_i}$ 
    & Average single-cell LFP power spectrum \\
    $\overline{C_{\phi}}(\mathbf{r},f)$ & $\frac{1}{N(N-1)}\sum\limits_{i=1}^N\sum \limits_{\substack{j=1 \\ j\neq i}}^{N} 
    C_{\phi_i \phi_j}(\mathbf{r},f)~\hfill\refstepcounter{equation}(\theequation)\label{eq:mean-cross-spectrum}$ & 
    Average cross-spectrum between single-cell LFPs \\
    $\overline{\kappa_{\phi}}(\mathbf{r},f)$ & 
    $ \overline{C_{\phi}}(\mathbf{r},f)/\overline{P_{\phi}}(\mathbf{r},f)~\hfill\refstepcounter{equation}(\theequation)\label{eq:coherence}$
    & Average LFP coherence between cells \\
    $P_{\phi}(\mathbf{r},f)$ & 
    $N \overline{P_{\phi}}(\mathbf{r},f)+ N(N-1)\overline{C_{\phi}}(\mathbf{r},f)\hfill\refstepcounter{equation}(\theequation)\label{eq:psd}$
    &  Power spectral density (PSD) of the compound LFP\\
    $P^{0}_{\phi}(\mathbf{r},f)$ &  
     $N \overline{P_{\phi}}(\mathbf{r},f)\hfill\refstepcounter{equation}(\theequation)\label{eq:psd_uncorrelated}$
    & Power spectral density (PSD) of the compound LFP signal omitting pairwise cross-correlations \\ \hline
  \end{tabularx}
  \vspace{\tabspace}
  \caption{Data analysis of model output signals. 
  }
  \label{tab:10}
\end{table}


\begin{table}[!htbp]
\small
  \begin{tabularx}{\linewidth}{|p{0.15\linewidth}|p{0.25\linewidth}|X|}
    \hline\modelhdr{3}{}{Post-processing}\\\hline
    \parameterhdr{3}{}{Parameters}  \\\hline  
    \bf Symbol & \bf Value & \bf Description \\\hline 
    $T_{\text{trans}}$ & $200\,$ms & Start-up transient  \\
    $dt_{\psi}$ & $1\,$ms & Signal resolution \\\hline
    \parameterhdr{3}{}{Power spectral density settings}  \\\hline  
    \bf Method & \texttt{plt.mlab.psd$^\ast$} & Welch's average periodogram \\ \hline
    \bf Symbol & \bf Value & \bf Description \\\hline 
    $T_\psi$ & $5{,}000\,$ms & Signal length \\
    NFFT & $256$  & Number of data points used in each block for the FFT \\
    Fs & $1\,$kHz & Sampling rate \\
    noverlap & $128$ & Number of overlapping data points between blocks   \\
    window & \texttt{plt.mlab.window\_hanning$^\ast$} & Window filter (\texttt{$^\ast$~plt} denote \texttt{matplotlib}) \\\hline
  \end{tabularx}
  \caption{Post-processing and data analysis parameters.}
  \label{tab:11}
\end{table}

\end{document}